\documentclass[12pt
]{article}

\usepackage{graphicx}
\usepackage{amsfonts}
\usepackage{amsmath}
\usepackage[usenames]{color}
\usepackage{amssymb}
\usepackage[mathscr]{eucal} 

\usepackage[all]{xy}

\def\Z{\mathbb{Z}}
\def\Q{\mathbb{Q}}
\def\R{\mathbb{R}}
\def\C{\mathbb{C}}
\def\P{\mathbb{P}}

\begin{document}
\baselineskip 0.6cm
\newcommand{\vev}[1]{ \left\langle {#1} \right\rangle }
\newcommand{\bra}[1]{ \langle {#1} | }
\newcommand{\ket}[1]{ | {#1} \rangle }
\newcommand{\Dsl}{\mbox{\ooalign{\hfil/\hfil\crcr$D$}}}
\newcommand{\nequiv}{\mbox{\ooalign{\hfil/\hfil\crcr$\equiv$}}}
\newcommand{\nsupset}{\mbox{\ooalign{\hfil/\hfil\crcr$\supset$}}}
\newcommand{\nni}{\mbox{\ooalign{\hfil/\hfil\crcr$\ni$}}}
\newcommand{\EV}{ {\rm eV} }
\newcommand{\KEV}{ {\rm keV} }
\newcommand{\MEV}{ {\rm MeV} }
\newcommand{\GEV}{ {\rm GeV} }
\newcommand{\TEV}{ {\rm TeV} }

\def\diag{\mathop{\rm diag}\nolimits}
\def\tr{\mathop{\rm tr}}

\def\Spin{\mathop{\rm Spin}}
\def\SO{\mathop{\rm SO}}
\def\O{\mathop{\rm O}}
\def\SU{\mathop{\rm SU}}
\def\U{\mathop{\rm U}}
\def\Sp{\mathop{\rm Sp}}
\def\SL{\mathop{\rm SL}}

\def\change#1#2{{\color{blue}#1}{\color{red} [#2]}\color{black}\hbox{}}


\begin{titlepage}

\begin{flushright}
UT-08-13\\
LTH 791 \\
IPMU 08-0024
\end{flushright}

\vskip 1cm
\begin{center}

{\large \bf New Aspects of Heterotic--F Theory Duality}
 
\vskip 1.2cm

Hirotaka Hayashi$^1$, Radu Tatar$^2$, \\
 Yukinobu Toda$^3$, Taizan Watari$^{1,3}$ and Masahito Yamazaki$^1$

\vskip 0.4cm
{\it $^1$Department of Physics, University of Tokyo, Tokyo 113-0033, Japan  
\\[2mm]

$^2$Division of Theoretical Physics, Department of Mathematical
 Sciences, The University of Liverpool, Liverpool, L69 3BX, England,
 U.K. \\[2mm]

$^3$Institute for the Physics and Mathematics of the Universe, University of Tokyo, Kashiwano-ha 5-1-5, 277-8592, Japan
}
\vskip 1.5cm

\abstract{
In order to understand both up-type and down-type Yukawa 
couplings, F-theory is a better framework than the 
perturbative Type IIB string theory.
The duality between the Heterotic and F-theory is a powerful 
tool in gaining more insights into F-theory description of 
low-energy chiral multiplets. 
Because chiral multiplets from bundles $\wedge^2 V$ and 
$\wedge^2 V^\times$ as well as those from a bundle $V$ are all 
involved in Yukawa couplings in Heterotic compactification, 
we need to translate descriptions of all those kinds of 
matter multiplets into F-theory language through the duality.
We find that chiral matter multiplets in F-theory are global 
holomorphic sections of line bundles on what we call 
covering matter curves. The covering matter curves are 
formulated in Heterotic theory in association with normalization 
of spectral surface, while they are where $M2$-branes wrapped 
on a vanishing two-cycle propagate in F-theory. 
Chirality formulae are given purely in terms of 
primitive four-form flux.
In order to complete the translation, the dictionary of the 
Heterotic--F theory duality has to be refined in some aspects.
A precise map of spectral surface and complex structure moduli 
is obtained,  
and with the map, we find that divisors specifying the line bundles 
correspond precisely to codimension-3 singularities in F-theory. 
 } 

\end{center}
\end{titlepage}


\tableofcontents

\section{Introduction}

For a description of our real world, we need both up-type quark 
Yukawa couplings 
\begin{equation}
 \Delta W = {\bf 10}^{ab} \; {\bf 10}^{cd} \; H({\bf 5})^e \; 
  \epsilon_{abcde}
\end{equation}
and down-type quark (and charged lepton) Yukawa couplings 
\begin{equation}
 \Delta W = \bar{\bf 5}_a \; {\bf 10}^{ab} \; \bar{H}(\bar{\bf 5})_b . 
\end{equation}
Here, we used a notation of effective theory with 
$\SU(5)_{\rm GUT}$ symmetry and ${\cal N} = 1$ supersymmetry. 
Perturbative super Yang--Mills interactions of open strings 
of Type IIA / IIB string theory may be able to give rise 
to the latter, but it is difficult to generated 
the up-type Yukawa couplings with $\SU(5)_{\rm GUT}$ indices 
contracted by an epsilon tensor.\footnote{Theories without $\SU(5)_{\rm GUT}$ 
unification do not have this problem. It should be remembered, 
however, that Supersymmetric Standard Models without unification 
would need an extra explanation for apparent gauge coupling 
unification, and Pati--Salam type theories need some mechanism 
to make sure that quark doublets and lepton doublets have totally 
different electroweak mixing, although they belong to the same 
irreducible representation of the Pati--Salam gauge group.} 
Heterotic $E_8 \times E_8'$ string theory, 
$G_2$-holonomy compactification of 11-dimensional supergravity 
and F-theory compactification, however, are capable of generating 
both types of Yukawa couplings \cite{TW1}. 

There are some motivations to develop theoretical tools 
to extract physics out of $G_2$-holonomy compactification of 11-dimensional 
supergravity or Calabi--Yau four-fold compactification of F-theory. 
\begin{itemize}
 \item Perturbative Heterotic $E_8\times E_8'$ string theory 
predicts all of the GUT scale, the Planck scale and the value of 
unified gauge coupling constant \cite{witten-strong}, but not 
all of them turn out right within the perturbative regime. 
The dilaton expectation value should 
be in the strongly-coupled regime to fit the data.
Certainly the Heterotic M-theory \cite{horavawitten} can cover 
strong-coupling region of the moduli space of the Heterotic 
$E_8 \times E'_8$ string theory, but that is not the only 
possibility. F-theory, for example, also describes some 
parts of the strong coupling region of the moduli space 
of the Heterotic string theory \cite{Vafa-Het-F}.
 \item As long as we insist that both the up-type and down-type 
       Yukawa couplings be obtained, we do not gain much freedom 
       by replacing the Heterotic $E_8 \times E'_8$ string theory 
       by $G_2$-holonomy compactification of eleven-dimensional
       supergravity or elliptic Calabi--Yau four-fold compactification
       of F-theory. In order to generate both types of Yukawa couplings, 
       an underlying gauge symmetry of $E_r$ ($r=7,8$) 
       is necessary \cite{TW1}. Thus, there may be not much room to 
       expect qualitatively different physics of quarks and leptons  
       in vacua obtained by 11-dimensional supergravity or F-theory. 
       Even at the qualitative level, however, one obtains greater 
       freedom in constructing other sectors of the real world.
       It will be much easier in F-theory than in the Heterotic 
       theory to construct models of gauge mediated supersymmetry 
       breaking, for example. 
 \item Flux compactification techniques \cite{flux} can be used 
       to discuss observables in our visible sector in F-theory. 
       In Heterotic string theory, it is really hard for now 
       to discuss stabilization of vector bundle moduli. 
       In Heterotic--F-theory duality, vector bundle moduli of 
       Heterotic theory correspond to a part of complex structure 
       moduli of F-theory \cite{MV1, MV2}, and flux compactification 
       of F-theory will stabilize such moduli as well as the rest of 
       the complex structure moduli.
\end{itemize}
These motivations provide enough reasons to study F-theory, although  
they are not particularly in favor of F-theory over $G_2$-holonomy 
compactification of eleven-dimensional supergravity.
Reference \cite{TW1} discussed qualitative pattern of Yukawa matrices expected in 
$G_{2}$-holonomy compactification with unbroken $SU(5)_{{\rm GUT}}$
symmetry, and found that there is generically a problem in the texture of 
up-type Yukawa matrix. This situation adds a motivation to develop a 
formulation to study Yukawa couplings in F-theory.

The most basic question in string phenomenology is who we 
are---what are quarks and leptons. These matter chiral multiplets 
in supersymmetric compactification are identified with independent 
elements of bundle-valued cohomology groups in the Heterotic, 
Type I and Type IIB string theory. The net chirality of matter multiplets
in chiral representations is expressed in terms of topological 
numbers such as Euler characteristics of vector bundles or 
pairing of D-brane charges in K-theory.
In Type IIA string compactification on a Calabi--Yau orientifold, 
we know that one chiral multiplet is localized at each 
D6--D6 intersection \cite{BDL}, and this local picture is extended 
to compactification of 11-dimensional supergravity on a $G_2$-holonomy 
``manifold'' with $A$--$D$--$E$ singularity \cite{Acharya-Witten}. 
We can provide satisfactory answers to the question above in all these theories. 
Surprisingly, though, such an effort to identify quarks and 
leptons in F-theory language went only halfway in 1990's, 
and almost came to a halt (at least to our knowledge), until 
the recent results of \cite{DW,BHV}. 

The identification of quarks and leptons, or of chiral matter multiplets in
general, has been such a challenging problem in F-theory, because 
an intrinsic formulation of F-theory has not been fully 
developed yet. The elementary degrees of freedom in F-theory 
can be described by $(p,q)$ strings or $M2$-branes of 
11-dimensional supergravity. It may be possible,  
to identify chiral matter multiplets on 
3+1 dimensions with some of their fluctuation modes. 
In practice, however, it is extremely difficult to 
disentangle complicated geometry of triple intersection  
of $(p,q)$ 7-branes, or to maintain distinction between 
left-handed and right-handed fermions in Calabi--Yau 
4-fold compactification of 11-dimensional supergravity 
down to 2+1 dimensions. Instead, the duality between 
the Heterotic string and F-theory \cite{
Vafa-Het-F,MV1,MV2,FMW,6authors,Het-F-4D, CD}
will be the most powerful tool in studying F-theory. 
 
This article is along the line of this approach; 
the Heterotic string theory and the Heterotic--F-theory duality 
are used to study F-theory. 
The Heterotic string theory compactified on 
an elliptically fibered Calabi--Yau 3-fold 
$\pi_Z: Z \rightarrow B_2$ is dual to F-theory compactified 
on an elliptically fibered Calabi--Yau 4-fold $\pi_X: X \rightarrow B_3$
whose base 3-fold $B_3$ is a $\P^1$ fibration over $B_2$.
The various matter multiplets in low-energy effective theory 
are identified with $H^1(Z; \rho(V))$ in Heterotic string 
description, where $\rho(V)$ is a vector bundle $V$ 
in representation $\rho$. Cohomology groups on a fibered space
can be calculated first on the fiber geometry, and later 
on the base geometry; except for certain cases (which will be covered 
in section~\ref{sec:trivial}), 
\begin{equation}
 H^1(Z; \rho(V)) \simeq H^0(B_2; R^1\pi_Z{*} \rho(V)), 
\end{equation}
and the direct images $R^1\pi_{Z*} \rho(V)$ have their 
support only on curves in $B_2$. In the Heterotic--F duality, 
these support curves correspond to 7-brane intersections, 
and the sheaves on the curves should be those on the 7-brane 
intersection curves. Chiral matter multiplets are identified 
with global holomorphic sections of such sheaves (except for 
certain cases). Thus, by using the Heterotic--F duality, 
we can obtain the sheaves whose sections are identified 
with quarks and leptons. 
Direct images $R^1\pi_{Z*} \rho(V)$ are, therefore, the 
information we would like to obtain from the Heterotic 
string theory.

Direct images of bundles in the fundamental representation 
$\rho(V) = V$ were obtained in 1990's \cite{Curio, DI}. 
Those of bundles in the anti-symmetric representation 
$\rho(V) = \wedge^2 V$ have not been clearly described 
as sheaves so far in the last decade, apart from some 
developments in \cite{Penn5, BMRW} in the context of Heterotic theory 
compactification. Calculation of the direct images of $\wedge^2 V$, 
therefore, is one of the central themes in this article.
This task is carried out in sections \ref{sec:Idea} and \ref{sec:Examples}.
This is by no means a minor problem. Both $\bar{\bf 5}$ and 
$\bar{H}(\bar{\bf 5})$ multiplets arise from $\wedge^2 V$ of 
an $\SU(5)$ bundle $V$, and $H({\bf 5})$ from $\wedge^2 V^\times$, 
where $V^\times$ is the dual bundle of $V$. Without understanding 
the geometry associated with $R^1\pi_{Z*} \wedge^2 V$ and 
$R^1\pi_{Z*} \wedge^2 V^\times$, there is no way to understand 
the Yukawa couplings of quarks and leptons in F-theory.

We introduce a new notion,\footnote{Essentially the same object 
was already introduced in \cite{BMRW}.} covering matter curve, 
(roughly speaking) in order to deal with singularities that appear 
along matter curves.
The direct image $R^1\pi_{Z*} \wedge^2 V$ is represented as 
a pushforward of a locally free rank-1 sheaf 
$\widetilde{\cal F}_{\wedge^2 V}$ on the covering matter 
curve for all the cases we have study in section~\ref{sec:Examples}, 
i.e. for ${\rm rank} \; V = 3,4,5,6$. (We should also note here that 
a minor assumption is made on structure of $R^1\pi_{Z*} \wedge^2 V$
around a particular type of singularity for the ${\rm rank} \; V = 4$
case.) Divisors determining the locally free rank-1 sheaves 
are determined in terms of data defining spectral surfaces.

In section~\ref{sec:Het2F}, the description of the sheaves 
$R^1\pi_{Z*} \rho(V)$ are translated into language of F-theory.
The dictionary between the Heterotic and F-theory quantities
was almost established in 1990's, but it 
has to be refined in some aspects. Improvements include 
\begin{itemize}
 \item a precise map (\ref{eq:dict-f0}--\ref{eq:dict-g6}) 
between the moduli of spectral surface in Heterotic theory 
description and complex structure moduli in F-theory. 
 \item a refinement of the correspondence between the 
discrete twisting data of vector bundles in Heterotic theory 
description and four-form fluxes in F-theory description. 
We are basically following the line of the idea laid out 
in \cite{CD, DW}. 
\end{itemize}
Using the precise map between the two moduli space, 
we find that most of the components of the divisors 
determining the sheaves ${\cal F}_{\rho(V)}$ (and all 
that we identified) correspond to codimension-3 singularities 
in $B_3$.

The last section provides our conclusion and describes chiral matter multiplets 
in effective theory purely in a language of F-theory.
All the results obtained in earlier sections are put together 
and provide a description of matter multiplets in a form 
that does not even assume an existence of Heterotic dual. 
Brief comments on Yukawa couplings are also found there. 

Appendices \ref{sec:push} and \ref{sec:app2Examples} 
cover mathematical subjects that are necessary 
in sections~\ref{sec:Idea} and \ref{sec:Examples}, respectively.

\section{Spectral Cover Construction and Direct Images}
\label{sec:review}

Heterotic string theory compactified on an elliptic 
Calabi--Yau 3-fold is dual to F-theory compactified on 
$K3$-fibered elliptic Calabi--Yau 4-fold. Once massless 
chiral multiplets are described in Heterotic string theory, 
the description can be passed on to F-theory, using 
the duality. In this section, we will review a powerful 
way to describe them that is known since late 1990's, 
mainly for the purpose of setting up notations used in this article. 

{\bf Elliptic Fibration}

Heterotic string theory has an F-theory dual description, if 
it is compactified on an elliptically fibered manifold. 
We consider an elliptic fibered Calabi--Yau 3-fold $Z$
\begin{equation}
 \pi_Z : Z \rightarrow B_2
\end{equation}
over a base 2-fold, so that ${\cal N} = 1$ supersymmetry is left 
in low-energy effective theory below the Kaluza--Klein scale.
An elliptic fibration $Z$ over $B_2$ is given by a Weierstrass 
equation,
\begin{equation}
 y^2 = x^3 + f_0 x + g_0. 
\label{eq:Weierstrass-Het}
\end{equation}
Here, $f_0$ and $g_0$ are sections of line bundles 
${\cal L}_H^{\otimes 4}$ and ${\cal L}_H^{\otimes 6}$ on $B_2$, 
respectively, 
and ${\cal L}_H \simeq {\cal O}(- K_{B_2})$ for $Z$ to be a 
Calabi--Yau 3-fold. The coordinates $(x,y)$  transform as sections of 
${\cal L}_H^{\otimes 2}$ and ${\cal L}_H^{\otimes 3}$, respectively.
The zero section $\sigma: B_2 \hookrightarrow Z$ maps $B_2$ to the locus 
of infinity points, $(x,y) = (\infty, \infty)$.

{\bf Spectral Cover Construction}

Compactification of the Heterotic $E_8 \times E'_8$ string theory 
involves a pair of vector bundles $(V_0, V_\infty)$ 
on a Calabi--Yau 3-fold $Z$. 
Spectral cover construction \cite{Donagi-taniguchi,FMW,FMW2} describes 
vector bundles on an elliptic fibered Calabi-Yau 3-fold $Z$. 
Let us consider a rank-$N$ vector bundle $V$ on $Z$. 
Spectral surface $C_V \in | N \sigma + \pi_Z ^* \eta|$ is 
a smooth hypersurface of $Z$ that is a degree $N$ cover over $B_2$, 
where $\eta$ is a divisor on $B_2$.
When a line bundle ${\cal N}_V$ on $C_V$ is given, 
a rank-$N$ vector bundle $V$ is given by the Fourier--Mukai transform
\begin{equation}
V=p_{2*} (p^*_1({\cal N}_V) \otimes {\cal P}_{B_2}),
\label{eq:FM-transf} 
\end{equation}
where $p_{1,2}$ are maps associated with a fiber product 
\begin{equation}
\vcenter{\xymatrix{
 & C_V \times_{B_2} Z \ar[dl]_{p_1} \ar[dr]^{p_2}&  \\
 C_V \ar[dr]_{\pi_C} & & Z \ar[dl]^{\pi_Z} \\
 & B_2 & 
}}
\label{eq:fib-prod}
\end{equation}
$q : = \pi_C \; \circ \; p_1 = \pi_Z \; \circ \; p_2$, and 
${\cal P}_{B_2}$ is the Poincar\'e line bundle 
${\cal O}_{C_V \times_{B_2} Z}(\Delta - \sigma_1 -  \sigma_2 + q^*
K_{B_2})$ with $\sigma_1=\sigma \times Z, \sigma_2=Z\times \sigma$ and 
$\Delta$ is a diagonal divisor of $Z \times Z$ restricted on 
$C_V \times_{B_2} Z$. 
The data $(C_V, {\cal N}_V)$, i.e. the spectral surface and a line bundle on it, 
determines a vector bundle $V$.

The characteristic classes of vector bundles constructed that way 
are expressed in terms of spectral data $(C_V, {\cal N}_V)$.
The first Chern class of the vector bundle $V$ is given by \cite{FMW}
\begin{equation}
 c_1(V)  =  \pi_Z^* \pi_{C*}\left( c_1({\cal N}_V) -  \frac{1}{2} r \right)
\label{eq:c1}
\end{equation}
where $r : = \omega_{C/{B_2}} : = K_{C_V} - \pi_C^* K_{B_2}$ is the 
ramification divisor on $C_V$ of $\pi_C: C_V \rightarrow B_2$, 
and $c_1(V)$ is a pullback of a 2-form on the base 2-fold $B_2$. 
With the notation 
\begin{equation}
 \gamma : = c_1({\cal N}_V)-\frac{1}{2} r, 
\label{eq:defgamma}
\end{equation} 
we have 
\begin{equation}
 c_1(V)=\pi^*_Z \pi_{C*} \gamma. 
\label{eq:c1V}
\end{equation}
We will sometimes use $c_1(V)$ 
in the sense of $\pi_{C*} \gamma$.
The second Chern character is 
\begin{equation}
 {\rm ch}_2(V) = - \sigma \cdot \eta + \pi_Z^* \omega,
\end{equation}
where $\omega$ is some 4-form on $B_2$ \cite{FMW}. 

We do not restrict our attention to cases with vanishing 
first Chern class $c_1(V)$. By considering vector bundles $V$ 
whose structure group is $\U(N)$, rather than $\SU(N)$, we will 
be able to perform a consistency check in calculating 
$R^1 \pi_{Z*} (\wedge^2 V)$ by examining $c_1(V)$ dependence. 
We maintain our discussion to be valid for $\U(N)$ bundles 
also because there are some phenomenological motivations 
to think of Heterotic string compactification with a bundle 
whose structure group is within 
$\U(N_1) \times \U(N_2) \subset \SU(5)$ \cite{TW1}.  

We now present a few technical remarks about the nature 
of vector bundles given by spectral cover construction. 
Such bundles cannot be completely generic $\U(N)$ bundles.
For example, the first Chern class $c_1(V) = c_1 ({\rm det}\; V)$ 
is always given by a pullback of a 2-form on $B_2$ to $Z$ (see \eqref{eq:c1V}). 
In other words, the first Chern class of ${\rm det}\; V$ is trivial 
in the fiber direction.
This is not a serious limitation when we are analysing Heterotic 
compactification in an attempt to understand F-theory better. 
In Heterotic string compactification, vector bundles have to be 
stable, and the stability condition (Donaldson--Uhlenbeck--Yau equation) is
\begin{equation}
 \int_Z c_1(V) \wedge J \wedge J = 0
\label{eq:UY}
\end{equation}
at tree level, where $J$ is the K\"{a}hler form of $Z$.
When the $T^2$-fiber is small, description in the Heterotic theory 
becomes less reliable, but a dual F-theory description becomes better. 
This is the situation we are interested in. 
In such a limit, the size of $T^2$-fiber becomes much smaller than that of the base, and the dominant contribution of (\ref{eq:UY}) is from 
two $J$'s in the two directions along $B_2$, and $c_1(V)$ in the fiber 
direction. Thus, the sole dominant contribution has to vanish, and hence
stable vector bundles should not have non-vanishing $c_1(V)$ along the 
fiber direction.\footnote{As long as $\int_{B_2} \pi_{C*}\gamma \wedge J = 0$ 
is satisfied on $B_2$, vector bundles with non-vanishing $c_1(V)$ 
can be stable.} 
Spectral cover construction, therefore, is fine for our purpose 
in this article, although it cannot describe a bundle 
with a non-vanishing first Chern class along the $T^2$-fiber direction. 

For $\U(N)$ bundles given by spectral cover construction,  
${\rm det} \; V$ are actually trivial along the elliptic fiber 
direction, not just degree zero.
The spectral surface $C_V \hookrightarrow Z$ is (on a local patch of
$B_2$) defined by the zero locus of an equation 
\begin{equation}
s = a_0 + a_2 x + a_3 y + a_4 x^2 + a_5 x y + \cdots + 
 a_N \left(x^{N/2} {\rm ~or~} x^{(N-3)/2} y \right)= 0, 
\label{eq:Cv-def-eq}
\end{equation}
where $a_r$ are sections of 
${\cal O}(\eta) \otimes {\cal L}_H^{\otimes (-r)} \simeq 
{\cal O}(r K_{B_2} + \eta)$ on $B_2$. 
The last term is $x^{N/2}$ or $x^{(N-3)/2} y$ depending on 
whether $N$ is even or odd. 
On a given fiber $E_b := \pi_Z^{-1}(b)$, $s$ determines 
an elliptic function, with $N$ zero points $\{ p_i\}_{i = 1, \cdots, N}$ 
(for $\U(N)$ bundles) and a rank-$N$ pole at $e_0$, 
zero section $\sigma$ on $E_b$.
Since the group-law sum of the zero points of an elliptic function 
is the same as that of the poles, 
\begin{equation}
 \boxplus_i p_i = e_0, 
\label{eq:traceless-Wilson}
\end{equation}
where $\boxplus$ stands for the summation according to the group 
law of an elliptic curve.

{\bf Direct Images and Matter Curves}

If an $\SU(N)$ vector bundle $V$ is turned on within one of 
$E_8$ gauge group of the Heterotic $E_8 \times E'_8$ string 
theory, symmetry group is reduced to $H \subset E_8$ that commutes 
with the $\SU(N)$ in effective theory below the Kaluza--Klein scale. 
The chiral multiplets in low-energy effective theory are identified 
with $H^1(Z; \rho(V))$. The correspondence between the representations $\rho(V)$
of $V$ and those of the unbroken symmetry group $H$
is summarized in Table~\ref{tab:Het-matter}. 
\begin{table}[t]
\begin{center}
  \begin{tabular}{c|c|c|c|c|c}
 structure group of $V$ & $\SU(2)$ & $\SU(3)$ & $\SU(4)$ &
  $\SU(5)_{\rm bdl}$ & $\SU(6)$  \\
\hline
 unbroken symmetry $H$ & $E_7$ & $E_6$ & $\SO(10)$ & $\SU(5)_{\rm GUT}$ & 
                 $\SU(3) \times \SU(2)$ \\
\hline
 from $V$ & 
 {\bf 56} & {\bf 27} & {\bf 16} & {\bf 10} & $(\bar{\bf 3}, {\bf 2})$ \\
 from $V^\times$ & 
 (vct.-like) & $\overline{{\bf 27}}$ & $\overline{{\bf 16}}$ & 
 $\overline{{\bf 10}}$ & $({\bf 3}, {\bf 2})$ \\ 
 from $\wedge^2 V$ &
  --- & --- & {\bf 10} & $\bar{\bf 5}$ & $({\bf 3},{\bf 1})$ \\
 from $\wedge^2 V^\times$ & 
  --- & --- & (vct.-like) & ${\bf 5}$ & $(\bar{\bf 3},{\bf 1})$ \\
 from $\wedge^3 V$ & --- & --- & --- & --- & $({\bf 1},{\bf 2})$ \\
 from $\pi^*_Z E$ & {\bf adj.} & {\bf adj.} & {\bf adj.} & {\bf adj.} 
& {\bf adj.} 
 \end{tabular}
\caption{\label{tab:Het-matter} If an $\SU(N)$ vector bundle $V$ 
is turned on within an $E_8$ gauge group, $E_8$ symmetry is broken down 
to $H$, and chiral matter multiplets come out from various irreducible 
components of $E_8$-${\bf adj.}$ decomposed under $\SU(N) \times H$. 
Irreducible components $(\rho(V), repr.)$ are denoted by $repr.$ 
in this table. ``(vct.-like)'' in this table indicates that a given 
$(\rho(V)^\times, repr.)$ is the same as $(\rho(V),repr.)$ and self 
Hermitian conjugate in $E_8$ adjoint representation. 
The symmetry $H$ may be further broken down by turning on a bundle 
$E$ on the base manifold $B_2$. The structure group of $E$ can be 
chosen, for example, in the $\U(1)_\chi$ direction in $H=\SO(10)$
so that the symmetry is broken down to $\SU(5)_{\rm GUT}$, or 
in the $\U(1)_Y$ direction in $H=\SU(5)_{\rm GUT}$ in order to 
break the $\SU(5)_{\rm GUT}$ unified symmetry to 
$\SU(3)_C \times \SU(2)_L$ possibly with $\U(1)_Y$ as well.
Matter multiplets from $\pi_Z^* E$ are characterized as cohomology 
groups on 4-cycles (7-branes) in F-theory, 
while those from $V$, $\wedge^2 V$ and $\wedge^3 V$ are as cohomology 
on 2-cycles (intersection of 7-branes) in F-theory.}
\end{center}
\end{table}

For a Calabi--Yau 3-fold $Z$ that is an elliptic fibration 
over a 2-fold $B_2$, cohomology groups $H^1(Z; \rho(V))$ can 
be calculated by Leray spectral sequence. One calculates the cohomology 
in the fiber direction first, $R^i\pi_{Z*} \rho(V)$ ($i=0,1$), 
and then the cohomology in the base directions. If $R^0\pi_{Z*} \rho(V)$ 
vanishes everywhere on $B_2$, which is often the case, then 
\begin{equation}
 H^1(Z; \rho(V)) \simeq H^0(B_2; R^1\pi_{Z*} \rho(V)).
\end{equation}
If one is interested only in the net chirality, i.e. the difference between the number of chiral multiplets and anti-chiral 
multiplets in a given representation,
\begin{eqnarray}
\chi(\rho(V)) & := & h^1(Z; \rho(V)) - h^1(Z; \rho(V)^\times),
                                            \nonumber \\
              & = & h^1(Z; \rho(V)) - h^2(Z; \rho(V)), \\
              & = & - \chi(\rho(V)^\times), \nonumber
\label{eq:chi-def}
\end{eqnarray}
then one has 
\begin{eqnarray}
 \chi(\rho(V)) & = & - \chi(Z; \rho(V)), \nonumber \\
               & = & - \chi(B_2; R^0\pi_{Z*} \rho(V)) 
                     + \chi(B_2; R^1\pi_{Z*} \rho(V)), \\
               & \rightarrow & \chi(B_2; R^1\pi_{Z*} \rho(V)) \qquad
		\qquad ({\rm if~} R^0\pi_{Z*} \rho(V) = 0).
\end{eqnarray}

Suppose that the vector bundle $V$ is given by spectral cover
construction from $(C_V, {\cal N}_V)$. Let us consider a 
Fourier--Mukai transform of $\rho(V)$:
\begin{equation}
 R^1p_{1*} \left[ p_2^* (\rho(V)) \otimes {\cal P}_B^{-1} \otimes {\cal O}(-q^* K_{B_2})  \right], 
\label{eq:FM-rho(V)}
\end{equation}
which is a sheaf on $Z$, and $p_1$ and $p_2$ here are maps in 
\begin{equation}
\vcenter{\xymatrix{
 & Z \times_{B_2} Z \ar[dl]_{p_1} \ar[dr]^{p_2}&  \\
 Z \ar[dr]_{\pi_Z} & & Z \ar[dl]^{\pi_Z} \\
 & B_2 & 
}}
\label{eq:fib-prod-ZZ}
\end{equation}
and $q = \pi_Z \circ p_1 = \pi_Z \circ p_2$. This sheaf 
is supported only on a codimension-1 subvariety $C_{\rho(V)}$. 
Unless $C_{\rho(V)}$ contains the zero section $\sigma$ as 
an irreducible component, $\rho(V)$ does not contain a trivial 
bundle when it is restricted on a fiber $E_b$ of a generic point 
$b\in B_2$. Thus, $R^1 \pi_Z{*} \rho(V)$ vanishes on a generic 
point on $B_2$; it survives  only along a curve 
\begin{equation}
 \bar{c}_{\rho(V)} = C_{\rho(V)} \cdot \sigma.
\end{equation}
in $B_2$ (see also the appendix \ref{sec:push}). 
Curves $\bar{c}_{\rho(V)}$ for various representations 
$\rho(V)$ are called matter curves, because cohomology groups 
are localized.

The localization of cohomology groups (or matter multiplets that appear 
in low-energy effective theory) on matter curves is not just 
an artifact of mathematical calculation. It also has physics meaning. 
For small elliptic fiber, where F-theory description becomes better, 
zero modes of Dirac equation in a given representation $\rho(V)$ 
have Gaussian profile around a locus where Wilson lines in the 
elliptic fiber directions vanish, just like the case explained 
for the $T^3$-fibration in \cite{Acharya-Witten}. 
Localized massless matter multiplets in Heterotic theory 
description correspond to those on 7-brane intersection 
curves in Type IIB / F-theory description.

Suppose that the sheaf (\ref{eq:FM-rho(V)}) on $Z$ is given by 
a pushforward of a sheaf ${\cal N}_{\rho(V)}$ on $C_{\rho(V)}$:
\begin{equation}
 R^1p_{1*} \left[ p_2^* (\rho(V)) \otimes {\cal P}_B^{-1} \otimes {\cal O}(-q^* K_{B_2})  \right] = i_{C_{\rho(V)} *} \; ({\cal N}_{\rho(V)}), 
\label{eq:asPushForward}
\end{equation}
where $i_{C_{\rho(V)}}: C_{\rho(V)} \hookrightarrow Z$.
Then, the direct images $R^1\pi_{Z*} \rho(V)$ are given 
by pushforwards of sheaves on matter curves \cite{Curio, DI, Penn5}: 
\begin{eqnarray}
 R^1\pi_{Z*} (\rho(V)) & = & 
   i_{\rho(V) *}\; {\cal F}_{\rho(V)}, \\
 {\cal F}_{\rho(V)} & = & j^*_{\rho(V)} {\cal N}_{\rho(V)} 
   \otimes {\cal O}(i^*_{\rho(V)} K_{B_2});
\end{eqnarray}
here, $i_{\rho(V)}: \bar{c}_{\rho(V)} = \sigma \cdot
C_{\rho(V)} \hookrightarrow \sigma \simeq B_2$, 
and $j_{\rho(V)}: \bar{c}_{\rho(V)} = \sigma \cdot C_{\rho(V)} 
\hookrightarrow C_{\rho(V)}$.
Chiral multiplets in low-energy effective theory are characterized 
as global holomorphic sections of the sheaves ${\cal F}_{\rho(V)}$ 
on the matter curves:
\begin{equation}
 H^1(Z; \rho(V)) \simeq H^0(B_2; R^1\pi_{Z*} \rho(V)) 
   \simeq H^0(\bar{c}_{\rho(V)}; {\cal F}_{\rho(V)}).
\end{equation}
The net chirality (\ref{eq:chi-def}) is now expressed by 
Euler characteristic on the matter curve:
\begin{equation}
 \chi(\rho(V)) = \chi(B_2; R^1\pi_{Z*} \rho(V))
  = \chi(\bar{c}_{\rho(V)}; {\cal F}_{\rho(V)}).
\label{eq:GRR-twice}
\end{equation}

{\bf Matter from Bundles in the Fundamental Representation}

In the above discussion we have assumed that the sheaf (\ref{eq:FM-rho(V)}) on $Z$ is given by a pushforward of a sheaf on $C_{\rho(V)}$. This is actually a highly non-trivial statement. Even if a sheaf ${\cal E}$ on an algebraic variety $X$ is 
supported on a closed subvariety $i_Y: Y \hookrightarrow X$, it is 
not true in general that ${\cal E}$ is a pushforward of 
a sheaf ${\cal F}$ on $Y$; ${\cal E} = i_{Y*} {\cal F}$. 
It is true that ${\cal E} = i_{Y*} {\cal F}$ for some ${\cal F}$ on $Y$
as a sheaf of Abelian group, but not necessarily as a sheaf 
of ${\cal O}_X$-module.
Thus, the discussion after (\ref{eq:asPushForward}) is not necessarily 
applied immediately for bundles in any representation. 

For bundles $V$ in the fundamental representation, 
however, their Fourier--Mukai transforms are pushforward of the original 
line bundles ${\cal N}_V$ (see section~\ref{sec:Idea} and appendix~\ref{sec:push}). Thus, the discussion all the way down 
to (\ref{eq:GRR-twice}) is applicable.
The matter curves $\bar{c}_{V} =C_V \cdot \sigma$ belong
to a topological class 
\begin{equation}
 \bar{c}_V \in |N K_{B_2} + \eta|
\end{equation}
because $C_V \in |N \sigma + \pi^*_Z \eta|$, and 
$\sigma \cdot \sigma = - \sigma \cdot c_1({\cal L}_H) = 
\sigma \cdot K_{B_2}$ \cite{FMW}.

$R^1 \pi_{Z*} V$ is given by a pushforward 
of a sheaf on $\bar{c}_V$
\begin{equation}
 {\cal F}_V = j^*_V {\cal N} \otimes i^*_V {\cal O}(K_{B_2})
  = {\cal O}\left(i_V^* K_{B_2} + \frac{1}{2}j^*_V r + j^*_V
	     \gamma\right)
\label{eq:F4V}
\end{equation}
as a sheaf of ${\cal O}_{B_2}$-module. 
Since the canonical divisor $K_{C_V}$ is also the divisor 
$C_V|_{\bar{c}_V}$
in a Calabi--Yau 3-fold, 
\begin{eqnarray}
 i^*_V K_{B_2} + \frac{1}{2} j^*_V r &= &
 i^*_V K_{B_2} + \frac{1}{2}j^*_V \left( K_{C_V} - \pi_C^* K_{B_2}\right)
= \frac{1}{2} \left(i^*_V K_{B_2} + C_V|_{\bar{c}_V} \right)
 \nonumber \\
&=&\frac{1}{2}\left( i^*_V K_{B_2}  +  N_{\bar{c}_V | B_2}  \right)
 = \frac{1}{2} K_{\bar{c}_V},
\label{eq:KcbarV}
\end{eqnarray}
where adjunction formula was used for 
$i_V: \bar{c}_V \hookrightarrow B_2$ \cite{Curio}.
Thus, the sheaf can be rewritten as 
\begin{eqnarray}
 {\cal F}_V & = & 
   {\cal O}\left( \frac{1}{2}K_{\bar{c}_V} + j^*_V \gamma \right), \\
 {\cal F}_{V^\times} & = & 
   {\cal O}\left( \frac{1}{2} K_{\bar{c}_V} - j^*_V \gamma \right);
\end{eqnarray}
here we determined ${\cal F}_{V^\times}$ by replacing $\gamma$ 
by $- \gamma$ \cite{DI}. It is easy to see that these sheaves satisfy 
\begin{equation}
{\cal F}_{V^\times} = {\cal O}(K_{\bar{c}_V}) \otimes {\cal F}_V^{-1}. 
\end{equation}

Massless chiral multiplets from the bundles $V$ and $V^\times$ 
are now given by independent global holomorphic sections of 
${\cal F}_V$ and ${\cal F}_{V^\times}$, respectively. 
If one is interested only in the difference between the numbers 
of those chiral multiplets, the net chirality is obtained by 
Riemann--Roch theorem \cite{Curio, DI}:
\begin{eqnarray}
 \chi(V) & = & \chi(\bar{c}_V; {\cal F}_V), \\
   & = & \left[1 - g(\bar{c}_V)\right] 
   + {\rm deg} \left(K_{B_2} + \frac{1}{2} j^*_V r \right) + 
   \int_{\bar{c}_V} j^*_V \gamma, \\
   & = & \left[1 - g(\bar{c}_V)\right] 
       + \frac{1}{2} {\rm deg}\, K_{\bar{c}_V} 
    + \int_{\bar{c}_V} j^*_V \gamma, \\
  & = & \int_{\bar{c}_V} j^*_V \gamma = \bar{c}_V \cdot \gamma.
\label{eq:chi-V-F}
\end{eqnarray}
It is reasonable that the final result is proportional to 
$\gamma$, because we know that $\chi(V) = - \chi(V^\times)$, 
and $V \leftrightarrow V^\times$ corresponds to 
$\gamma \leftrightarrow - \gamma$ and ${\cal P}_B \leftrightarrow 
{\cal P}_B^{-1}$ \cite{DI}.

\section{Bundles Trivial in the Fiber Direction}
\label{sec:trivial}

In this section we briefly discuss the cohomology groups 
$H^i(Z; \pi_{Z}^* E)$, where $Z$ is an elliptic fibration 
$\pi_{Z}: Z \rightarrow B_2$, and we consider a bundle 
given by a pullback of a bundle $E$ on $B_2$.
Bundles given by $\pi_{Z}^*$ are trivial in the fiber direction, 
and hence $R^0\pi_{Z*} (\pi^*_Z E)$ on $B_2$ does not vanish, 
and $R^1 \pi_{Z*} (\pi^*_Z E)$ is not supported on a curve in $B_2$,
either. Thus, the cohomology groups of the bundles $\pi^*_Z E$
are not described in the same way as those of such bundles as 
$V$, $\wedge^2 V$ and $\wedge^3 V$. We need to express
$H^i(Z; \pi^*_Z E)$ ($i=1,2$) in terms of cohomology groups 
of $R^p \pi_{Z*} (\pi^*_Z E)$ ($p=0,1$), so that those 
expressions are interpreted in F-theory.

This issue has been discussed in the footnote~13 of \cite{Het-F-4D}. 
(See also \cite{DW}.)
Here, we add a minor comment to the description given there.

First, note that
\begin{eqnarray}
 R^0 \pi_{Z*} \left( \pi^*_Z E \right) & \simeq & E,  \label{eq:R0pi-rk2}\\
 R^1 \pi_{Z*} \left( \pi^*_Z E \right) & \simeq & E \otimes {\cal
  L}^{-1}_H   \simeq E \otimes {\cal O}(K_{B_2}),  \label{eq:R1pi-rk2}
\end{eqnarray}
where the Calabi--Yau condition of $\pi_Z : Z \rightarrow B_2$ 
is used in the last equality. 
Thus, 
\begin{eqnarray}
 H^0(Z; \pi^*_Z E) & \simeq & H^0(B_2; E), \label{eq:ss-0}\\
 \left[ H^0(Z; \pi_Z^* E^\times ) \right]^\times \simeq 
   H^3(Z; \pi^*_Z E)  & \simeq & 
 H^2(B_2; E \otimes {\cal O}(K_{B_2})) \simeq 
 [H^0(B_2; E^\times)]^\times.
\end{eqnarray}
Since these cohomology groups correspond to massless gauginos 
at low energy, one can assume that those groups are trivial 
when one is concerned with matter multiplets. 
Using the spectral sequence, one can see that the two other 
cohomology groups $H^{r}(Z; \pi^*_Z E)$ ($r = 1,2$) satisfy  
\begin{eqnarray}
& & 0 \rightarrow H^1(B_2 ; E) \rightarrow H^1(Z; \pi^*_Z E)
  \rightarrow H^0(B_2; E \otimes {\cal O}(K_{B_2})) \rightarrow 
              H^2(B_2; E), \label{eq:ss-1} \\ 
& & H^0(B_2; E \otimes{\cal O}(K_{B_2})) \rightarrow H^2(B_2; E) 
  \rightarrow H^2(Z; \pi^*_Z E) \rightarrow 
  H^1(B_2; E \otimes{\cal O}(K_{B_2})) \rightarrow 0. 
\label{eq:ss-2}
\end{eqnarray}
In the spectral sequence calculation of cohomology groups, 
$E_2^{p,q} = H^p(B_2; R^q \pi_{Z*} \pi^*_Z E)$, and 
$d_2: E^{p,q}_2 \rightarrow E^{p+2,q-1}_2$ for 
$(p,q) = (0,1)$ determines the map 
\begin{equation}
 d_2 : H^0(B_2; E \otimes {\cal O}(K_{B_2})) \rightarrow 
       H^2(B_2; E) \simeq 
  \left[ H^0(B_2; E^\times \otimes {\cal O}(K_{B_2}))\right]^\times
 \label{eq:d2}
\end{equation}
used in (\ref{eq:ss-1}, \ref{eq:ss-2}).

It thus follows that 
\begin{eqnarray}
 h^1(Z; \pi^*_Z E) & = & h^1(B_2; E) + {\rm ker} \; d_{2}, 
  \label{eq:bulk-matter-1-Het}\\
 h^2(Z; \pi^*_Z E) & = & h^1(B_2; E \otimes {\cal O}(K_{B_2})) 
   + {\rm coker} \; d_2, \\
                   & = & h^1(B_2; E^\times)   + {\rm coker} \; d_2, 
  \label{eq:bulk-matter-2-Het}
\end{eqnarray}
where $d_2$ is the one in (\ref{eq:d2}). 
If $d_2$ is trivial (including cases where either 
$h^0(B_2; E \otimes {\cal O}(K_{B_2})) = 0$ or 
$h^0(B_2; E^\times \otimes {\cal O}(K_{B_2})) = 0$), 
the results in \cite{Het-F-4D} follow:
\begin{eqnarray}
 h^1(Z; \pi^*_Z E) & = & h^1(B_2; E) + 
   h^0(B_2; E \otimes {\cal O}(K_{B_2})), \\
 h^1(Z; \pi^*_Z E^\times) = h^2(Z; \pi^*_Z E) & = & 
   h^1(B_2; E \otimes {\cal O}(K_{B_2})) + h^2(B_2; E), \\
  & = & h^1(B_2; E^\times) + h^0(B_2; E^\times \otimes {\cal O}(K_{B_2})).
\end{eqnarray}

For a general $d_2$, 
(\ref{eq:bulk-matter-1-Het}, \ref{eq:bulk-matter-2-Het}) are 
the right expressions for the number of massless matter 
multiplets from $\pi^*_Z E$. This means that some degrees of freedom in 
$H^0(B_2; E \otimes {\cal O}(K_{B_2}))$ and 
$H^0(B_2; E^\times \otimes {\cal O}(K_{B_2}))$ 
are paired up and do not remain in the low-energy spectrum. 
One might phrase this phenomenon as those degrees of freedom 
having ``masses.'' It should be noted that all the degrees of freedom 
in $H^1(B_2; E)$ and $H^1(B_2; E^\times)$ do not have such 
``masses.'' We do not study the detail of the map $d_2$
based on explicit examples. Such ``masses'' may be understood 
as a kind of obstruction in geometry. We leave these interesting 
questions as open problems for the future.

The structure group of a bundle $E$ can be chosen so that 
the unbroken symmetry $H$ in Table~\ref{tab:Het-matter} 
is reduced to whatever one likes, say $\SU(5)_{\rm GUT}$ or 
$\SU(3) \times \SU(2)$. The irreducible decomposition of 
${\bf adj.} H$ under the structure group of $E$ and the 
true unbroken symmetry may contain a pair of vector-like 
representations, $(\rho(E), repr.)$--$(\rho(E)^\times, repr.^\times)$.
For such a pair, the net chirality is calculated by 
\begin{eqnarray}
 \chi(\rho(E)) & :=  & 
 h^1(Z; \pi^*_Z \rho(E)) - h^1(Z; \pi^*_Z \rho(E)^\times ), \\
 & = & - \chi(Z; \pi^*_Z \rho(E)), \\    
 & = & - \chi(B_2; \rho(E)) + \chi(B_2; \rho(E) \otimes {\cal
  O}(K_{B_2})), \\
 & = & - \int_{B_2} c_1(TB_2) \wedge c_1 (\rho(E)).
\label{eq:E-chi-K}
\end{eqnarray} 
Rank of the map $d_2$ in (\ref{eq:d2}) does not matter here. 

The chirality formula (\ref{eq:E-chi-K}) can also be obtained 
from the discussion reviewed in the previous section~\cite{TW1}. The bundle $\pi^*_Z \rho(E)$ 
is regarded as a Fourier--Mukai transform of 
$(C_{\rho(E)},{\cal N}_{\rho(E)}) = (\sigma, \rho(E))$. 
Thus, the matter curve is formally given by 
$C_{\rho(E)} \cdot \sigma$ which belongs to a class 
of $K_{B_2}$. Since the ramification divisor of 
$\pi_{C}: C \rightarrow B_2$ is trivial, 
one finds (i) from the argument in (\ref{eq:KcbarV}) 
that $K_{B_2}$ is half the canonical divisor of the ``matter curve'' 
$\bar{c}_{\rho(E)} \sim K_{B_2}$ in $B_2$, and (ii) that 
${\cal N}_{\rho(E)} \otimes {\cal O}(r/2)^{-1} = \rho(E)$.
Therefore, 
\begin{equation}
 \chi(\rho(E)) = \int_{K_{B_2}} c_1(E) = 
  - \int_{B_2} c_1(TB_2) \wedge c_1(\rho(E)),
\end{equation}
reproducing (\ref{eq:E-chi-K}).

\section{Analysis of $R^1 \pi_{Z*} \wedge^2 V$}
\label{sec:Idea}

Not all the chiral multiplets in low-energy effective theory 
are identified with cohomology groups of bundle $V$ (or $V^\times$)
in the fundamental (anti-fundamental) representation, as we see 
in Table~\ref{tab:Het-matter}. In order to obtain description of 
all kinds of matter multiplets in F-theory, we also need to 
determine the sheaves $R^1 \pi_{Z*} \rho(V)$ for bundles associated with $\rho(V) = \wedge^2 V$ and $\wedge^3 V$.
As we have emphasized in Introduction, 
the Higgs multiplets and $\bar{\bf 5} = (\bar{D}, L)$ in the 
$\SU(5)_{\rm GUT}$-${\bf 5}+\bar{\bf 5}$ representations originate 
from the bundle $\wedge^2 V$, and the Higgs multiplet in the 
$\SO(10)$-${\bf 10}={\bf vec.}$ representation from $\wedge^2 V$.
Thus, it is important to determine $R^1\pi_{Z*} \wedge^2 V$ 
in order to understand Yukawa couplings of quarks and leptons 
in F-theory language.

For the bundles $V$ (or $V^\times$), the generic element of 
a topological class of spectral surface $|N \sigma + \pi_Z^* \eta|$
is smooth,\footnote{Some conditions have to be imposed on the divisor 
$\eta$. See \cite{OPP}.} 
and the transverse coordinate of $C_V$ in $Z$ can be chosen 
at any points on $C_V$. This property can be used to show 
that the Fourier--Mukai transform of $V$ is given by a pushforward 
of a sheaf on $C_V$ as a sheaf of ${\cal O}_Z$ module 
(see the appendix \ref{sec:push}).
Furthermore, the rank of fiber of the Fourier--Mukai transform 
never jumps on $C_V$ and the sheaf is the locally-free rank-1 sheaf 
${\cal N}_V$ itself.

For the bundles $\wedge^2 V$ (or $\wedge^2 V^\times$), on the other 
hand, $C_{\wedge^2 V}$ is not necessarily smooth, even if $C_V$ is. 
Here, we denote by $C_{\wedge^2 V}$ the support of Fourier--Mukai 
transform (\ref{eq:FM-rho(V)}) of $\rho(V)=\wedge^2 V$.
Suppose that $C_V |_{E_b}$ for a point $b \in B_2$ 
consists of $N$ points $\{ p_i\}_{i = 1, \cdots N}$. 
Then, $C_{\wedge^2 V}|_{E_b}$ is given by 
$\{ p_i \boxplus p_j\}_{1 \leq i < j \leq N}$. 
At a generic point $b \in B_2$, the $N(N-1)/2$ points 
$p_i \boxplus p_j$ $(i < j)$ in elliptic fiber $E_b$ are 
all different, and $C_{\wedge^2 V}$ is a smooth degree $N(N-1)/2$ 
cover. For these points, the arguments of the appendix \ref{sec:push} 
can be used to show that 
there a locally free rank-1 sheaf ${\cal N}_{\wedge^2 V}$
exists on $C_{\wedge^2 V}$ (locally around smooth points 
in $C_{\wedge^2 V}$), and the Fourier--Mukai transform of 
$\wedge^2 V$ is represented as the pushforward of 
${\cal N}_{\wedge^2 V}$ as a sheaf of ${\cal O}_Z$-module.
But, on a codimension-1 locus of $C_{\wedge^2 V}$, 
$C_{\wedge^2 V}$ may become singular \cite{DW}, 
and a little more attention must be paid.

We will describe a rough sketch of how to determine 
$R^1 \pi_{Z*} \wedge^2 V$ in this section, beginning 
with how to deal with such singularities. Details 
of $R^1 \pi_{Z*} \wedge^2 V$ are deferred to the next 
section. Since some crucial aspects of 
$R^1\pi_{Z*} \wedge^2 V$ depend very much on the rank of $V$, 
we will provide detailed description of $R^1\pi_{Z*} \wedge^2 V$ 
for the rank of $V$ between 2 and 6 in the next section. 
Once we see how to deal with $R^1 \pi_{Z*} \wedge^2 V$, it is rather 
straightforward to find how to deal with $R^1 \pi_{Z*} \wedge^3 V$. 
We will only discuss $R^1 \pi_{Z*} \wedge^3 V$ in 
section~\ref{ssec:rk6}.

\subsection{Resolving Double-Curve Singularity of $C_{\wedge^2 V}$}
\label{ssec:dbl-curve}

$C_{\wedge^2 V}$ is described locally as $N(N-1)/2$ 
surfaces that $p_i \boxplus p_j$ ($i < j$) scan.
$C_{\wedge^2 V}$ has a double-curve singularity if 
$p_i \boxplus p_j$ $(i<j)$ and $p_k \boxplus p_l$ 
($k < l$, $\{i,j\} \cap \{ k,l\} = \phi$) become equal. 
It is not obvious how to choose a coordinate in $Z$ 
that is normal to $C_{\wedge^2 V}$ along the double-curve 
locus, and the argument in the appendix is not readily applicable.

In a local neighborhood of the double curve, 
$C_{\wedge^2 V}$ consists of two irreducible components, 
$C_{(ij)}$ and $C_{(kl)}$, and their intersection is the 
double-curve singularity. 
$C_{(ij)}$ and $C_{(kl)}$ are surfaces scanned in $Z$ 
by $p_i \boxplus p_j$ and $p_k \boxplus p_l$. 
$\rho(V) = \wedge^2 V$ can be regarded locally as direct sum of 
${\cal O}(C_{(ij)} - \sigma)$, ${\cal O}(C_{(kl)} - \sigma)$ 
and others. Its Fourier--Mukai transform in (\ref{eq:FM-rho(V)})
is given by a sum of the above two summands.
The Fourier--Mukai transform of the two summands 
${\cal O}(C_{(ij)} - \sigma)$ and 
${\cal O}(C_{(kl)} - \sigma)$ is expressed locally as 
\begin{eqnarray}
 R^1 p_{1*} \left[{\cal O}(C_{(ij)} - \sigma) \otimes {\cal P}_B^{-1}
   \otimes {\cal O}(- q^* K_{B_2}) \right] & = & 
  i_{C_{(ij)} * } {\cal O}_{C_{(ij)}}, \\
 R^1 p_{1*} \left[{\cal O}(C_{(kl)} - \sigma) \otimes {\cal P}_B^{-1}
   \otimes {\cal O}(- q^* K_{B_2}) \right] & = & 
  i_{C_{(kl)} * } {\cal O}_{C_{(kl)}}. 
\end{eqnarray}
Here, $i_{C_{\wedge^2 V}}: C_{\wedge^2 V} \hookrightarrow Z$ 
(which is different from previously defined $i_{\wedge^2 V}:
\bar{c}_{\wedge^2 V} \hookrightarrow \sigma$), and 
\begin{eqnarray}
 \nu_{C_{ij}}: C_{(ij)} \hookrightarrow C_{\wedge^2 V}, & \qquad &  
 i_{C_{(ij)}} = i_{C_{\wedge^2 V}} \circ \nu_{C_{ij}},  \\
 \nu_{C_{kl}}: C_{(kl)} \hookrightarrow C_{\wedge^2 V}, & & 
 i_{C_{(kl)}} = i_{C_{\wedge^2 V}} \circ \nu_{C_{kl}}.  
\end{eqnarray}
Therefore, the Fourier--Mukai transform of $\wedge^2 V$ is 
\begin{eqnarray}
 R^1 p_{1*} \left[p_2^* (\wedge^2 V) \otimes {\cal P}_B^{-1} \otimes 
  {\cal O}(-q^* K_{B_2}) \right] \simeq 
  i_{C_{\wedge^2 V} * } \left( 
   \nu_{C_{ij} *} {\cal O}_{C_{(ij)}} \oplus 
   \nu_{C_{kl} *} {\cal O}_{C_{(kl)}} \right)
\end{eqnarray}
locally along a double-curve singularity. Thus, 
it is given by a pushforward of a sheaf ${\cal N}_{\wedge^2 V}$ 
on $C_{\wedge^2 V}$ as a sheaf of ${\cal O}_Z$-module. 
The sheaf ${\cal N}_{\wedge^2 V}$ is the object 
inside the parenthesis on the right hand side.

The sheaf ${\cal N}_{\wedge^2 V}$ is not locally free along 
the double-curve singularity. The rank of fiber jumps up there. 
But we already know that the sheaf ${\cal N}_{\wedge^2 V}$ 
is given by a pushforward of locally-free rank-1 sheaf via 
\begin{equation}
 \nu_{C_{\wedge^2 V}}: \widetilde{C}_{\wedge^2 V} = 
 C_{(ij)} \coprod C_{(kl)} 
 \rightarrow C_{(ij)} \cup C_{(kl)} = C_{\wedge^2 V}.
\label{eq:resolveCV2-local}
\end{equation} 
The map $\nu_{C_{\wedge^2 V}}$ is determined by 
$\nu_{C_{ij}} \coprod \nu_{C_{kl}}$. Note that 
$\widetilde{C}_{\wedge^2 V} := C_{(ij)} \coprod C_{(kl)}$ 
is the resolution of double-curve singularity in $C_{\wedge^2 V}$. 
Therefore, the discussion so far means that there exists a 
locally free rank-1 sheaf $\widetilde{\cal N}_{\wedge^2 V}$ on 
the resolved $\widetilde{C}_{\wedge^2 V}$ such that 
\begin{equation}
 {\cal N}_{\wedge^2 V} = \nu_{C_{\wedge^2 V}* } 
\widetilde{\cal N}_{\wedge^2 V}.
\label{eq:NandN}
\end{equation}

We have seen that the sheaf ${\cal N}_{\wedge^2 V}$ exists on 
$C_{\wedge^2 V}$ and (\ref{eq:asPushForward}) is satisfied as 
a sheaf of ${\cal O}_Z$ module. 
Thus, the discussion around equations  
(\ref{eq:asPushForward}--\ref{eq:GRR-twice}) is applied 
for the bundles $\rho(V) = \wedge^2 V$ and $\wedge^2 V^\times$ 
as well. In particular, the sheaf on the matter curve 
$\bar{c}_{\wedge^2 V}$ is given by 
\begin{equation}
 {\cal F}_{\wedge^2 V} = j_{\wedge^2 V}^* {\cal N}_{\wedge^2 V}
  \otimes i_{\wedge^2 V}^* {\cal O}(K_{B_2}). 
\end{equation}

We introduce the notion of covering matter curve, which turns out 
to be very important in characterizing matter multiplets in F-theory.
The covering matter curve $\tilde{\bar{c}}_{\wedge^2 V}$ 
is defined as the set-theoretic inverse image 
of the matter curve $\bar{c}_{\wedge^2 V}$ in 
$\widetilde{C}_{\wedge^2 V}$. That is, 
$\tilde{\bar{c}}_{\wedge^2 V} := \nu_{C_{\wedge^2 V}}^{-1}
(\bar{c}_{\wedge^2 V})$. Since the matter curve 
$\bar{c}_{\wedge^2 V}$ is also regarded as a divisor 
$\sigma|_{C_{\wedge^2 V}}$ in $C_{\wedge^2 V}$, the covering 
matter curve is also regarded as a divisor 
$\nu_{C_{\wedge^2 V}}^*(\sigma)$ on $\widetilde{C}_{\wedge^2 V}$.
Using a locally rank-1 sheaf $\widetilde{\cal N}_{\wedge^2 V}$ 
on $\widetilde{C}_{\wedge^2 V}$, a locally free rank-1 sheaf 
$\widetilde{\cal F}_{\wedge^2 V}$ can be defined on the covering 
matter curve:
\begin{equation}
 \widetilde{\cal F}_{\wedge^2 V} = \tilde{\jmath}_{\wedge^2 V}^*
  \widetilde{\cal N}_{\wedge^2 V} \otimes 
  \tilde{\imath}_{\wedge^2 V}^* {\cal O}(K_{B_2}), 
\end{equation}
where $\tilde{\jmath}_{\wedge^2 V}: \tilde{\bar{c}}_{\wedge^2 V} 
\hookrightarrow \widetilde{C}_{\wedge^2 V}$, 
$\nu_{\bar{c}_{\wedge^2 V}} := 
\nu_{C_{\wedge^2 V}}|_{\tilde{\bar{c}}_{\wedge^2 V}}$, 
and 
$\tilde{\imath}_{\wedge^2 V}:= i_{\wedge^2 V} \circ \nu_{\bar{c}_{\wedge^2 V}}:
\tilde{\bar{c}}_{\wedge^2 V} \hookrightarrow \sigma \simeq B_2$.
The sheaf ${\cal F}_{\wedge^2 V}$ on the matter curve 
$\bar{c}_{\wedge^2 V}$ is given by 
\begin{equation}
 {\cal F}_{\wedge^2 V} = \nu_{\bar{c}_{\wedge^2 V} *} 
  \widetilde{\cal F}_{\wedge^2 V}.
\end{equation}

Although we have dealt with double-curve singularities 
on $C_{\wedge^2 V}$, there can still be other types of singularities 
on $C_{\wedge^2 V}$. For example, there may be codimension-2 singularities 
on $C_{\wedge^2 V}$.
Thus, the argument in section~\ref{ssec:dbl-curve} is not regarded 
as a complete proof of the existence of ${\cal N}_{\wedge^2 V}$ 
or the existence of $\widetilde{\cal N}_{\wedge^2 V}$ and 
its locally-free rank-1 nature.
For practical purposes, however, we only need to know 
$R^1 \pi_{Z*} \wedge^2 V$ along the matter curves. 
Codimension-1 singularities such as double curve on $C_{\wedge^2 V}$ 
may be encountered somewhere along the matter curve 
$\bar{c}_{\wedge^2 V}$ \cite{DW}, but codimension-2 singularities of 
$C_{\wedge^2 V}$ are seldom exactly on the matter curve. 
Thus, an analysis of codimension-2 singularities of $C_{\wedge^2 V}$ 
is not required for the generic case. We will see, however, that 
codimension-2 singularities inevitably show up on the matter curve 
$\bar{c}_{\wedge^2 V}$ when ${\rm rank} \; V = 4, 6$.
We will deal with such exceptional cases separately 
in sections~\ref{ssec:rk4} and \ref{ssec:rk6}.

\subsection{Determining $\widetilde{{\cal N}}_{\wedge^2 V}$ 
in Terms of ${\cal N}_V$}

Even after we find that a sheaf ${\cal N}_{\wedge^2 V}$ exists and 
(\ref{eq:asPushForward}) is satisfied as a sheaf of ${\cal O}_Z$ 
module, we still face a theoretical challenge. 
How is ${\cal N}_{\wedge^2 V}$ (or $\widetilde{\cal N}_{\wedge^2 V}$)
expressed in terms of the original spectral data $(C_V, {\cal N}_V)$? 
Pioneering work was done in \cite{Penn5, BMRW}. 
Our presentation in the 
following is basically along their idea,\footnote{One of the 
authors (TW) thanks Ron Donagi for explaining the idea of 
\cite{Penn5} (March, 2006).} but we introduce 
a little modification for a couple of reasons.
First, we will obtain sheaves $\tilde{\cal N}_{\wedge^2 V}$ 
and $\widetilde{\cal F}_{\wedge^2 V}$ on the covering matter curve  
$\tilde{\bar{c}}_{\wedge^2 V}$, instead of ${\cal F}_{\wedge^2 V}$ 
on the matter curve $\bar{c}_{\wedge^2 V}$. By doing so, 
much clearer description of the direct image 
$R^1\pi_{Z*} \wedge^2 V$ is obtained. The other reason 
for modification is that we are not assuming that 
${\cal N}_V|_D$ is invariant under 
$\tau$ that flips the sign of the coordinate $y$. 
$D$ is a curve on $C_V$; we will explain it later.

Since (\ref{eq:asPushForward}) for $\rho(V) = \wedge^2 V$ is the 
definition of ${\cal N}_{\wedge^2 V}$, it follows that 
\begin{equation}
 {\cal N}_{\wedge^2 V} = i^*_{C_{\wedge^2 V}} R^1p_{1*} 
\left[p_2^* (\wedge^2 V) \otimes {\cal P}_{B_2}^{-1} \otimes 
 {\cal O}(-q^* K_{B_2}) \right].
\end{equation}
What we really need is its restriction on $\bar{c}_{\wedge^2 V}$, 
and hence 
\begin{eqnarray}
  {\cal F}_{\wedge^2 V} & = & 
 {\cal N}_{\wedge^2 V}|_{\bar{c}_{\wedge^2 V}} 
   \otimes i^*_{\wedge^2 V} {\cal O}(K_{B_2}), \nonumber \\ 
 & = &  (i_{C_{\wedge^2 V}} \circ j_{\wedge^2 V})^* 
 R^1p_{1*}\left[p_2^* (\wedge^2 V) \otimes {\cal P}_{B_2}^{-1} \otimes 
 {\cal O}(-q^* K_{B_2}) \right] \otimes i^*_{\wedge^2 V} {\cal O}(K_{B_2})
 ,  \nonumber \\
 & = &    (i_{C_{\wedge^2 V}} \circ j_{\wedge^2 V})^* 
 R^1 p_{1*} \left[p_2^* (\wedge^2 V) \right], 
  \nonumber  \\
 & = & R^1 p_{1Y*} \left[ \wedge^2 V|_{Y}\right]; 
\end{eqnarray}
here, $Y := \bar{c}_{\wedge^2 V} \times_{B_2} Z =  
\pi_Z^{-1}(\bar{c}_{\wedge^2 V})$. 
In the third equality, we used the property that ${\cal P}_{B_2}$ 
is trivial when it is restricted to a zero section \cite{FMW}, and 
in the last equality the base change theorem associated with the commutative diagram 
\begin{equation}\vcenter{\xymatrix{
  Y := \bar{c}_{\wedge^2 V} \times_{B_2} Z  \ar[d]_{p_{1Y}} \ar@{^{(}->} [0,2]&  & 
    Z \times_{B_2} Z \ar[d]^{p_1}&\\
   \bar{c}_{\wedge^2 V} \ar@{^{(}->}[r]_{j_{\wedge^2 V}}  &  C_{\wedge^2 V} \ar@{^{(}->}[r]_{i_{C_{\wedge^2 V}}} 
&  Z  \\
}}.
\end{equation}
This is the standard procedure used in \cite{Curio, DI, Penn5}.

The rank-$N$ bundle $V|_Y$ is given by a Fourier--Mukai 
transform of ${\cal N}_V|_{C_V \cdot Y}$:
\begin{equation}
\vcenter{\xymatrix{
 & (C_V \cdot Y) \times_{\bar{c}_{\wedge^2 V}} Y \ar[dl]_{p_1} \ar[dr]^{p_2} & \\
C_V \cdot Y \ar[dr]_{\pi_C|_{C_V \cdot Y}}&
& Y \ar[dl]^{\pi_Y}\\
 & \bar{c}_{\wedge^2 V} & 
}}, \qquad 
V|_Y = p_{2*} \left({\cal P}_{B_2} \otimes 
                    p_1^* ({\cal N}_V|_{C_V \cdot Y})\right).
\end{equation}
The spectral curve $C_V \cdot Y$ is a degree-$N$ cover over 
$\bar{c}_{\wedge^2 V}$.

Let $C_V|_{E_b}$ be a collection of $N$ points 
$\{ p_i\}_{i = 1,\cdots, N}$. 
For a point $b \in \bar{c}_{\wedge^2 V} \subset C_{\wedge^2 V}$, 
some pairs of the $N$ points, e.g., $p_k$ and $p_l$, satisfy 
$p_k \boxplus p_l = e_0$. Such points in $C_V \cdot Y$ 
form a curve $D$, and others form a curve $D'$.
\begin{equation}
C_V \cdot Y = D + D'. 
\label{eq:DD'}
\end{equation}
By the definition of $D$, the following diagram 
commutes \cite{BMRW},\footnote{In \cite{BMRW}, 
$D$ corresponds to our $C_V \cdot Y$, and $D'$ 
to our $D$. The covering matter curve $\tilde{\bar{c}}_{\wedge^2 V}$
introduced in this article is essentially the same as 
$D'/\tau$ in \cite{BMRW}.
See footnote \ref{fn:normalization} for more about the relation 
between $D'/\tau$ in \cite{BMRW} and $\tilde{\bar{c}}_{\wedge^2 V}$ 
here.}  
\begin{equation}
 \vcenter{\xymatrix{
  D  \ar[dr]_{\pi_D} \ar[r]^{\tilde{\pi}_{D}}& \tilde{\bar{c}}_{\wedge^2 V} \ar[d]^{\nu_{\bar{c}_{\wedge^2 V}}} 
\\
    & \bar{c}_{\wedge^2 V} 
 }}
\end{equation}
and $\tilde{\pi}_D$ is a degree-2 cover, and 
$\pi_D$ is a restriction of $\pi_C$ on $D$.
If $b \in \bar{c}_{\wedge^2 V} \hookrightarrow \sigma$ 
is on the double-curve singularity of $C_{\wedge^2 V}$, 
then there are four points $p_{i,j,k,l}$, satisfying 
$p_i \boxplus p_j = e_0$ and $p_k \boxplus p_l = e_0$.
In the covering matter curve, the inverse image of $b$, 
that is, $\nu_{\bar{c}_{\wedge^2 V}}^{-1}(b)$, 
consists of two points. Two points $p_{i,j} \in D$ are 
mapped by $\tilde{\pi}_D$ to one of the two points 
in $\nu_{\bar{c}_{\wedge^2 V}}^{-1}(b)$, and $p_{k,l} \in D$ 
to the other. Although all the four points are mapped 
to $b \in \bar{c}_{\wedge^2 V}$ by $\pi_D$, 
$\tilde{\pi}_D$ remains strictly a degree-2 cover 
everywhere on $\tilde{\bar{c}}_{\wedge^2 V}$.

The Fourier--Mukai transform of ${\cal N}_V|_D$ on a 
degree-2 cover spectral curve 
$\tilde{\pi}_D: D \rightarrow \tilde{\bar{c}}_{\wedge^2 V}$
gives a rank-2 bundle $W_2$:
\begin{equation}
\vcenter{\xymatrix{
 & D \times_{\tilde{\bar{c}}_{\wedge^2 V}} \tilde{Y} \ar[dl]_{p_1} \ar[dr]^{p_2}& \\
D \ar[dr]_{\tilde{\pi}_D} & & \tilde{Y} \ar[dl]^{p_{1\tilde{Y}}}\\
 & \tilde{\bar{c}}_{\wedge^2 V} & 
}}, \qquad 
 W_2 =  p_{2*} \left({\cal P}_B|_{D \times_{\tilde{\bar{c}}_{\wedge^2
		V}} Y} \otimes  p_1^* ({\cal N}_V |_D ) \right),
\end{equation}
where $\tilde{Y}:= \tilde{\bar{c}}_{\wedge^2 V}
\times_{\bar{c}_{\wedge^2 V}} Y$. 
The pushforward of this rank-2 bundle $W_2$ through projection 
$\nu_Y: \tilde{Y} = \tilde{\bar{c}}_{\wedge^2 V} 
\times_{\bar{c}_{\wedge^2 V}} Y \rightarrow Y$ defines 
a subsheaf of $V|_Y$.

For a point $b \in \bar{c}_{\wedge^2 V} \subset C_{\wedge^2 V}$ 
that is not on the double-curve locus, $H^1(E_b; \wedge^2 V|_{E_b})$ 
comes from $H^1(E_b; \wedge^2 (\nu_{Y*} W_2)|_{E_b}) 
= H^1(E_b; \wedge^2 W_2|_{E_b})$. 
For a point $b \in \bar{c}_{\wedge^2 V}$ 
on the double-curve singularity of $C_{\wedge^2 V}$, however, 
there are two independent contributions corresponding to 
$H^1(E_b; \wedge^2 W_2 |_{E_{\tilde{b}}})$ for two 
points $\tilde{b} \in \nu_{\bar{c}_{\wedge^2 V}}^{-1}(b)$. 
We introduced the covering matter curve $\tilde{\bar{c}}_{\wedge^2 V}$
in order to resolve these two contributions.
The locally free rank-1 sheaf 
$\widetilde{\cal N}_{\wedge^2 V}|_{\tilde{\bar{c}}_{\wedge^2 V}}$ 
(and $\widetilde{\cal F}_{\wedge^2 V}$, consequently) 
is obtained by assigning them to the corresponding two points 
$\tilde{b}$ on $\tilde{\bar{c}}_{\wedge^2 V}$.
Therefore, 
\begin{eqnarray}
 \widetilde{\cal F}_{\wedge^2 V} & = & 
  \widetilde{\cal N}_{\wedge^2 V}|_{\tilde{\bar{c}}_{\wedge^2 V}}
  \otimes \tilde{\imath}_{\wedge^2 V}^* 
  {\cal O}(K_{B_2}), \nonumber \\
  & = & R^1p_{1\tilde{Y}*} \left[\wedge^2 W_2 \right].
\end{eqnarray}

The line bundle $\wedge^2 W_2$ is trivial in the fiber direction of 
$p_{1\tilde{Y}}$. Thus, it is regarded as a Fourier--Mukai transform 
of $(C_{\wedge^2 W_2}, {\cal N}_{\wedge^2 W_2}) = 
(\sigma, \widetilde{\cal N}_{\wedge^2 V}|_{\tilde{\bar{c}}_{\wedge^2 V}})$.
It then follows that 
\begin{equation}
 \wedge^2 W_2 = p_{1\tilde{Y}}^* ({\cal N}_{\wedge^2 W_2}).
\label{eq:A}
\end{equation}
Thus, 
\begin{equation}
 \widetilde{\cal F}_{\wedge^2 V} = 
{\cal N}_{\wedge^2 W_2} \otimes {\cal L}_H^{-1} = 
{\cal N}_{\wedge^2 W_2} \otimes 
 \tilde{\imath}^*_{\wedge^2 V} {\cal O}(K_{B_2}).
\end{equation}

Now it is useful to remember that the first Chern class of 
the line bundle $\wedge^2 W_2$ is 
\begin{eqnarray}
 c_1(\wedge^2 W_2) = c_1(W_2) & = &
p_{1\tilde{Y}}^* \tilde{\pi}_{D*} \left(c_1({\cal N}_V|_D) - 
 \frac{1}{2}R \right), \\
  & = & p_{1\tilde{Y}}^* \tilde{\pi}_{D*} \left(\gamma|_D + 
 \frac{1}{2}(r|_D - R) \right), 
\label{eq:B}
\end{eqnarray}
just like in (\ref{eq:c1}). 
Here, $R: = K_D - \tilde{\pi}_D^* K_{\tilde{\bar{c}}_{\wedge^2 V}}$ is 
the ramification divisor on $D$ associated with the projection 
$\tilde{\pi}_D : D \rightarrow \tilde{\bar{c}}_{\wedge^2 V}$. 
Thus,\footnote{Equations (3.56) and (3.57) in \cite{BMRW} would 
be consistent with (\ref{eq:F4V2}), if the sign of the $R/2$ terms 
in the equations were opposite.} by dropping $p_{1\tilde{Y}}^*$ 
from (\ref{eq:A}) and (\ref{eq:B}),
\begin{equation}
 \widetilde{\cal F}_{\wedge^2 V} = {\cal N}_{\wedge^2 W_2} \otimes 
\tilde{\imath}^*_{\wedge^2 V}{\cal O}(K_{B_2}) = 
 {\cal O}\left(\tilde{\imath}^*_{\wedge^2 V} K_{B_2} 
  + \tilde{\pi}_{D*}\left(\gamma|_D+\frac{1}{2}(r|_D - R) \right)
			    \right).
\label{eq:F4V2}
\end{equation}
%

\section{In-Depth Analysis of Associated Bundles}
\label{sec:Examples}

In this section, we will study direct images $R^1\pi_{Z*} \wedge^2 V$
for ${\rm rank} \; V = 2,3,4,5,6$, and $R^1\pi_{Z*} \wedge^3 V$ 
for ${\rm rank} \; V = 6$. Apart from $\wedge^2 V$ for a rank-2 bundle
$V$, all those associated bundles are non-trivial in the fiber 
direction, and those direct images have their supports on 
matter curves. The description of sheaves on the matter curves 
obtained in this section in Heterotic description are translated 
into F-theory language in later sections. 

Before we commit ourselves to individual cases, we quote some 
results from \cite{Penn5} that are useful in this section 
regardless of the rank of $V$. 
First, the spectral surface $C_{\wedge^2 V}$ belongs to a class 
\begin{equation}
C_{\wedge^2 V} \in \left|\frac{N(N-1)}{2} \sigma + (N-2)\pi^*_Z \eta \right|. 
\label{eq:CVofV2}
\end{equation}
The coefficient $N(N-1)/2$ of the first term is the rank of 
$\wedge^2 V$, and $(N-2)$ for the second term twice the Dynkin index 
of the rank-2 anti-symmetric tensor representation of $\SU(N)$. 
When the matter curve is given by $C_{\wedge^2 V} \cdot \sigma$, 
(there is an exception; see sections~\ref{ssec:rk4})
\begin{equation}
 C_{\wedge^2 V} \cdot \sigma \in 
  \left|\frac{N(N-1)}{2} K_B + (N-2) \eta\right| \subset B_2.
\label{eq:cVofV2}
\end{equation}
A curve $D$ in section~\ref{sec:Idea} 
belongs to a topological class 
\begin{equation}
 D \in |\sigma \cdot \left[N(N-1) K_{B_2} + 2(N-2)\eta \right] 
 + \pi_{Z}^* \left[\eta \cdot (3K_{B_2} + \eta) \right]|. 
\label{eq:D-topology}
\end{equation}
See \cite{Penn5} for why this is the case. 

\subsection{Rank-2 Vector Bundles}
\label{ssec:rk2}

The $\wedge^2 V$ bundle for a rank-2 bundle $V$ is exceptional 
in this section, because $\wedge^2 V = {\rm det} \; V$ is trivial 
in the fiber direction for $V$ given by spectral cover construction.
If the structure group of $V$ is $\SU(2)$, we have nothing non-trivial 
to say because $\wedge^2 V = {\cal O}_Z$. If the structure group is $\U(2)$, 
then $\wedge^2 V = \pi_Z^* E$ for some line bundle $E$ on the 
base manifold $B_2$. Thus, this case is a special case of what we 
discussed in section~\ref{sec:trivial}.

$\U(2)$ bundles have appeared in phenomenological applications. 
In \cite{TW1}, for example, a rank-5 bundle $U_3 \oplus U_2$ 
was considered for $\SU(5)_{\rm GUT}$ unified theories, where 
$U_3$ is a rank-3 bundle with the structure group $\U(3)$, and 
$U_2$ a rank-2 bundle with $\U(2)$. By considering a vector 
bundle in such a semi-stable limit, and some controlled deformation 
from this limit, one can bring dimension-4 and dimension-5 
proton decay problems under control \cite{TW1, TW2, KNW}.
The up-type Higgs multiplet was identified with 
$H^2(Z; \wedge^2 U_2^\times)$ and its F-theory dual.
Thus, it is not without phenomenological motivation to 
provide an F-theory description of $\wedge^2 V$ of 
a $\U(2)$ bundle $V$.

Since $\wedge^2 V$ is trivial in the fiber direction, 
anything written in section~\ref{sec:trivial} apply here, 
with 
\begin{equation}
 E = {\cal O}(\pi_{C*} \gamma_2),
\end{equation}
where $\gamma_2$ is $\gamma$ in (\ref{eq:defgamma}) for the bundle 
$\U_2$.
This line bundle $E$ has a structure group U(1) in 
the commutant of $\SU(2)$ in $E_8$, which is $E_7$.
In the case an $\SU(5)$ bundle $U_3 \oplus U_2$ is considered 
in Heterotic theory compactification, for example, then 
the line bundle 
\begin{equation}
 \wedge^2 U_2 = {\rm det}\; U_2 = ({\rm det} \; U_3)^{-1} = 
 \pi_{Z}^* E
\end{equation}
has its U(1) structure group in the commutant 
of $\SU(3) \times \SU(2)$ in $E_8$, which is now 
$\SU(6) \supset \SU(5)_{\rm GUT}$. The structure group 
is now the $\U(1)$ direction in $\SU(6)$ that also commutes 
with $\SU(5)_{\rm GUT}$.

\subsection{Rank-3 Vector Bundles}
\label{ssec:rk3}

Next, let us consider a rank-3 vector bundle $V$ with the data 
$(C_V,{\cal N}_V)$. 
Note that $\wedge^2 V \simeq V^{\times} \otimes (\det \; V)$.
Since ${\rm det} \; V$ line bundle is trivial in the fiber 
direction, the spectral surface of $\wedge^2 V$ is the same 
as that of $V^\times$. If the spectral surface of $V$ 
is given by a zero locus of 
\begin{equation}
 s = a_0 + a_2 x + a_3 y,
\label{eq:CV-rk3-defeq}
\end{equation}
then the spectral surface $C_{V^\times} = C_{\wedge^2 V}$ 
is given by the zero locus of 
\begin{equation}
 s^\times = a_0 + a_2 x - a_3 y,
\label{eq:CV2-rk3-defeq}
\end{equation}
flipping the sign of terms containing one $y$ from (\ref{eq:CV-rk3-defeq}).
Thus, $C_{\wedge^2 V}$ belongs to the class (\ref{eq:CVofV2}) with 
$N=3$. 

The matter curve for $\wedge^2 V$ is given by $a_3 = 0$. 
This is because (\ref{eq:CV2-rk3-defeq}) determines three points 
in each elliptic fiber, and one of the three points approaches 
the zero section $\sigma$ as $a_3 \rightarrow 0$. This is also 
because the two points determined by (\ref{eq:CV-rk3-defeq}) 
share the same value of $x$, $-a_0/a_2$, if $a_3 = 0$; two points 
on an elliptic curve $p_j = (x,y)$ and $p_k = (x,-y)$ are 
inverse elements of each other in terms of group law on the 
elliptic curve, that is, $p_j \boxplus p_k = e_0$. 
The matter curve $\bar{c}_{\wedge^2 V}$, specified by $a_3 = 0$,
belongs to a class (\ref{eq:cVofV2}) for $N=3$.

Since $\wedge^2 V = V^\times \otimes {\rm det} \; V$, and 
${\rm det} \; V = \pi_Z^* {\cal O}(\pi_{C*} \gamma)$, 
it is straightforward to obtain the sheaf 
${\cal F}_{\wedge^2 V}$ on the matter curve 
$\bar{c}_{\wedge^2 V} = \bar{c}_V$. Applying 
$\otimes {\cal O}(\pm \pi_{C*} \gamma)$ to ${\cal F}_{V^\times}$
and ${\cal F}_V$, 
\begin{eqnarray}
 {\cal F}_{\wedge^2 V} & = & {\cal O}\left(
  i^* K_{B_2} + \frac{1}{2} j^* r - j^* \gamma
  + i^* \pi_{C*} \gamma  \right), \label{eq:F4V2-rk3} \\
 {\cal F}_{\wedge^2 V^\times} & = & {\cal O}\left(
  i^* K_{B_2} + \frac{1}{2} j^* r + j^* \gamma
  - i^* \pi_{C*} \gamma  \right), 
\end{eqnarray}
where $i: \bar{c}_{\wedge^2 V} = \bar{c}_V \hookrightarrow B_2$, 
and $j: \bar{c}_{\wedge^2 V} = \bar{c}_V \hookrightarrow C_V$.
It is thus unnecessary to use the idea presented in
section~\ref{sec:Idea} in determining the ${\cal F}_{\wedge^2 V}$ 
for rank-3 bundles $V$. In the rest of section~\ref{ssec:rk3}, 
however, we use the idea to reproduce this result, so that we 
get accustomed to using the idea in practice. 

In the fiber $E_b$ of an arbitrary point 
$b \in \bar{c}_V = \bar{c}_{\wedge^2 V} \subset B_2$, 
$C_V|_{E_b}$ consists of three points, 
one in the zero section $p_i =e_0 = \sigma \cdot E_b$
and two others satisfying $p_j \boxplus p_k = e_0$.
Thus, the irreducible decomposition (\ref{eq:DD'}) 
becomes 
\begin{equation}
 C_V \cdot Y = D + \bar{c}_V.
\end{equation}
The curve $D$ is already a degree-2 cover on 
$\bar{c}_V = \bar{c}_{\wedge^2 V}$, and we do not need 
to introduce a covering curve $\tilde{\bar{c}}_{\wedge^2 V}$ 
for rank-3 bundles $V$.

Among various components of divisors specifying the rank-1 sheaf 
${\cal F}_{\wedge^2 V}$ in (\ref{eq:F4V2}), $\pi_{D*} \gamma$
and $\pi_{D*} (r|_D - R)/2$ can be treated separately. 
Because of the irreducible decomposition we have seen above, 
\begin{equation}
 \pi_{D*} \gamma = i^* \pi_{C*} \gamma - j^* \gamma,
\label{eq:gamma-V2-rk3}
\end{equation}
and hence the $\gamma$-dependent part of (\ref{eq:F4V2-rk3}) 
is reproduced.

\begin{figure}[tbp]
 \begin{center}
\begin{tabular}{ccccc}
\includegraphics[width=.25\linewidth]{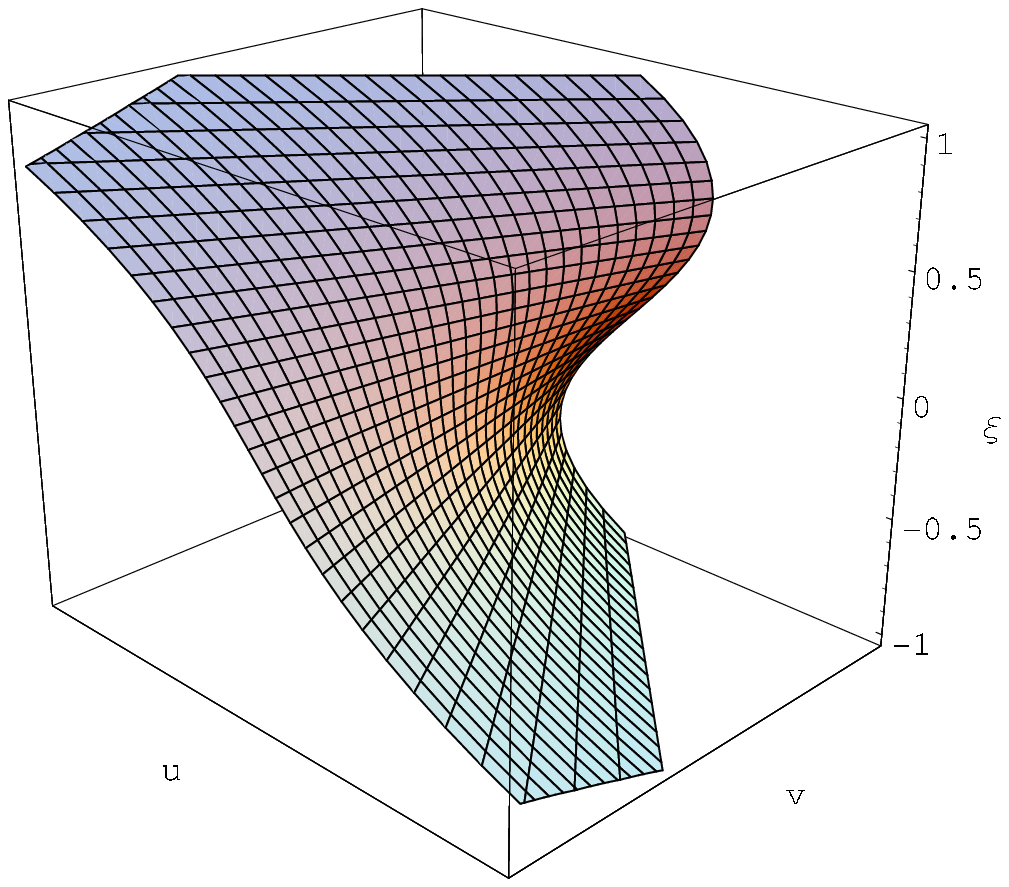} & &
\includegraphics[width=.25\linewidth]{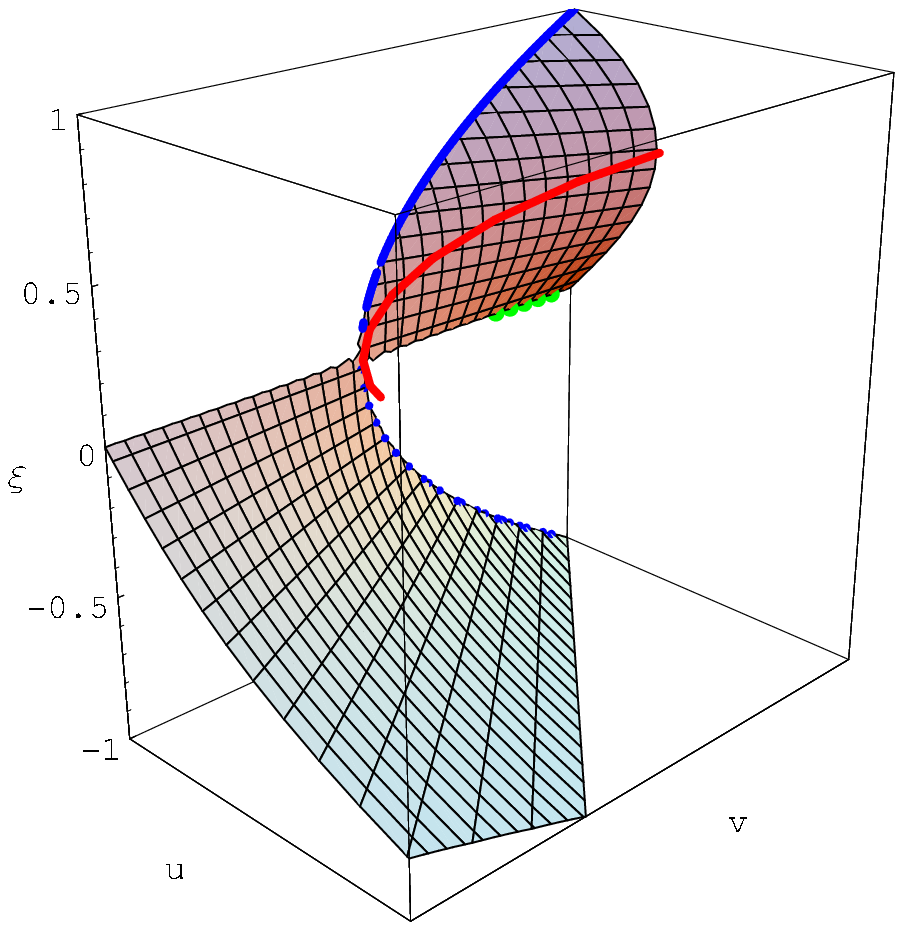} & &
\includegraphics[width=.25\linewidth]{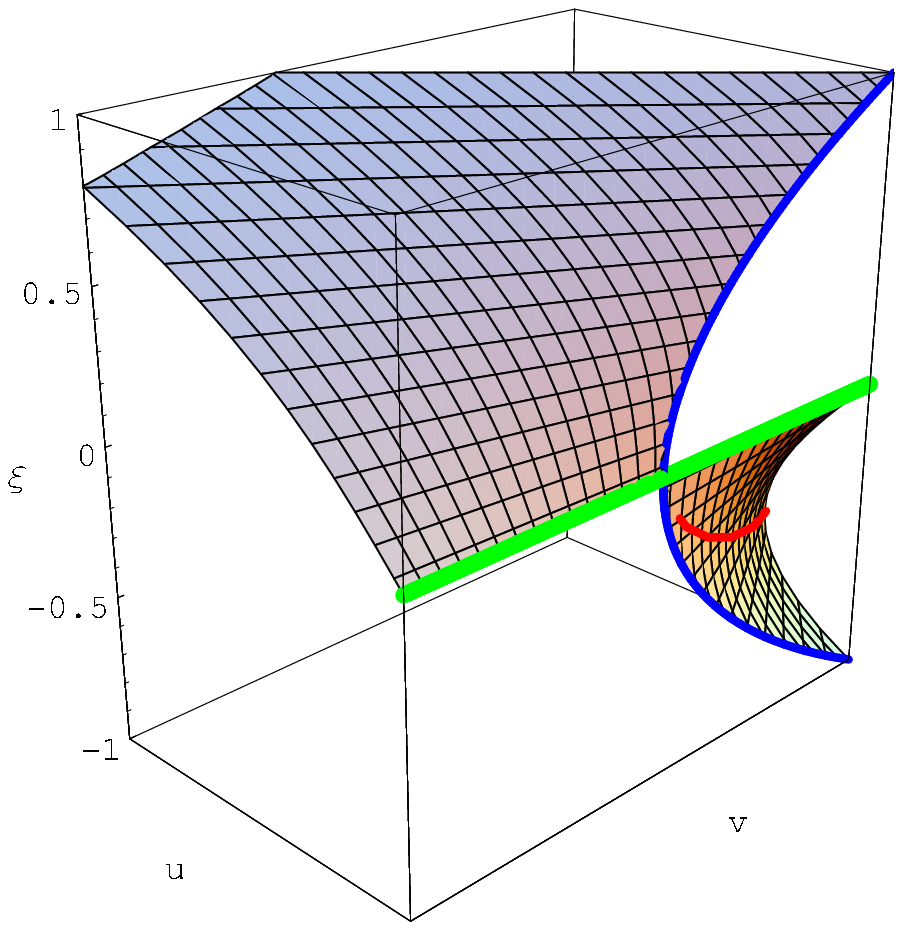} \\
(i) & & (ii) & & (iii)
\end{tabular} 
\caption{\label{fig:rk3-CV}
(i) is a local picture of spectral surface $C_V$ around a point 
where $D$ and $r$ intersect as (a) in the classification in the text.
Among the three local coordinates $(\xi, u, v)$ ($x$ is used as axes
  label in the figures instead of $\xi$), $(u,v)$ are for the base 
2-fold $B_2$, and $\xi$ is for the elliptic fiber. 
$(u,v)=(0,0)$ corresponds to a type (a) point in 
$\bar{c}_{\wedge^2 V} = \bar{c}_V$, and $\xi=0$ to the zero section 
$\sigma$. (ii) and (iii) cut out $u \geq 0$ and $u \leq 0$ parts of (i), 
so that curves $\bar{c}_V = \bar{c}_{\wedge^2 V}$ (thick green), 
$D$ (thin blue) and $r$ (thin red) are clearly visible. 
See the appendix~\ref{ssec:r-D} for more details.}
\end{center}
\end{figure}

The remaining task is to examine $\pi_{D*}(r|_D - R)/2$. Because 
the spectral surface $C_V$ is ramified over $\sigma$ whenever 
$D \subset C_V$ is on $\bar{c}_{\wedge^2 V}$, we begin with 
classifying the intersection points of the two divisors 
$r$ and $D$ in $C_V$. For rank-3 bundles $V$, there are 
two types of $r$--$D$ intersection points on $C_V$:
\begin{itemize}
 \item [(a)] $p_j = p_k = e_0$, where $p_i = e_0$, too,
 \item [(b)] $p_j = p_k = e'$, where $e'$ denotes 
one of three points of order two in an elliptic curve $E_b$ 
(i.e., $e' \boxplus e' = e_0$).
\end{itemize}
Figure~\ref{fig:rk3-CV} shows the behavior of the spectral surface $C_V$ 
around a $D$--$r$ intersection point of type (a). From the figure, 
one can read off that 
\begin{equation}
 {\rm deg} \; r|_D = D \cdot r = 2, \qquad {\rm deg} \; R = 1
\end{equation}
at each type (a) $D$--$r$ intersection point. 
Explicit calculation of ${\rm deg} \; r|_D$ at type (a) 
intersection points is found in the appendix~\ref{ssec:r-D}.
On the other hand, at a type (b) $D$--$r$ intersection, one can see 
that 
\begin{equation}
 {\rm deg} \; r|_D = 1, \qquad {\rm deg} \; R = 1.
\end{equation}
Therefore, ${\rm deg} \; (r|_D - R) = 1$ remains valid at each type (a) 
$D$--$r$ intersection points, while ${\rm deg} \; (r|_D - R) = 0$ 
at type (b) $D$--$r$ intersections. Thus, by denoting collection of 
the image of all the type (a) $D$--$r$ intersection points by 
$\pi_D$ as $b^{(a)}$, we find that 
\begin{equation}
 \frac{1}{2} \pi_{D*}(r|_D - R) = \frac{1}{2} b^{(a)}. 
\label{eq:rDR-rk3}
\end{equation}

One can see that (a) and (b) exhaust all the $r$--$D$ intersection points.
First, type (a) intersection points mapped by $\pi_D$ to 
$\bar{c}_V = \bar{c}_{\wedge^2 V}$ are characterized by 
$a_3 = 0$ and $a_2 = 0$. Thus, there are 
\begin{equation}
 {\rm deg} \; b^{(a)} = (3K_{B_2} + \eta) \cdot (2K_{B_2} + \eta)
\label{eq:r-D-typeA-rk3}
\end{equation}
of them. 
Type (b) intersection points are characterized by the 
intersection of the curve $D$ and a locus of order-2 points, 
$\sigma'$. $\sigma'$ (denoted by $\sigma_2$ in \cite{Penn5}) 
is topologically $\sigma' \sim 3 \sigma - 3 K_{B_2}$ \cite{AC1}.
Using the topological form of $D$ in (\ref{eq:D-topology}) for 
$N=3$, we find that there are 
\begin{equation}
 D \cdot \sigma' = \left[\sigma \cdot (6 K_{B_2} + 2 \eta) + 
 \pi_Z^* \eta \cdot (3K_{B_2}+\eta) \right] \cdot 
  (3 \sigma - 3 \pi_Z^*K_{B_2}) = 
 \sigma \cdot 3 \eta \cdot (3K_{B_2} + \eta)
\label{eq:r-D-typeB-rk3}
\end{equation}
type (b) $r$--$D$ intersection points. 
Now remembering that ${\rm deg} \; r|_D = 2$ at each 
type (a) intersection point and $=1$ at type (b) 
intersection points, it is easy to see that 
the intersection number 
\begin{eqnarray}
 D \cdot r & = & \left[\sigma \cdot (6 K_{B_2} + 2 \eta) + 
 \pi_Z^* \eta \cdot (3K_{B_2}+\eta) \right] \cdot 
 \left[3 \sigma + \pi_Z^* (\eta - K_{B_2})\right], \\
 & = & (3 K_{B_2} + \eta) \cdot \sigma \cdot 
  \left(2 \left(2K_{B_2} + \eta \right)+ 3\eta\right)
\end{eqnarray}
is accounted for by the type (a) and type (b) intersection points.
We used $r \sim K_{C_V} - \pi^*_C K_{B_2} \sim 
N \sigma + \pi_C^* (\eta - K_{B_2})$ for $N=3$.

Once we show that 
\begin{equation}
 j^* r = r|_{\bar{c}_V} = b^{(a)}, 
\label{eq:jr=ba}
\end{equation}
then (\ref{eq:F4V2}), (\ref{eq:gamma-V2-rk3}) and (\ref{eq:rDR-rk3})
reproduce (\ref{eq:F4V2-rk3}). To see this relation between 
$r|_{\bar{c}_V}$ and $b^{(a)}$, it is sufficient to count the number 
of $\bar{c}_V$--$r$ intersection points in $C_V$. This is because 
${\rm deg}\; j^* r = 1$ at each type (a) intersection point 
(as one can see intuitively from Figure~\ref{fig:rk3-CV}, or from 
explicit calculation in the appendix~\ref{ssec:r-D}).
The intersection number $\bar{c}_V \cdot r$ in $C_V$ is 
given by 
\begin{equation}
{\rm deg} \; j^* r =  \bar{c}_V \cdot r 
 = \sigma \cdot (N\sigma + \pi_C^*(\eta - K_{B_2}))
 = \sigma \cdot ((N-1) K_{B_2} + \eta)
 = (N K_{B_2} + \eta) \cdot ((N-1) K_{B_2} + \eta)
\label{eq:cv-r}
\end{equation}
for $N=3$, and hence is the same as (\ref{eq:r-D-typeA-rk3}).
We have finally seen that (\ref{eq:F4V2}) reproduces 
(\ref{eq:F4V2-rk3}) properly. 

Once the sheaf (and in particular, line bundle) for 
$\wedge^2 V$ is obtained, its net chirality follows immediately. 
Using the Riemann--Roch theorem on the matter curve 
$\bar{c}_{\wedge^2 V} = \bar{c}_V$, 
\begin{eqnarray}
 \chi(\wedge^2 V) & = & 
  1-g(\bar{c}_{V}) +  {\rm deg} \; {\cal F}_{\wedge^2 V}, \\
  & = & 1 - g(\bar{c}_V) + {\rm deg} \; {\cal F}_{V^\times} 
   + {\rm deg} \; i^* \pi_{C*} \gamma \\
 & = &  - \chi(V) + (\pi_{C*} \gamma) \cdot (3K_B + \eta).
\label{eq:conj-rel-rk3}
\end{eqnarray}
This calculation confirms, using only the sheaves on the matter curves, 
that a consistency relation (\ref{eq:conj-rel}) between $\chi(V)$ 
and $\chi(\wedge^2 V)$ is satisfied. 

\subsection{Rank-4 Vector Bundles}
\label{ssec:rk4}

Let us now study $R^1\pi_{Y*} \wedge^2 V$ for rank-4 bundles $V$. 
The spectral surface of a rank-4 bundle $V$ is a zero locus of 
\begin{equation}
 s = a_0(u,v) + a_2(u,v) x + a_3(u,v) y + a_4(u,v) x^2,
\label{eq:CV-rk4-defeq}
\end{equation}
where $(u,v)$ are local coordinates on the base manifold $B_2$, 
and $(x,y)$ describe the elliptic fiber. The matter curve 
$\bar{c}_V$ for the fundamental representation $V$ is given by 
$a_4 = 0$, since one of the solutions becomes 
$(x,y)=(\infty, \infty)=e_0$.
The matter curve for $\wedge^2 V$, $\bar{c}_{\wedge^2 V}$ is 
determined by the condition that $s$ in (\ref{eq:CV-rk4-defeq})
factorizes as 
\begin{equation}
 s = (A x + B )(P x + Q).
\end{equation}
Thus, $\bar{c}_{\wedge^2 V}$ denotes the locus $a_3 = 0$.
If $s$ factorized\footnote{The factorization of $s$ means that the 
structure group of the bundle---one that is read out from 
the spectral surface---is reduced from $\SU(4)$ to 
$\SU(2) \times \SU(2)$. Thus, the commutant of this 
``structure group'' is enhanced from $\SO(10)$ to $\SO(12)$. 
This enhanced symmetry determines the form of enhanced singularity 
along the matter curve $\bar{c}_{\wedge^2 V}$.} for a point $b \in B_2$,
a condition $A x + B = 0$ determines two points in $E_b$.
They are in a relation $p_i = (x,y)$ and $p_j = (x, -y)$, with 
$x = - B/A$. Thus, $p_i \boxplus p_j = e_0$, and hence 
$b \in \bar{c}_{\wedge^2 V}$.

Along the matter curve $\bar{c}_{\wedge^2 V}$, there is another 
pair of points in $C_V|_{E_b}$, 
$p_k = (x, y)$ and $p_l = (x,-y)$ with $x = -Q/P$ satisfying  
$p_k \boxplus p_l = e_0$. Thus, all the four points in each fiber 
of $C_V|_{E_b}$ for $b \in \bar{c}_{\wedge^2 V}$ belong to 
the component $D \subset C_V \cdot Y$. Thus, the irreducible
decomposition (\ref{eq:DD'}) becomes 
\begin{equation}
 C_V \cdot Y = D
\end{equation}
for rank-4 bundles $V$. $\pi_D: D \rightarrow \bar{c}_{\wedge^2 V}$ 
is now a degree-4 cover. 

The spectral surface $C_{\wedge^2 V}$ forms a double curve along 
the locus where it intersects with the zero section. One branch 
corresponds to $p_i \boxplus p_j$ and the other to $p_k \boxplus p_l$.
Once $Z$ is blown-up along $\bar{c}_{\wedge^2 V}$ and 
the double-curve singularity of $C_{\wedge^2 V}$ is resolved, 
each one of generic points of $\bar{c}_{\wedge^2 V}$ is doubled, 
one for $p_i \boxplus p_j$ and the other for $p_k \boxplus p_l$, 
and such points form the covering matter curve
$\tilde{\bar{c}}_{\wedge^2 V}$. 
A degree-2 cover $\tilde{\pi}_D: D \rightarrow 
\tilde{\bar{c}}_{\wedge^2 V}$ is defined naturally, but 
$\nu_{\bar{c}_{\wedge^2 V}}: \tilde{\bar{c}}_{\wedge^2 V} \rightarrow 
\bar{c}_{\wedge^2 V}$ is also a degree-2 cover everywhere\footnote{
The topological class of $\bar{c}_{\wedge^2 V} \in |3 K_{B_2} + \eta|$
is different from a naive expectation (\ref{eq:cVofV2}) for $N=2$ 
by a factor of two, because of this doubling was not taken into 
account there. } on
$\bar{c}_{\wedge^2 V}$. This is how 
$\nu_{\bar{c}_{\wedge^2 V}} \circ \tilde{\pi}_D = \pi_D$ becomes a degree-4 cover.

At some special points on the matter curve $\bar{c}_{\wedge^2 V}$, 
$(Ax + B) = 0$ and $(P x + Q) = 0$ determine the same pair of 
points in the fiber. This happens when 
\begin{equation}
 R^{(4)} := a_2^2 - 4 a_4 a_0 = (A Q - B P)^2 = 0
\label{eq:def-R4}
\end{equation}
on $\bar{c}_{\wedge^2 V}$.
One can further see (in the appendix~\ref{ssec:pinch}) 
that $p_i \boxplus p_j$ and $p_k \boxplus p_l$ are interchanged 
as a result of monodromy around a zero point of $R^{(4)}$ 
on a complex curve $\bar{c}_{\wedge^2 V}$. 
Thus, the covering matter curve $\tilde{\bar{c}}_{\wedge^2 V}$ 
is ramified on $\bar{c}_{\wedge^2 V}$ over zero locus of $R^{(4)}$. 
We see in the appendix~\ref{ssec:pinch} that $C_{\wedge^2 V}$ 
develops a codimension-2 singularity at a zero point of $R^{(4)}$. 
\begin{figure}[tbp]
 \begin{center}
\begin{tabular}{c|cc}
\includegraphics[width=.3\linewidth]{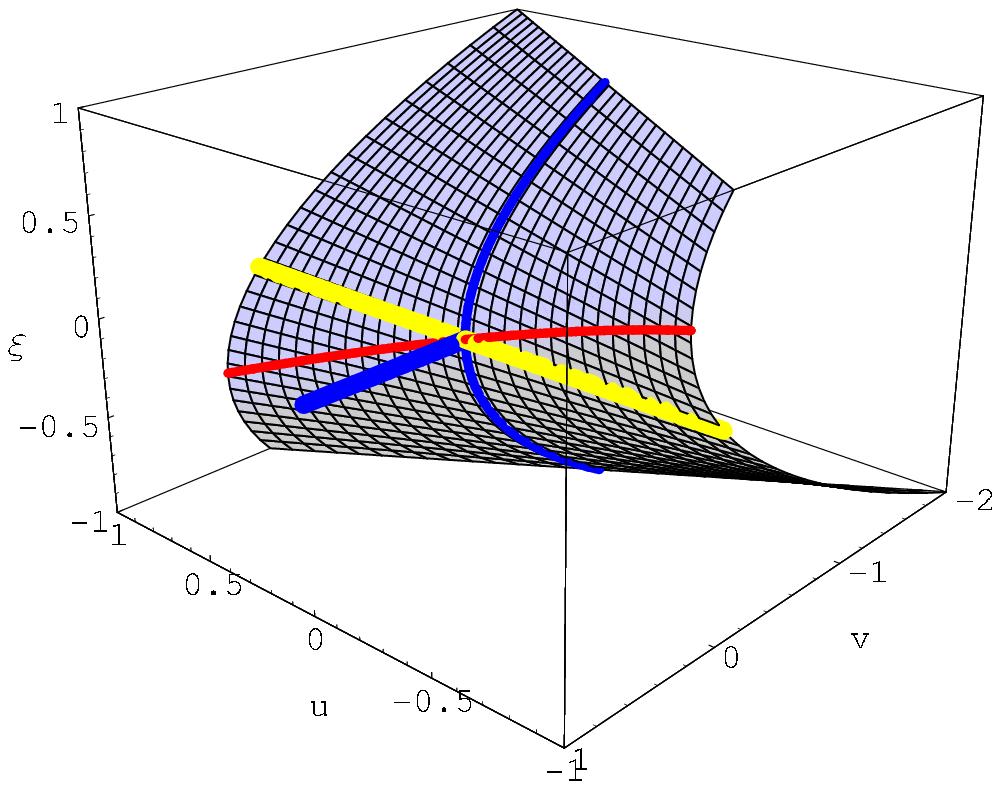}  &
\includegraphics[width=.23\linewidth]{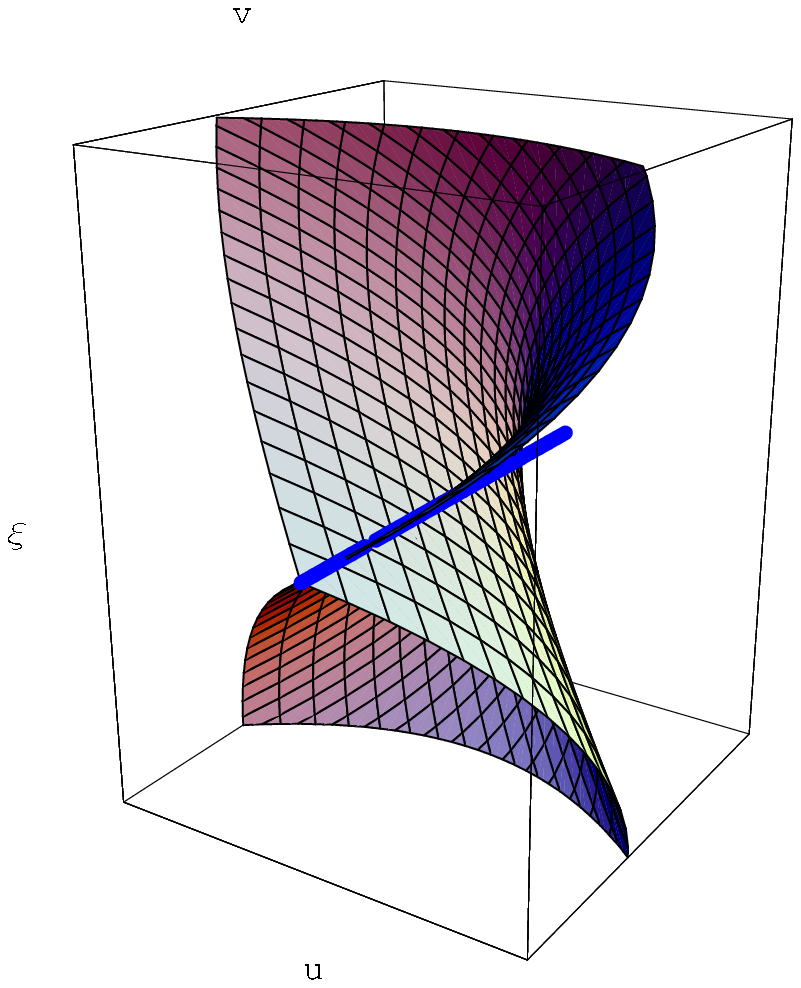} & 
\includegraphics[width=.23\linewidth]{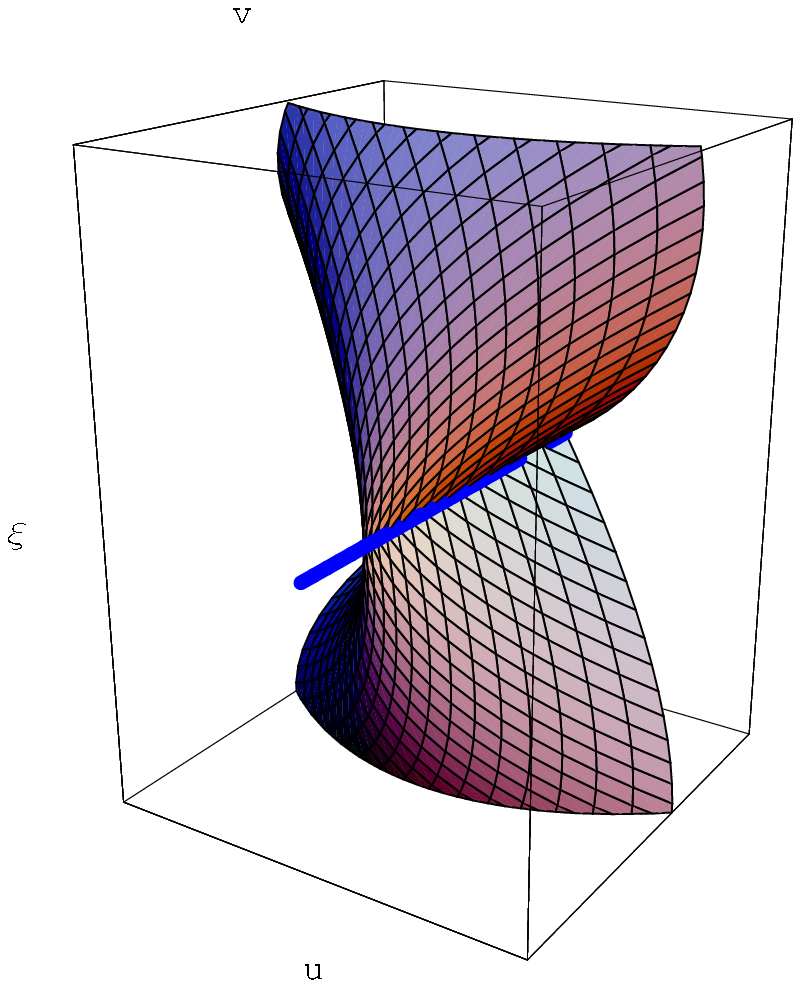} \\
$C_V$ at type (a) & $C_{\wedge^2 V}$ at type (c) & 
$C_{\wedge^2 V}$ at type (c)
\end{tabular}
\caption{\label{fig:rk4-CV}
Coordinates $(u,v)$ describe base manifold $B_2$, and $\xi$ 
the fiber direction, with a zero section corresponding to $\xi=0$.
The left panel shows a local picture of $C_V$ for a rank-4 
vector bundle around type (a) points on $\bar{c}_{\wedge^2 V}$.
Local coordinates $(u,v)$ on the base is chosen so that $a_3 = u$ 
and $a_4 = v$. $\bar{c}_V$ (thick yellow) and $\bar{c}_{\wedge^2 V}$ 
(thick blue) in the zero section are given by $a_4 = 0$ and $a_3 = 0$, 
respectively. Curves $D$ and $r$ on $C_V$ are also shown by 
thin blue and thin red curves, respectively. 
Two right panels show local pictures of $C_{\wedge^2 V}$ around 
type (c) points viewed from opposite directions. 
Local coordinates are chosen so that $a_3 = u$ (a direction transverse 
to $\bar{c}_{\wedge^2 V}$), and $a_2^2 - 4a_0 a_4 = v$ on the base manifold. 
The matter curve $\bar{c}_{\wedge^2 V}$ (thick blue) is a double curve 
in $C_{\wedge^2 V}$, and a more complicated singularity---called a pinch
  point---is developed at a type (c) point.    
}  
 \end{center}
\end{figure}
This codimension-2 singularity 
cannot be avoided generically on the matter curve 
$\bar{c}_{\wedge^2 V}$. We introduce a divisor 
\begin{equation}
 b^{(c)} := {\rm div} R^{(4)}
\label{eq:def-bc-rk4}
\end{equation}
on the matter curve. There are 
\begin{equation}
 {\rm deg} \; b^{(c)} = (3K_{B_2} + \eta) \cdot (4K_{B_2} + 2 \eta)
\label{eq:nbr-bc-rk4}
\end{equation}
such codimension-2 singularities of $C_{\wedge^2 V}$, which 
are also the branch points of the degree-2 cover 
$\nu_{\bar{c}_{\wedge^2 V}}: \tilde{\bar{c}}_{\wedge^2 V} \rightarrow \bar{c}_{\wedge^2 V}$.

Apart from these isolated points on $\bar{c}_{\wedge^2 V}$, 
the idea presented in section~\ref{sec:Idea} can be used to 
determine the sheaf $\widetilde{\cal F}_{\wedge^2 V}$ on 
$\tilde{\bar{c}}_{\wedge^2 V}$.
$\tilde{\pi}_D: D \rightarrow \tilde{\bar{c}}_{\wedge^2 V}$ determines 
a divisor $\tilde{\pi}_{D*} \gamma$ on the covering matter curve 
$\tilde{\bar{c}}_{\wedge^2 V}$. The curve $D$ is ramified over 
$\tilde{\bar{c}}_{\wedge^2 V}$ at two types of points, 
\begin{itemize}
 \item [(a)] $p_i = p_j = e_0$,
 \item [(b)] $p_i = p_j = e'$.
\end{itemize}
Thus, there is a potential contribution $\tilde{\pi}_{D*} (r|_D - R)/2$ 
to the divisor determining $\widetilde{\cal F}_{\wedge^2 V}$. 
It turns out, however, that ${\rm deg} \; (r|_D - R) = 0$ at 
each one of type (a) or type (b) points. 
As we will see shortly, 
the $r$--$D$ intersection points of type (a) and type (b) 
account for all the $r$--$D$ intersection that are not 
in the fiber of the zero locus of $R^{(4)}$.
Thus, there is no contribution to the divisor from 
$\tilde{\pi}_{D*} (r|_D - R)/2$ on $\tilde{\bar{c}}_{\wedge^2 V}$ 
away from the zero locus of $R^{(4)}$, and 
\begin{equation}
 \widetilde{\cal F}_{\wedge^2 V} = 
  {\cal O}_{\tilde{\bar{c}}_{\wedge^2 V}} (\tilde{\pi}_{D*} \gamma).
\end{equation}

The remaining $r$--$D$ intersection points on $C_V$ come from 
type (c) intersection points 
\begin{itemize}
 \item [(c)] $p_i = p_k =: p_+$ and $p_j = p_l =: p_-$.
\end{itemize}
These points are in the fiber of the zero locus of $R^{(4)}$.
The number of all those types of $r$--$D$ intersection points are
given by 
\begin{eqnarray}
 \# ({\rm a}) & = & D \cdot \sigma = 
  (4K_{B_2} + \eta) \cdot (3K_{B_2} + \eta), \\
 \# ({\rm b}) & = & D \cdot \sigma' = D \cdot 3( \sigma - K_{B_2}), \\
 \# ({\rm c}) & = & 2 \times {\rm deg} \; b^{(c)} 
  = 2\, \bar{c}_{\wedge^2 V} \cdot (4K_{B_2} + 2 \eta).
\end{eqnarray}
All these $r$--$D$ intersection points contribute to the 
intersection number $D \cdot r$ with unit multiplicity. 
We can now see that $D \cdot r$ is accounted for by 
these intersection points:
\begin{eqnarray}
 D \cdot r & = & D \cdot (4\sigma + \eta - K_{B_2}), \\  
   & = & D \cdot \sigma + D \cdot (3\sigma - 3 K_{B_2})
    + D \cdot (2K_{B_2} + \eta), \\
  & = & \# (\rm{ a}) + \# (\rm{ b}) + \# (\rm{ c}).
\end{eqnarray}

Let us now study the structure of $R^1\pi_{Z*} \wedge^2 V$ 
in a local neighborhood of a zero point of $R^{(4)}$.
We assume\footnote{\label{fn:pushforward}
It is not obvious whether $R^1\pi_{Z*} \wedge^2 V$
is represented as $i_{\wedge^2 V *}{\cal F}$ as a sheaf of 
${\cal O}_{B_2}$-module, although the support of $R^1\pi_{Z*} \wedge^2 V$ 
is $\bar{c}_{\wedge^2 V}$. See the appendix~\ref{sec:push} for more.
To show that this is the case, we need to see that the ideal sheaf of 
$\bar{c}_{\wedge^2 V}$ acts trivially on $R^1\pi_{Z*} \wedge^2 V$.
We have seen in the appendix~\ref{sec:push} and section~\ref{sec:Idea}
that this is true for generic points on the matter curve
$\bar{c}_{\wedge^2 V}$. Double-curve singularity on $C_{\wedge^2 V}$
does not pose a problem. However, we have not shown this 
in a neighborhood containing a zero locus of $R^{(c)}$. \\
As we see in the appendix~\ref{ssec:pinch}, $C_{\wedge^2 V}$ approaches 
the zero section as either $\xi \sim \pm (w_+ - w_-)$ or 
$\xi \sim \pm (w_+ + w_-)$. Since the normal coordinate of
$\bar{c}_{\wedge^2 V}$ is $u \propto (w_+ + w_-)(w_+ - w_-)$ 
around the zero locus of $R^{(4)}$, 
it sounds quite reasonable that the normal coordinate acts trivially 
on the generator of $R^1\pi_{Z*} \wedge^2 V$, just like the normal 
coordinate does in the case 
discussed in the appendix~\ref{sec:push}. But, we have not completed 
a proof, and we just leave this as an assumption.
} that $R^1 \pi_{Z\ast}\wedge^2 V$ is written as 
$i_{\wedge^2 V \ast}\mathcal{F}$ for some sheaf $\mathcal{F}$ 
on $\bar{c}_{\wedge^2 V}$.  Then, 
\begin{align}
\mathcal{F} &\cong i_{\wedge^2 V \ast}^{\ast}R^1 \pi_{Z\ast}\wedge^2 V \\
&\cong R^1 \pi_{Y\ast}(\wedge^2 V|_{Y}),
\label{eq:begin}
\end{align}
where we used the base change formula in the second isomorphism.

\begin{figure}[t]
\begin{center}
\includegraphics[width=.8\linewidth]{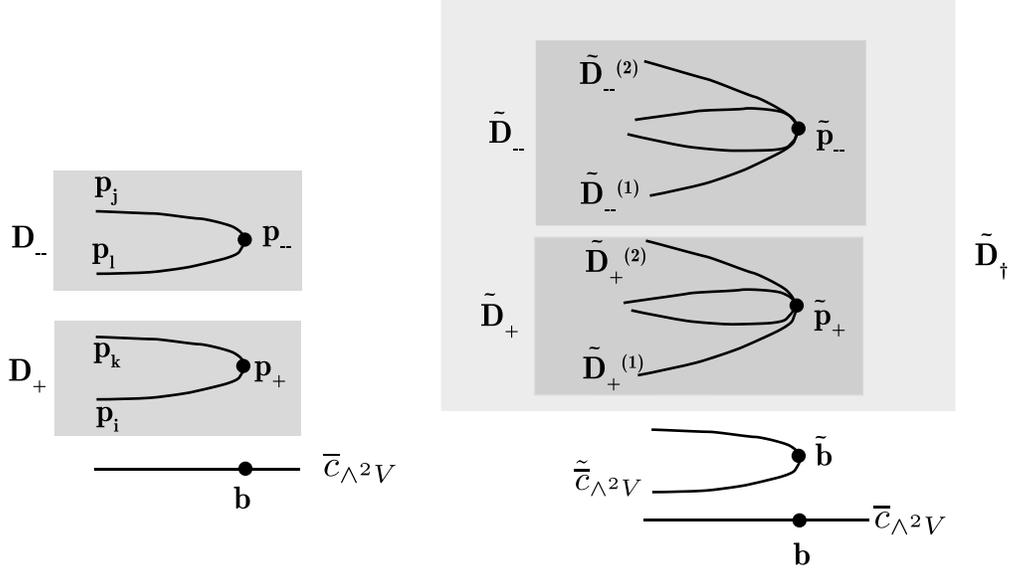}   
 \caption{\label{fig:D's}
A schematic picture showing relations between various curves and points 
that are used in the text.}
\end{center}
\end{figure}

Let $b\in \bar{c}_{\wedge^2 V}$ be a zero locus of $R^{(4)}$. 
Locally (in the analytic topology) around $b$, the curve $D$ 
is decomposed into a disjoint union of $D_{+}$ and $D_{-}$. 
We consider the following diagram, 
\begin{equation}
\vcenter{\xymatrix{
& \tilde{D}_{\dag}\times _{\tilde{\bar{c}}_{\wedge^2 V}}\tilde{Y}
\ar[dl]_{p_1} 
\ar[dr]^{p_2}\ar[r]^{\nu_{\tilde{D}_{\dag}}} & 
D\times_{\bar{c}_{\wedge^2 V}}Y \ar[dr]^{p_2} & \\
\tilde{D}_{\dag} \ar[dr]_{\pi_{\tilde{D}}}
&&\tilde{Y}\ar[dl]^{\pi_{\tilde{Y}}} \ar[r]^{\nu_{Y}} & Y \ar[dl]^{\pi_Y} 
\\
& \tilde{\bar{c}}_{\wedge^2 V} \ar[r]_{\nu_{\bar{c}_{\wedge^2 V}}} 
& \bar{c}_{\wedge^2 V}. & 
}}
\end{equation}
Here $\tilde{Y}= \tilde{\bar{c}}_{\wedge^2 V} 
\times_{\bar{c}_{\wedge^2 V}} Y$ (as we have already introduced 
in section~\ref{sec:Idea}), and 
$\tilde{D}_{\dag}=\tilde{\bar{c}}_{\wedge^2 V} 
\times_{\bar{c}_{\wedge^2 V}}D$. 
We have the decompositions, 
$$\tilde{D}_{\dag}=\tilde{D}_{+}\coprod \tilde{D}_{-},
\quad \tilde{D}_{\pm}=\tilde{D}_{\pm}^{(1)} \cup 
\tilde{D}_{\pm}^{(2)},$$
where $\tilde{D}_{\pm}=\tilde{\bar{c}}_{\wedge^2 V} 
\times_{\bar{c}_{\wedge^2 V}}D_{\pm}$ and 
$\tilde{D}_{\pm}^{(i)}$ for $i=1, 2$ are irreducible components of 
$\tilde{D}_{\pm}$.
Note that each $\tilde{D}_{\pm}^{(i)}$ is a section of 
$\pi_{\tilde{Y}}$, and $\tilde{D}_{\pm}^{(1)}$, 
$\tilde{D}_{\pm}^{(2)}$ intersect at one point transversally,
say $\tilde{p}_{\pm} \in \tilde{D}_{\pm}^{(1)}\cap \tilde{D}_{\pm}^{(2)}$. 
Moreover we may assume 
$\tilde{D}_{+}^{(1)} \boxplus \tilde{D}_{-}^{(2)}$
and $\tilde{D}_{-}^{(1)} \boxplus \tilde{D}_{+}^{(2)}$
are zero sections of $\pi_{\tilde{Y}}$. 
See Figure~\ref{fig:D's}.
Since $\nu_{\bar{c}_{\wedge^2 V}}$ is a Galois cover with Galois 
group $G=\mathbb{Z}/2\mathbb{Z}$, we have 
$$R^1\pi_{Y_\ast}(\wedge^2 V|_{Y}) \cong 
\left( \nu_{\bar{c}_{\wedge^2 V}\ast}\nu_{\bar{c}_{\wedge^2 V}}^{\ast}
R^1\pi_{Y_\ast}(\wedge^2 V|_{Y}) \right)^{G}.$$
By the base change formula, we have 
\begin{align}
\nu_{\bar{c}_{\wedge^2 V}}^{\ast}R^1\pi_{Y_\ast}(\wedge^2 V|_{Y})
\cong R^1\pi_{\tilde{Y}\ast}(\wedge^2 \nu_{Y}^{\ast}(V|_{Y})),
\end{align}
and
\begin{align}
\nu_{Y}^{\ast}V|_{Y}
&\cong p_{2\ast}(p_1^{\ast}{\cal N}_V|_{\tilde{D}_{\dag}}\otimes 
\mathcal{P}_{\tilde{\bar{c}}_{\wedge^2 V}}) \\
&\cong p_{2\ast}(p_1^{\ast}{\cal N}_V|_{\tilde{D}_{+}}\otimes 
\mathcal{P}_{\tilde{\bar{c}}_{\wedge^2 V}}) \oplus 
p_{2\ast}(p_1^{\ast}{\cal N}_V|_{\tilde{D}_{-}}\otimes 
\mathcal{P}_{\tilde{\bar{c}}_{\wedge^2 V}}), 
\end{align}
Here $\mathcal{P}_{\tilde{\bar{c}}_{\wedge^2 V}}$, 
${\cal N}_V|_{\tilde{D}_{\ast}}$ for $\ast=\dag, \pm$
are pullbacks of $\mathcal{P}_B$, $N_V|_{\tilde{D}_{\ast}}$
via $\nu_{\tilde{D}_{\dag}}$ and the second projection
$\tilde{D}_{\ast} \to D$ respectively. 
Let us set $W_{\pm}=p_{2\ast}(p_1^{\ast}{\cal N}_V|_{\tilde{D}_{\pm}}
\otimes \mathcal{P}_{\tilde{\bar{c}}_{\wedge^2 V}})$. 
We have
\begin{align}
\nu_{\bar{c}_{\wedge^2 V}}^{\ast}R^1\pi_{Y_\ast}(\wedge^2 V|_{Y})
\cong
R^1 \pi_{\tilde{Y}\ast}(W_{+}\otimes W_{-}).
\end{align}

It is useful in calculating $R^1\pi_{\tilde{Y}*} (W_+ \otimes W_-)$
to note that 
\begin{align}
0 \to \mathcal{O}_{\tilde{D}_{\pm}} \to \mathcal{O}_{\tilde{D}_{\pm}^{(1)}}
\oplus \mathcal{O}_{\tilde{D}_{\pm}^{(2)}} \to
 \mathcal{O}_{\tilde{p}_{\pm}}  \to 0
\end{align}
are exact. Applying $\otimes {\cal N}_{V}|_{\tilde{D}_{\dag}}$
and Fourier-Mukai transforms, we obtain the exact sequences, 
\begin{align}
0 \to W_{\pm} \to W_{\pm}^{(1)}\oplus W_{\pm}^{(2)} \to W_{\pm}^{(0)} \to 0.
\end{align} 
Here $W_{\pm}^{(\ast)}$ for $\ast=0, 1, 2$ are Fourier--Mukai transforms, 
\begin{align}
W_{\pm}^{(\ast)}=
p_{2\ast}(p_1^{\ast}{\cal N}_V|_{\tilde{D}_{\pm}^{(\ast)}}\otimes 
\mathcal{P}_{\tilde{\bar{c}}_{\wedge^2 V}}),
\end{align}
where $\tilde{D}_{\pm}^{(0)}=\tilde{p}_{\pm}$. 
Note that $W_{\pm}^{(\ast)}$ for $\ast=1, 2$ is a line bundle 
on $Y$ and $W_{\pm}^{(0)}$ is a line bundle on the fiber 
$\tilde{Y}_b=\pi_{\tilde{Y}}^{-1}\nu_{\bar{c}_{\wedge^2 V}}^{-1}(b)$. 

Applying $\otimes W_{-}^{(1)}$ to the above sequence yields the 
exact sequence, 
\begin{align}
0 \to W_{+}\otimes W_{-}^{(1)} \to (W_{+}^{(1)}
\otimes W_{-}^{(1)})\oplus (W_{+}^{(2)}
\otimes W_{-}^{(1)}) \to W_{+}^{(0)}
\otimes W_{-}^{(1)} \to 0.
\end{align}
$W_+^{(0)} \otimes W_-^{(1)}$ is a trivial line bundle 
on $\tilde{Y}_b$, and $W_+^{(2)} \otimes W_-^{(1)}$ 
is also a line bundle that is trivial in the elliptic 
fiber direction. 
Thus, by applying $R^i \pi_{\tilde{Y}\ast}$, we have the long exact sequence, 
\begin{equation}
%
\vcenter{\xymatrix@R=10pt@M=4pt@H+=22pt{
0 \ar[r] &  R^0 \pi_{\tilde{Y}*} (W_{+}\otimes W_{-}^{(1)}) \ar[r] & 
  {\cal O}_{\tilde{\bar{c}}_{\wedge^2 V}} 
   \left( \tilde{b} + \tilde{\pi}_{D*} \gamma \right)  \ar@{->>}[r]
 & \mathcal{O}_{\tilde{b}}   
 \ar`[rd]^<>(0.5){}`[l]`[dlll]`[d][dll] & 
\\
 &  R^1\pi_{\tilde{Y}\ast}(W_{+}\otimes W_{-}^{(1)}) \ar[r] 
& {\cal O}_{\tilde{b}} \oplus 
 {\cal O}_{\tilde{\bar{c}}_{\wedge^2 V}} 
    \left( \tilde{b} + \tilde{\pi}_{D*} \gamma +
     \tilde{\imath}^*_{\wedge^2 V} K_{B_2} \right)  \ar[r]
&  \mathcal{O}_{\tilde{b}}  \ar[r]
& 0, 
}}
\end{equation}
where $\tilde{b} := \nu^{-1}_{\bar{c}_{\wedge^2 V}}(b)$, 
$ \tilde{\imath}_{\wedge^2 V} := i_{\wedge^2 V} 
\circ \nu_{\bar{c}_{\wedge^2 V}}$, and we used 
\begin{eqnarray}
  R^0\pi_{\tilde{Y}*} (W_+^{(2)} \otimes W_-^{(1)}) & \cong & 
     \tilde{\pi}_{D_+*} ({\cal N}_V|_{D_+}) \otimes 
     \tilde{\pi}_{D_-*} ({\cal N}_V|_{D_-}) \cong 
   {\cal O}_{\tilde{\bar{c}}_{\wedge^2 V}} 
      (\tilde{b} + \tilde{\pi}_{D*} \gamma), \label{eq:R0-A}\\
  R^1\pi_{\tilde{Y}*} (W_+^{(2)} \otimes W_-^{(1)}) & \cong & 
   {\cal O}_{\tilde{\bar{c}}_{\wedge^2 V}} 
      (\tilde{b} + \tilde{\pi}_{D*} \gamma) \otimes {\cal L}_H^{-1}
   \cong {\cal O}_{\tilde{\bar{c}}_{\wedge^2 V}}
       (\tilde{b} + \tilde{\pi}_{D*} \gamma + 
         \tilde{\imath}_{\wedge^2 V}^* K_{B_2});
\end{eqnarray}
the ramification divisor $r$ on $C_V$ intersects with $D_\pm$ 
at $p_{\pm}$, and 
$\tilde{\pi}_{D_\pm} = \tilde{\pi}_D|_{D_\pm}: D_{\pm} \rightarrow 
\tilde{\bar{c}}_{\wedge^2 V}$ maps $p_\pm$ to 
$\tilde{b} \in \tilde{\bar{c}}_{\wedge^2 V}$. This is why we have 
a divisor $\tilde{b}$ in (\ref{eq:R0-A}).
%
%
We thus conclude that 
\begin{eqnarray}
 R^0 {\pi}_{\tilde{Y}*} (W_+ \otimes W_-^{(1)}) & \cong & 
   {\cal O}_{\tilde{\bar{c}}_{\wedge^2 V}} (\tilde{\pi}_{D*} \gamma), \\
 R^1 {\pi}_{\tilde{Y}*} (W_+ \otimes W_-^{(1)}) & \cong & 
   {\cal O}_{\tilde{\bar{c}}_{\wedge^2 V}} (\tilde{b} + 
     \tilde{\pi}_{D*} \gamma + \tilde{\imath}^*_{\wedge^2 V} K_{B_2}).
\end{eqnarray}
By the same argument, we also have the same results for 
$R^i \pi_{\tilde{Y}*} (W_+ \otimes W_-^{(2)})$ ($i=0,1$).
%

Finally we have the exact sequence, 
 \begin{align}
 0\to W_{+}\otimes W_{-} \to (W_{+}\otimes W_{-}^{(1)})\oplus 
 (W_{+}\otimes W_{-}^{(2)}) \to W_{+}\otimes W_{-}^{0} \to 0.
 \end{align}
Note that $W_{+}\otimes W_{-}^{0}$ is a rank two degree-zero sheaf
on an elliptic curve $\tilde{Y}_b$ given in \cite{Atiyah}.
Thus, we have the associated long exact sequence, 
%
%
%
%
\begin{equation}
\vcenter{\xymatrix@R=10pt@M=4pt@H+=22pt{
& \mathcal{O}_{\tilde{\bar{c}}_{\wedge^2 V}}(\tilde{\pi}_{D*}\gamma)
\oplus \mathcal{O}_{\tilde{\bar{c}}_{\wedge^2 V}}(\tilde{\pi}_{D*} \gamma)
 \ar@{->>}[r] & \mathcal{O}_{\tilde{b}}  \ar[r] &
R^1\pi_{\tilde{Y}\ast}(W_{+}\otimes W_{-})  
\ar`[rd]^<>(0.5){}`[l]`[dlll]`[d][dll] 
& \\
&
  \oplus^2
  \mathcal{O}_{\tilde{\bar{c}}_{\wedge^2 V}}(
  \tilde{b} + \tilde{\pi}_{D*} \gamma + 
          \tilde{\imath}^*_{\wedge^2 V} K_{B_2} ) \ar[r]
 & \mathcal{O}_{\tilde{b}} \ar[r] &  0.&
}}
\end{equation}
%
Therefore, we obtain 
\begin{eqnarray}
 R^1\pi_{\tilde{Y}*} (W_+ \otimes W_-) & \cong & 
   {\rm Ker} \left(
  {\cal O}(\tilde{b} + \tilde{\pi}_{D*} \gamma 
           + \tilde{\imath}^*_{\wedge^2 V} K_{B_2}) \oplus 
  {\cal O}(\tilde{b} + \tilde{\pi}_{D*} \gamma 
           + \tilde{\imath}^*_{\wedge^2 V} K_{B_2})
  \rightarrow {\cal O}_{\tilde{b}} \right), \nonumber \\
  & = & \left\{ (f,g)|f,g \in {\cal O}(\tilde{b} + \tilde{\pi}_{D*} \gamma 
  + \tilde{\imath}^*_{\wedge^2 V} K_{B_2}), 
    \quad f|_{\tilde{b}}=g|_{\tilde{b}} \right\}
\end{eqnarray}
Under the above isomorphism, we can easily see that the action 
of $G$ on $\nu_{\bar{c}_{\wedge^2 V}\ast} 
R^1\pi_{\tilde{Y}\ast}(W_{+}\otimes W_{-})$
is given by $(f(\tilde{u}), g(\tilde{u})) \mapsto 
(g(-\tilde{u}), f(-\tilde{u}))$, where $\tilde{u}$ is the local
coordinate of $\tilde{\bar{c}}_{\wedge^2 V}$ around $\tilde{b}$. 
Hence we have 
\begin{align}
\mathcal{F}_{\wedge^2 V} & \cong 
\left(\nu_{\bar{c}_{\wedge^2 V}\ast}
 R^1\pi_{\tilde{Y}\ast}(W_{+}\otimes W_{-})\right)^{G}\\
 & \cong \nu_{\bar{c}_{\wedge^2 V}*} 
 {\cal O}_{\tilde{\bar{c}}_{\wedge^2 V}}(\tilde{b} + \tilde{\pi}_{D*}
 \gamma + \tilde{\imath}^*_{\wedge^2 V} K_{B_2}).
\end{align}
Therefore, after making the assumption discussed in 
footnote~\ref{fn:pushforward}, we find that 
${\cal F}_{\wedge^2 V}$ on $\bar{c}_{\wedge^2 V}$
is given by a pushforward of a locally free rank-1 sheaf\footnote{
This result follows immediately from (\ref{eq:F4V2}), had we made 
a stronger assumption that there exists a {\em locally free rank-1} sheaf 
$\widetilde{\cal N}_{\wedge^2 V}$ on $\widetilde{C}_{\wedge^2 V}$ 
such that ${\cal N}_{\wedge^2 V} := \nu_{\bar{c}_{\wedge^2 V}*} 
\widetilde{\cal N}_{\wedge^2 V}$ is used in (\ref{eq:asPushForward})
and (\ref{eq:asPushForward}) is satisfied as a sheaf of 
${\cal O}_Z$-module. We opted to adopt a weaker assumption in this 
article, and provided a derivation after (\ref{eq:begin}), instead.
} 
\begin{equation}
 \widetilde{\cal F}_{\wedge^2 V} = {\cal O}\left(
  \tilde{\imath}^*_{\wedge^2 V} K_{B_2} + \tilde{b}^{(c)} 
  + \tilde{\pi}_{D*} \gamma \right) 
\label{eq:Ftil4V2-rk4}
\end{equation}
on $\tilde{\bar{c}}_{\wedge^2 V}$ via $\nu_{\bar{c}_{\wedge^2 V}}$
everywhere on $\bar{c}_{\wedge^2 V}$. Here, 
$\tilde{b}^{(c)}$ denotes a divisor 
$\nu_{\bar{c}_{\wedge^2 V}}^{-1} b^{(c)}$, collecting all the points 
that we have denoted as $\tilde{b}$ up to now.

Matter chiral multiplets from the $\wedge^2 V$ bundle 
are now identified with 
\begin{equation}
 H^1(Z; \wedge^2 V) \simeq H^0(\tilde{\bar{c}}_{\wedge^2 V}; 
   \widetilde{\cal F}_{\wedge^2 V}) 
  \simeq H^0(\bar{c}_{\wedge^2 V}; {\cal F}_{\wedge^2 V}).
\end{equation}
${\cal F}_{\wedge^2 V} = \nu_{\bar{c}_{\wedge^2 V} * } 
\widetilde{\cal F}_{\wedge^2 V}$ is a locally-free rank-2 sheaf 
(rank-2 vector bundle) on $\bar{c}_{\wedge^2 V}$.

The genus of the covering matter curve is given by 
\begin{equation}
 g(\tilde{\bar{c}}_{\wedge^2 V}) = 1 + 2 (g(\bar{c}_{\wedge^2 V}) - 1)
  + \frac{1}{2} {\rm deg} \; b^{(c)},
\end{equation}
since $\nu_{\bar{c}_{\wedge^2 V}}: \tilde{\bar{c}}_{\wedge^2 V} 
\rightarrow \bar{c}_{\wedge^2 V}$ is a degree-2 cover with 
$(1/2) {\rm deg} \; b^{(c)}$ branch cuts. Thus, it is also expressed 
as 
\begin{equation}
 g(\tilde{\bar{c}}_{\wedge^2 V}) = 1 + 
   (3K_{B_2} + \eta) \cdot (4 K_{B_2} + \eta) 
  + \frac{1}{2}(3K_{B_2} + \eta) \cdot (4K_{B_2} + 2 \eta),
\label{eq:genus-formula-kr4}
\end{equation}
and it follows that 
\begin{equation}
 {\rm deg} \; K_{\tilde{\bar{c}}_{\wedge^2 V}} = 
 2 \times (3K_{B_2} + \eta) \cdot (6K_{B_2} + 2\eta).
\end{equation}

On the other hand, one can also calculate the following independently:
\begin{eqnarray}
 {\rm deg} \; \left(\tilde{\imath}^*_{\wedge^2 V} K_{B_2} 
                    + \tilde{b}^{(c)} \right)
 & = & 2 (3K_{B_2} + \eta) \cdot K_{B_2} + (3K_{B_2} + \eta) \cdot
  (4K_{B_2} + 2 \eta), \nonumber \\
 & = & (3K_{B_2} + \eta) \cdot (6K_{B_2} + 2 \eta) 
= \frac{1}{2} {\rm deg} \; K_{\tilde{\bar{c}}_{\wedge^2 V}}.
\label{eq:relate-rk4}
\end{eqnarray}
Because of this non-trivial relation between the genus of the 
covering curve and the degree of the divisor above, we obtain 
through Riemann--Roch theorem\footnote{The same result is obtained 
by applying the Riemann--Roch theorem to 
$\chi(\bar{c}_{\wedge^2 V}; {\cal F}_{\wedge^2 V})$. One needs to use
\begin{equation}
 c_1({\cal F}_{\wedge^2 V}) = 2 i^*_{\wedge^2 V} K_{B_2} + 
  \frac{1}{2} b^{(c)} + \pi_{D*} \gamma, 
\end{equation}
and a relation analogous to (\ref{eq:relate-rk4})
\begin{equation}
 {\rm deg} \; \left(2 i^*_{\wedge^2 V} K_{B_2} + \frac{1}{2} b^{(c)} \right)
  = {\rm deg} \; K_{\bar{c}_{\wedge^2 V}}.
\end{equation}
} that 
\begin{eqnarray}
 \chi(\wedge^2 V) & = & 
 \chi(\tilde{\bar{c}}_{\wedge^2 V}; \widetilde{\cal F}_{\wedge^2 V}) 
 = (1 - g(\tilde{\bar{c}}_{\wedge^2 V})) 
  + {\rm deg} \; \left(\tilde{\imath}^*_{\wedge^2 V} K_{B_2} 
                       + \tilde{b}^{(c)}\right)
  + \int_{\tilde{\bar{c}}_{\wedge^2 V}} \tilde{\pi}_{D*} \gamma 
  \nonumber \\
  & = & \int_{\tilde{\bar{c}}_{\wedge^2 V}} \tilde{\pi}_{D*} \gamma 
   = \int_{\bar{c}_{\wedge^2 V}} \pi_{D*} \gamma 
   = (3K_{B_2} + \eta) \cdot \pi_{C*} \gamma.
\label{eq:chi-rk4-V2-F}
\end{eqnarray}
This chirality formula in terms of (covering) matter curve and 
$\gamma$ was rather anticipated from the beginning. We know 
that $\chi(\wedge^2 V) = - \chi(\wedge^2 V^\times)$, and 
the difference between $V$ and $V^\times$ comes from changing 
the sign of $\gamma$. For an $\SU(4)$ bundle $V$, 
$\wedge^2 V \simeq \wedge^2 V^\times$ and the net chirality should 
vanish. We can confirm this in the formula above, because 
$\pi_{C*} \gamma = 0$ for an $\SU(4)$ bundle $V$. 
For a $\U(4)$ bundle $V$, its chirality formula (\ref{eq:chi-rk4-V2-F}) 
agrees with (\ref{eq:conj-rel}) in the appendix that is obtained 
without calculating direct images. All these consistency checks 
give us confidence that the locally free rank-1 sheaf 
(\ref{eq:Ftil4V2-rk4}) provides the right description 
for the matter multiplets from $\wedge^2 V$.

\subsection{Rank-5 Vector Bundles}
\label{ssec:rk5}

Spectral surface $C_V$ of rank-5 bundle $V$ is given by 
\begin{equation}
 s = a_0(u,v) + a_2(u,v) x + a_3(u,v) y + a_4(u,v) x^2 + a_5(u,v) xy = 0;
 \label{eq:CVeq-rk5}
\end{equation}
$(u,v)$ are local coordinates of a base 2-fold $B_2$, and 
$a_r$ ($r = 0,2,3,4,5$) are sections of ${\cal O}(r K_{B_2} + \eta)$.
The matter curve of the fundamental representation 
$\bar{c}_V = C_V \cdot \sigma$ is given by the zero locus of $a_5$, 
and hence belong to a topological class $|5K_{B_2} + \eta|$.
The matter curve of $\wedge^2 V$ is determined by requiring that 
the defining equation of the spectral surface factorizes\footnote{
``The structure group of the spectral surface'' is reduced 
from $\SU(5)$ to $\SU(2) \times \SU(3)$, and the commutant 
within $E_8$ enhanced from $\SU(5)_{\rm GUT}$ to $\SU(6)$. As we have 
already noted in section~\ref{ssec:rk4}, it is this ``structure group 
of the spectral surface'' that determines the enhanced singularity 
along the matter curve in F-theory dual description.} locally as 
\begin{equation}
 s = (A x + B)(P y + Q x + R).
\label{eq:factorize-rk5}
\end{equation}
Among the five points $\{ p_i, p_j, p_k, p_l, p_m \}$ 
satisfying (\ref{eq:factorize-rk5}) in a given elliptic fiber,  
two points $p_{i,j}$ satisfying $(A x + B) = 0$ satisfy 
$p_i \boxplus p_j = e_0$, as we discussed before 
in section~\ref{ssec:rk4}. One can see that this 
factorization condition is equivalent to 
\begin{equation}
 P^{(5)} := a_0 a_5^2 - a_2 a_3 a_5 + a_4 a_3^2 = 0
\label{eq:5bar-curve-eq}
\end{equation}
which was derived in \cite{FMW}.
Since the left-hand side is a section of ${\cal O}(10K_{B_2} + 3\eta)$,
$\bar{c}_{\wedge^2 V}$ belongs to a class $|10K_{B_2} + 3\eta|$, 
which corresponds to the $N=5$ case of (\ref{eq:cVofV2}). 

The two matter curves $\bar{c}_V$ and $\bar{c}_{\wedge^2 V}$ 
intersect in $B_2$ in general. There are two different types of 
intersection:
\begin{itemize}
 \item [(a)] $a_5=0$ and $a_4 = 0$, and hence $P^{(5)} = 0$,
 \item [(d)] $a_5=0$ and $a_3 = 0$, and hence $P^{(5)} = 0$. 
\end{itemize}
The two curves intersect with multiplicity 1 at any type (a)
intersection points, and with multiplicity 2 at any type (d) 
intersection points. This is a complete classification of 
the intersection points of the two matter curves, because 
\begin{eqnarray}
\# (\rm{ a}) + 2 \times \# (\rm{ d}) & = & 
  (5K_{B_2} + \eta) \cdot (4K_{B_2} + \eta) + 
 2 \times (5K_{B_2} + \eta) \cdot (3K_{B_2} + \eta), \nonumber \\
  & = & (5K_{B_2} +\eta) \cdot (10 K_{B_2} + 3 \eta)
 = \bar{c}_V \cdot \bar{c}_{\wedge^2 V} 
\end{eqnarray}
accounts for all the contributions to the intersection number. 
At the type (a) intersection points, five points in $C_V$ 
become $p_i = p_j = e_0$ and three general points. At the type (d)
intersection points, they become $p_m = e_0$, $p_i \boxplus p_j = e_0$
and $p_k \boxplus p_l = e_0$.  
\begin{figure}[tb]
 \begin{center}
  \begin{tabular}{ccc}
\includegraphics[scale=0.4
]{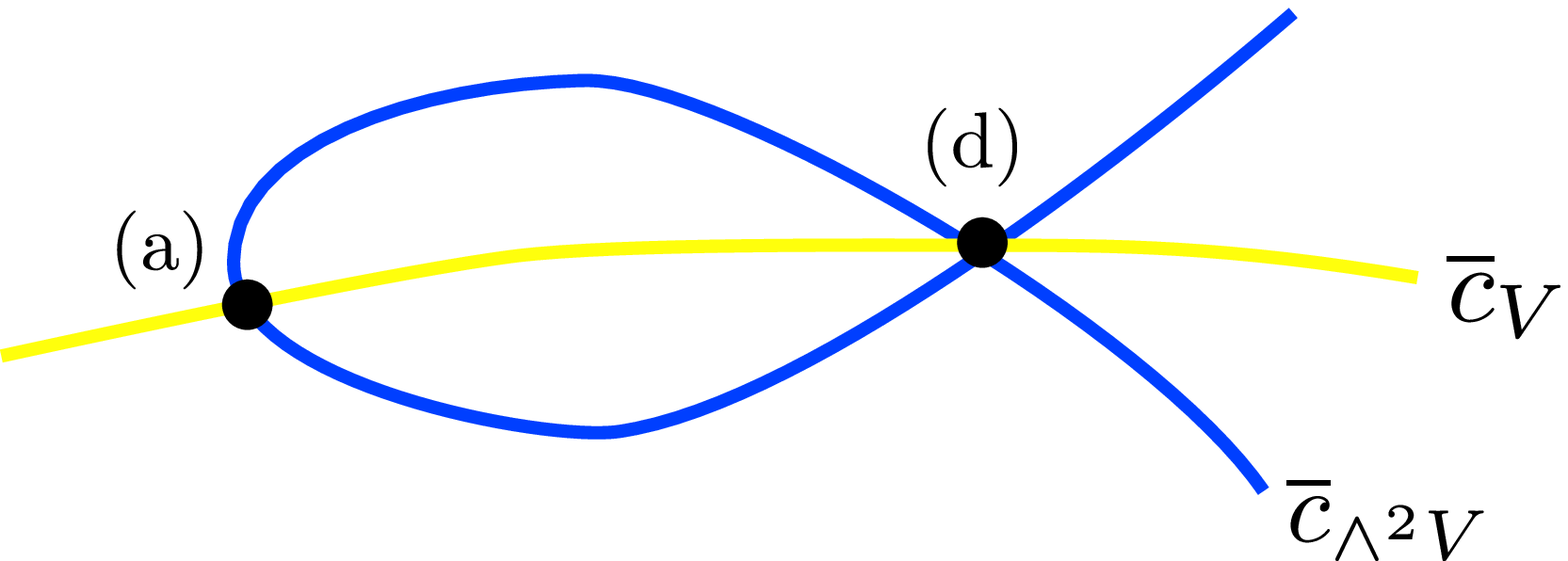} & ~~~&
\includegraphics[width=.3\linewidth]{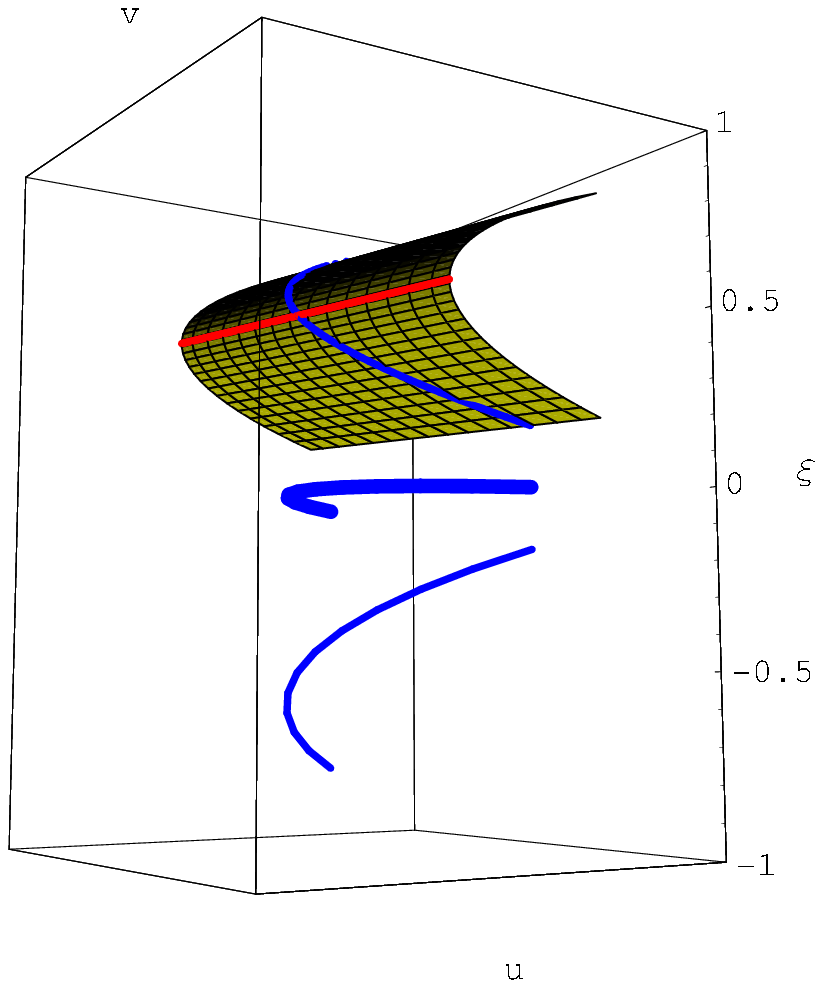}  \\ 
 (i)  & ~~~& (ii)
  \end{tabular}
\caption{\label{fig:rk5} The left panel (i) shows 
how the two matter curves $\bar{c}_V$ (thick yellow) and 
$\bar{c}_{\wedge^2 V}$ (thick blue) intersect in the zero section 
for cases with rank-5 bundles $V$. The right panel (ii) shows 
geometry of a curve $D$ (thin blue) and $\bar{c}_{\wedge^2 V}$ 
(thick blue) associated with a type (c1) point. 
Only a degree-2 part of degree-5 spectral cover surface $C_V$ 
is shown here. $C_V$ is a ramified cover over $B_2$ 
along a ramification divisor $r$ (thin red), but $D$ is not 
a ramified cover over $\bar{c}_{\wedge^2 V}$. 
}
 \end{center}
\end{figure}

The explicit form of $P^{(5)}$ reveals that $\bar{c}_{\wedge^2 V}$
forms a double point at each type (d) intersection point.
This is where a double-curve locus $p_i \boxplus p_j = p_k \boxplus p_l$
of $C_{\wedge^2 V}$ intersects with the zero section.
The projection $\pi_D: D \rightarrow \bar{c}_{\wedge^2 V}$ is a degree-2 cover at generic points on 
$\bar{c}_{\wedge^2 V}$, but the four points $p_{i,j,k,l}$ map 
to a type (d) point on $\bar{c}_{\wedge^2 V}$. 
The covering matter curve $\tilde{\bar{c}}_{\wedge^2 V}$ 
is obtained by blowing up the double points of the matter curve 
$\bar{c}_{\wedge^2 V}$, and the map $\tilde{\pi}_D: D\rightarrow 
\tilde{\bar{c}}_{\wedge^2 V}$ becomes a degree-2 cover.
The idea of section~\ref{sec:Idea} is applied, and 
a locally free rank-1 sheaf $\widetilde{\cal F}_{\wedge^2 V}$ 
on the covering matter curve $\tilde{\bar{c}}_{\wedge^2 V}$ is 
obtained. The sheaf ${\cal F}_{\wedge^2 V}$ on the matter curve 
$\bar{c}_{\wedge^2 V}$ is locally free and has rank 1 in 
a neighborhood of a generic point, but it is not locally free 
around the type (d) intersection points as the rank of the fiber 
jumps from 1 to 2.

We would like to better understand the locally free rank-1 sheaf 
(\ref{eq:F4V2}), by studying $\pi_{D*}(r|_D - R)$ 
in more detail. As we have discussed in section~\ref{ssec:rk3}, 
$C_V$ is ramified over $\sigma$ whenever $D$ is over
$\tilde{\bar{c}}_{\wedge^2 V}$. Thus, ${\rm supp} (r|_D - R) \subset 
{\rm supp}\, r|_D$ on $D$, and we begin with a classification of 
the $D$--$r$ intersection points on $C_V$.
This must include all the intersection points:
\begin{itemize}
 \item (a) $p_i = p_j = e_0$ ($p_i, p_j \in D$); 
 $\qquad \# ({\rm a}) = D \cdot \sigma$
 \item (b) $p_i = p_j = e'$ ($p_i, p_j \in D$); 
 $\qquad \# ({\rm b}) = D \cdot \sigma'$,
 \item (c) others.
\end{itemize}
The type (a) and type (b) intersection points exhaust 
all the cases where ${\rm deg} \; R \neq 0$. 
The type (a) $D$--$r$ intersection points are also the type (a) 
$\bar{c}_V$--$\bar{c}_{\wedge^2 V}$ intersection points.
As we have seen before for rank-4 bundles, 
${\rm deg}\; r|_D = {\rm deg}\; R = 1$ at both type (a) and type (b) 
$D$--$r$ intersections points, and no contribution to 
$\tilde{\pi}_{D*}(r|_D - R)/2$ arises from these points.\footnote{
Note added in version 4: it is now understood clearly \cite{Hayashi-2} 
why the type (a) and type (b) points do not contribute to the divisor 
of the line bundle (\ref{eq:F4V2}). Ramification behavior of
$C_{\wedge^2 V}$ is the key. See footnote 47 of \cite{Hayashi-2}.}

The images of the remaining $D$--$r$ intersections points---called 
type (c) points---via $\tilde{\pi}_{D*}$ define a divisor 
$\tilde{b}^{(c)}$ on $\tilde{\bar{c}}_{\wedge^2 V}$.
Since ${\rm deg} \; R = 0$ at the type (c) $D$--$r$ intersection 
points, $\tilde{\pi}_{D*} (r|_D - R)/2$ becomes $\tilde{b}^{(c)}/2$.
From the definition of the type (c) points, the number of such 
points is given by  
\begin{equation}
 {\rm deg} \; \tilde{b}^{(c)} = D \cdot r - D \cdot \sigma - D \cdot \sigma'
 = D \cdot ((N-4) \sigma + 2 K_{B_2} + \eta),
\label{eq:nbr-bc}
\end{equation}
with $N=5$. 

{\bf This paragraph was modified in version 4}:   
The type (c) $D$--$r$ intersection points are characterized 
as follows. In the right panel of Figure~\ref{fig:rk5}, the
two curves $D$ and $r$ intersect on the spectral surface $C_V$, 
but $D \rightarrow \bar{c}_{\wedge^2 V}$ is not a ramified cover.
We count the number of such $D$--$r$ intersection points in the
following, and show that the number agrees with (\ref{eq:nbr-bc}).
Let us denote the point at this type of $D$--$r$ intersection 
as $p_i$, and the other point on $D$ as $p_j$; $p_i \boxplus p_j = e_0$ 
by definition. On an elliptic fiber $E_b$ that has such 
$p_i$ and $p_j$ in it, the defining equation of the spectral surface 
(\ref{eq:CVeq-rk5}) has $p_i$ as a zero of order two. 
Assuming that $a_5 \neq 0$, the points $p_i$ and $p_j$ correspond to 
$(x_*, \pm y_*)$ with $x_* = -a_3/a_5$ and $y_*^2 = x_*^3 + f_0 x_* + g_0$.
Either $p_i$ or $p_j$ being a zero of order two of (\ref{eq:CVeq-rk5}) 
means that 
\begin{equation}
 \frac{d s}{dx} = (a_2 + 2 a_4 x + a_5 y) + 
 \frac{dy}{dx}\left(a_3 + a_5 x\right)
\label{eq:dbl-root}
\end{equation}
vanishes at either $p_i$ or $p_j$. The second term always 
vanishes for $x_* = - a_3/a_5$. Thus, the condition 
becomes $(a_2 + 2 a_4 x_*)^2 - a_5^2 y_*^2 = 0$.
Writing $x_*$ and $y_*$ in terms of $a_{0,2,3,4,5}$ and $f_0$ and $g_0$,  
we find that this condition is equivalent to 
\begin{equation}
 R^{(5)} :=  \left(a_2 - \frac{2 a_4 a_3}{a_5}\right)^2
   + a_5^2 \left( \left(\frac{a_3}{a_5}\right)^3 + f_0 \frac{a_3}{a_5} -
 g_0\right) = 0.
\label{eq:R5-def}
\end{equation}
$R^{(5)}$ restricted upon $\bar{c}_{\wedge^2 V}$ defines a divisor 
$\sim 4K_{B_2} + 2\eta$, and 
\begin{equation}
{\rm deg} R^{(5)}|_{\bar{c}_{\wedge^2 V}} 
= (4K_{B_2} + 2\eta) \cdot \bar{c}_{\wedge^2 V} = 
D \cdot (2K_{B_2} + \eta).  \nonumber 
\end{equation}
It should be noted, however, that $R^{(5)}$ in (\ref{eq:R5-def}) 
was derived under an assumption that $a_5 \neq 0$. Points on 
$a_5 = 0$ should not contribute to the type (c) points,\footnote{
The $a_5 = 0$ locus on the matter curve $\bar{c}_{\wedge^2 V}$ 
($P^{(5)} = 0$) are classified into two groups: type (a) and type (d).
The type (a) points are, by definition, different from type (c) points, 
and we have seen that they do not contribute to $\pi_{D*}(r|_D - R)$. 
Over the type (d) points, the spectral surface $C_V$ consists of 
five layers of $p_{i,j,k,l,m}$ without ramification, and there is no 
contribution to $\pi_{D_*}(r|_D - R)$.} and one needs to examine 
whether $R^{(5)}|_{\bar{c}_{\wedge^2 V}}$ naively applied to 
the $a_5 = 0$ locus gives rise to fake contributions or not.
This is carried out by taking a local coordinate on 
$\tilde{\bar{c}}_{\wedge^2 V}$ at around type (d) and type (a) points, 
respectively, and by examining whether $R^{(5)}$ has a pole or zero 
at these codimension-3 singularity points. After a bit of detailed
analysis,\footnote{At around a type (d) point, $a_5/a_4$ can be chosen 
as a local coordinate on each one of the two branches of
$\bar{c}_{\wedge^2 V}$. Along the curve $\bar{c}_{\wedge^2 V}$, close 
to the type (d) point, $a_3/a_5$ can be treated as a finite constant
value $-x_*$. Thus, $R^{(5)}/a_4^2$ neither has a pole or zero at 
type (d) points. \\
At around a type (a) point, $a_5/a_3$ can be chosen as a local
coordinate on $\bar{c}_{\wedge^2 V}$. Because of the defining equation 
$P^{(5)} = a_4 a_3^2 - a_2 a_5 a_3 + a_0 a_5^2 = 0$ of the curve
$\bar{c}_{\wedge^2 V}$, $a_4/a_5$ can be treated as a finite 
constant value $a_2/a_3$ on the curve $\bar{c}_{\wedge^2 V}$ close to 
the type (a) point. Thus, $R^{(5)}/a_3^2|_{\bar{c}_{\wedge^2 V}}$ 
has a pole of order 1 (that is, a fake contribution of $-1$) 
at every type (a) point, when it is applied
naively to the type (a) points.} we see that
$R^{(5)}|_{\bar{c}_{\wedge^2 V}}$ has a fake contribution of $-1$ at 
every type (a) points. Thus, the true number of type (c) points
characterized by (\ref{eq:dbl-root}) and $a_5 \neq 0$ is 
\begin{equation}
 {\rm deg} R^{(5)}|_{\bar{c}_{\wedge^2 V}} + \# (a) = 
  D \cdot (2K_{B_2} + \eta) + D \cdot \sigma, \nonumber 
\end{equation}
which exhausts all the type (c) points expected in (\ref{eq:nbr-bc}).
Thus, all the type (c) points are on the $a_5 \neq 0$ part of 
$P^{(5)} = 0$ matter curve $\bar{c}_{\wedge^2 V}$, and are characterized 
by (\ref{eq:R5-def}).\footnote{Since $a_5 \neq 0$ and $P^{5} = 0$ are
assumed, the definition of $R^{(5)}$ can be modified by
multiplying/dividing by $a_5$ or adding/subtracting by $P^{(5)}$. 
It is an option to take 
\begin{equation}
 R^{(5)}_{\rm mdfd} := a_5 R^{(5)} - 4 \frac{a_4}{a_5} P^{(5)} 
  = (a_2^2 - 4 a_4 a_0) a_5 + (a_3^3 + f_0 a_3 a_5^2 - g_0 a_5^3).
\nonumber 
\end{equation}
$R^{(5)}_{\rm mdfd}|_{\bar{c}_{\wedge^2 V}}$ has $+2$ fake contribution
from every type (d) point of $\bar{c}_{\wedge^2 V}$. 
We are benifited from (2.71) of \cite{DW-3}, in making an improvement 
in version 4 here. Since the authors of \cite{DW-3} assigned a scaling 
dimension $r$ to $a_r$ ($r=5,4,3,0$), the first three terms have all 
scaling dimension 9, whereas the last two term have higher dimensions. 
This is why the last two terms are missing in (2.71) of \cite{DW-3}, 
whereas they are retained here.}

To conclude, the locally free rank-1 sheaf 
$\widetilde{\cal F}_{\wedge^2 V}$ on $\tilde{\bar{c}}_{\wedge^2 V}$
is given by 
\begin{equation}
 \widetilde{\cal F}_{\wedge^2 V} = 
{\cal O}\left( \tilde{\imath}^*_{\wedge^2 V} K_{B_2} 
  + \frac{1}{2}\tilde{b}^{(c)} + \tilde{\pi}_{D*} \gamma\right),
\label{eq:Ftil4V2-rk5}
\end{equation}
where $\tilde{\imath}_{\wedge^2 V} = 
i_{\wedge^2 V} \circ \nu_{\bar{c}_{\wedge^2 V}}: 
\tilde{\bar{c}}_{\wedge^2 V} \rightarrow \sigma$, just like 
in section~\ref{ssec:rk4}. Table~\ref{tab:rk5-example} shows 
a couple of examples of geometric data of the matter curves 
for different choice of the divisor $\eta$.
\begin{table}[t]
\begin{center}
  \begin{tabular}{c||c|c|c|c|c|c}
   $\eta$ & $\bar{c}_V \sim $ & $\bar{c}_{\wedge^2 V} \sim$ &
  \#(\rm{a}) & \#(\rm{d}) & $g(\tilde{\bar{c}}_{\wedge^2 V})$ & 
\#(\rm{c}) \\
\hline
$10 D_b + 16 D_f$ & $D_f$ & $10 D_b + 18 D_f$ &2 & 4 & 104 & 298  \\
$11 D_b + 15 D_f$ & $D_b$ & $13 D_b + 15 D_f$ & 0 & 1 & 89 & 262  \\
$11 D_b + 16 D_f$ & $D_{b'}$ & $13 D_b + 18 D_f$ & 4 & 7 & 119 & 334  
  \end{tabular}
\caption{\label{tab:rk5-example} 
Examples: We chose $F_1$ (Hirzebruch surface) 
as the base manifold $B_2$; $D_b$ and $D_f$ are two independent divisors 
satisfying $D_b \cdot D_f = 1$, $D_b \cdot D_b = -1$ 
and $D_f \cdot D_f = 0$. We also use $D_{b'} \sim D_b + D_f$ in 
the table above. Three examples are chosen for a divisor $\eta$ 
on $B_2$, 
so that the matter curve $\bar{c}_V \in |5K_{B_2} + \eta|$ 
is effective, and $|\eta|$ is base-point free, conditions 
derived in \cite{OPP}. 
In all the three examples in this table, the matter curve $\bar{c}_V$ 
is isomorphic to $\P^1$ and generically smooth. 
On the other hand, the matter curve $\bar{c}_{\wedge^2 V}$ has  
\#~(\rm{d})~$> 0$ double points, and is not smooth in any one of 
the examples. These matter curves $\bar{c}_{\wedge^2 V}$ and 
their normalizations, $\tilde{\bar{c}}_{\wedge^2 V}$, have very large genus.
}
\end{center}
\end{table}

The covering matter curve is determined through 
\begin{eqnarray}
 2g(\tilde{\bar{c}}_{\wedge^2 V}) - 2 & = & 
  {\rm deg} \; K_{\tilde{\bar{c}}_{\wedge^2 V}}, \nonumber \\
  & = & {\rm deg} \; K_{\bar{c}_{\wedge^2 V}} - 2 \times \# (\rm{ d}), \\
  & = & \bar{c}_{\wedge^2 V} \cdot (11K_{B_2} + 3\eta) 
     - 2 (5K_{B_2} + \eta) \cdot (3K_{B_2} + \eta), \\
  & = & 80 K_{B_2}^2 + 47 K_{B_2} \cdot \eta + 7 \eta^2.
\end{eqnarray}
We used the fact in the second equality that the Euler number (genus)
of a curve increases by $+2$ (resp. $-1$) whenever a double point 
is blown up \cite{Hartshorne, GH}.
On the other hand, one can calculate the following:
\begin{eqnarray}
 {\rm deg} \; \left(\tilde{\imath}^*_{\wedge^2 V} K_{B_2} 
 + \frac{1}{2}\tilde{b}^{(c)}\right)
  & = & \bar{c}_{\wedge^2 V} \cdot K_{B_2} + 
        \frac{1}{2} D \cdot (\sigma + 2K_{B_2} + \eta), \\
  & = & 40 K_{B_2}^2 + \frac{47}{2} K_{B_2} \cdot \eta +
   \frac{7}{2}\eta^2 = 
  \frac{1}{2} {\rm deg} \; K_{\tilde{\bar{c}}_{\wedge^2 V}}.
\label{eq:relate-rk5}
\end{eqnarray}
Thus, by applying Riemann--Roch theorem, the net chirality 
is given by 
\begin{eqnarray}
 \chi(\wedge^2 V) & = &
 \chi(\tilde{\bar{c}}_{\wedge^2 V}; \widetilde{\cal F}_{\wedge^2 V}) 
  =  \left[1 - g(\tilde{\bar{c}}_{\wedge^2 V}) \right] 
  + {\rm deg} \; \left(\tilde{\imath}^*_{\wedge^2 V} K_{B_2} + 
                   \frac{1}{2} \tilde{b}^{(c)} \right)
  + \int_{\tilde{\bar{c}}_{\wedge^2 V}} \tilde{\pi}_{D*} \gamma,
  \nonumber \\
 & = & \int_{\tilde{\bar{c}}_{\wedge^2 V}} \tilde{\pi}_{D*} \gamma
  = D \cdot \gamma.
\label{eq:chi-rk5-V2-F}
\end{eqnarray}
It is reasonable, as we discussed right after (\ref{eq:chi-rk4-V2-F}), 
that the result is proportional to $\gamma$. See also a discussion 
after (\ref{eq:chi-rkN-V2-F}).

\subsection{Rank-6 Vector Bundles}
\label{ssec:rk6}

The spectral surface $C_V$ of a rank-6 bundle $V$ 
is the zero locus of 
\begin{eqnarray}
 s & = & a_0 + a_2 x + a_3 y + a_4 x^2 + a_5 x y + a_6 x^3, \nonumber \\
   & = & \tilde{a}_0 + \tilde{a}_2 x + a_3 y + a_4 x^2 + a_5 x y + a_6 y^2.
\label{eq:CV-rk6-defeq}
\end{eqnarray}
The coefficients $a_{0,2}$ and $\tilde{a}_{0,2}$ are related through 
\begin{equation}
 a_0 = \tilde{a}_0 + a_6 g_0, \qquad a_2 = \tilde{a}_2 + a_6 f_0.
\end{equation}
The matter curve of the fundamental representation $V$, $\bar{c}_V$, 
is given by $a_6 = 0$.

\subsubsection{$\wedge^2 V$}

\begin{figure}[tb]
 \begin{center}
  \begin{tabular}{ccc}
   \includegraphics[width=.5\linewidth]{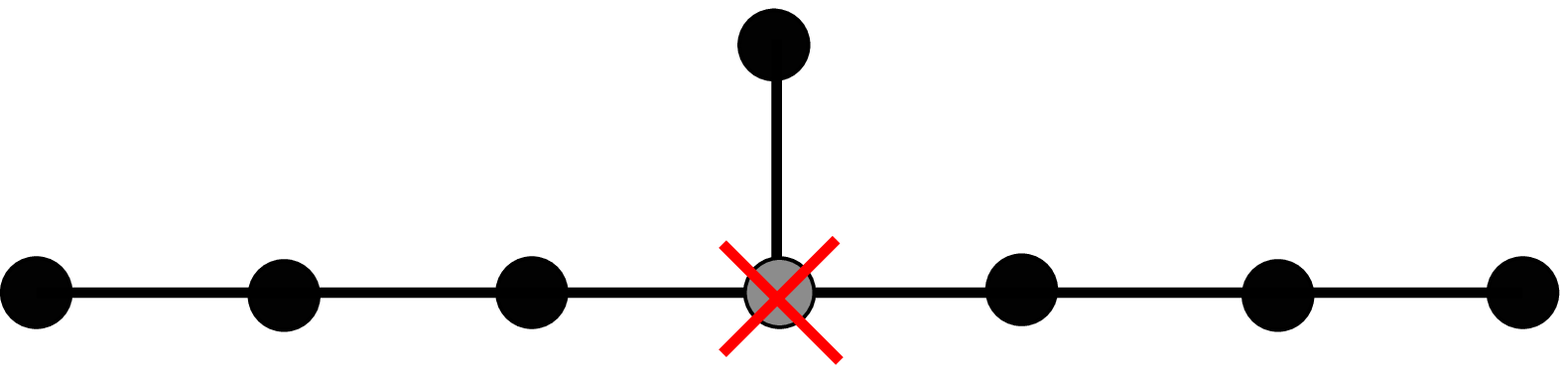} & ~~~ &
   \includegraphics[width=.3\linewidth]{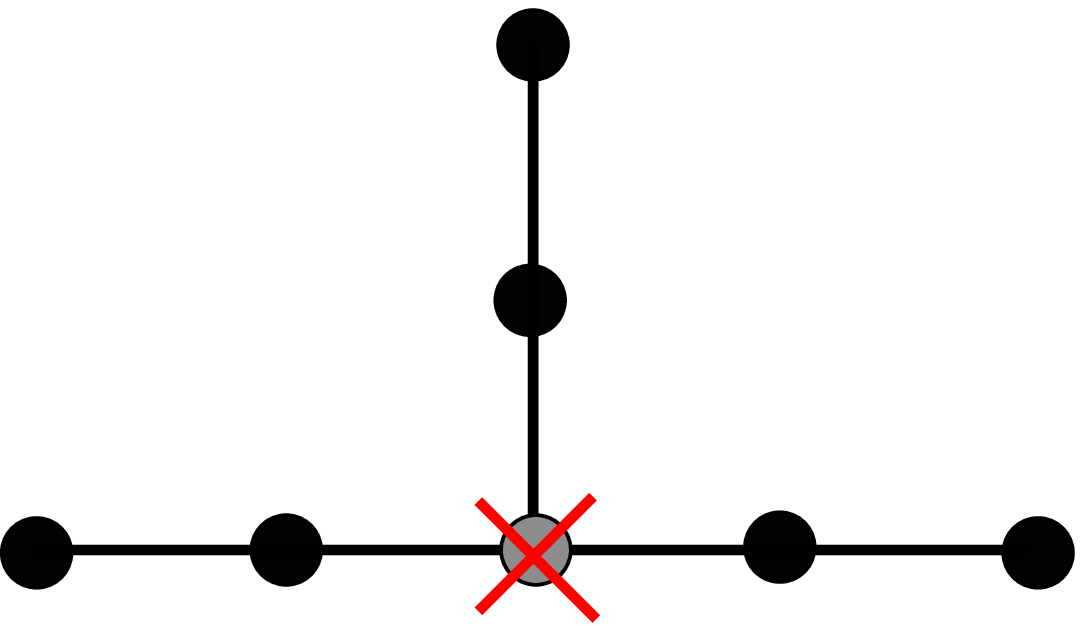} \\
   (i) & &(ii)
  \end{tabular}
\caption{\label{fig:E7E6} Extended Dynkin diagrams of (i) $E_7$ and
(ii) $E_6$. By removing one node from the diagrams, one can see 
that these groups have subgroups 
$\SU(4) \times \SU(2) \times \SU(4) \subset E_7$, and 
$\SU(3) \times\SU(3) \times \SU(3) \subset E_6$. }
 \end{center}
\end{figure}

$R^1 \pi_{Z*} \wedge^2 V$ can be studied for a rank-6 bundle, 
just like in the case for a rank-5 bundle. 
The matter curve $\bar{c}_{\wedge^2 V}$ is determined by 
requiring that the defining equation of the spectral surface 
(\ref{eq:CV-rk6-defeq}) factorizes\footnote{The structure group of the 
spectral surface is reduced from $\SU(6)$ to $\SU(2) \times \SU(4)$.
The commutant symmetry group in $E_8$ is enhanced from 
$\SU(2) \times \SU(3)$ to $\SU(2) \times \SU(4)$; note that 
$E_8 \supset \SU(2) \times E_7$, and 
$E_7 \supset \SU(4) \times \SU(2) \times \SU(4)$, as one can see 
by removing one node from the extended Dynkin diagram of $E_7$ 
(see Figure~\ref{fig:E7E6}~(i)). } 
locally as
\begin{equation}
 s = (Ax + B)(P x^2 + Q y + R x + S).
\end{equation}
This condition is equivalent to 
\begin{equation}
 P^{(6)} := a_0 a_5^3  - a_2 a_5^2 a_3 + a_4 a_5 a_3^2 - a_6 a_3^3 = 0,  
\label{eq:cV2-eq-rk6}
\end{equation} 
and this is the defining equation of the matter curve 
$\bar{c}_{\wedge^2 V}$. $P^{(6)}$ is a section of 
${\cal O}(15K_{B_2} + 4\eta)$, and the curve $\bar{c}_{\wedge^2 V}$ 
belongs to a class $|15 K_{B_2} + 4\eta|$; this corresponds 
to the $N=6$ case of (\ref{eq:cVofV2}).

There are two different types of $\bar{c}_V$--$\bar{c}_{\wedge^2 V}$ 
intersection points, because $P^{(6)}|_{a_6 = 0}$ factorizes.
\begin{itemize}
 \item [(a)] $a_6 = 0$ and $a_5 = 0$, 
 \item [(d)] $a_6 = 0$ and $a_0 a_5^2 - a_2 a_5 a_3 + a_4 a_3^2 
  = P^{(5)} = 0$.
\end{itemize}
We call them type (a) and type (d) intersection points, respectively.
The two curves intersect transversely at both types of intersection
points. These two types exhaust all kinds of the intersection:
\begin{eqnarray}
 \# (\rm{ a}) + \# (\rm{ d}) & = & (6K_{B_2} + \eta) \cdot (5K_{B_2} + \eta) 
  + (6K_{B_2} + \eta) \cdot (10K_{B_2} + 3 \eta), \nonumber \\
  & = & (6K_{B_2} + \eta) \cdot (15 K_{B_2} + 4 \eta) 
 = \bar{c}_V \cdot \bar{c}_{\wedge^2 V}.
\end{eqnarray}
The matter curve $\bar{c}_{\wedge^2 V}$ itself is smooth 
at these intersection points.

The matter curve $\bar{c}_{\wedge^2 V}$ has a triple point, wherever 
\begin{itemize}
 \item [(e)] $a_5 = a_3 = 0$.
\end{itemize}
The defining equation $P^{(6)}$ in (\ref{eq:cV2-eq-rk6}) 
have three different solutions for $a_3:a_5$ as a function of 
local values of $a_{0,2,4,6}$ in a local neighborhood of 
this type of point, and the three solutions correspond to 
three branches of $C_{\wedge^2 V}$ intersecting with the 
zero section. One corresponds to $p_i \boxplus p_j = e_0$, another 
to $p_k \boxplus p_l = e_0$, and the last one to 
$p_m \boxplus p_n = e_0$. Whenever the first two branches of the 
matter curve intersect in $\sigma \simeq B_2$, the last one also 
passes through the intersection point, because of the traceless 
condition (\ref{eq:traceless-Wilson}) of rank-6 bundles.
This is why $\bar{c}_{\wedge^2 V}$ for a rank-6 bundle has triple points. 
\begin{figure}[t]
\begin{center}
 \begin{tabular}{cc}
\includegraphics[width=.5\linewidth]{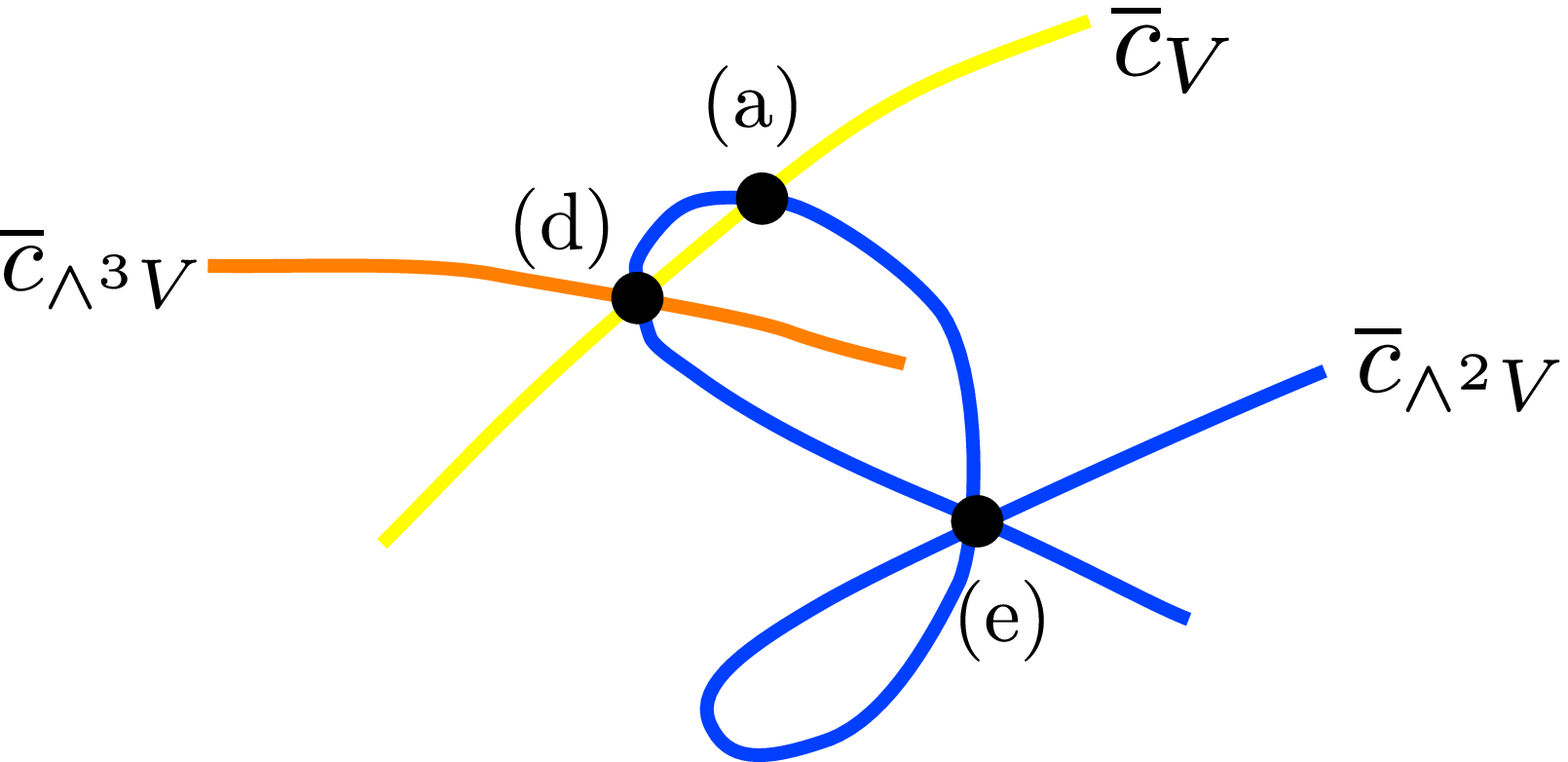} &
\includegraphics[width=.25\linewidth]{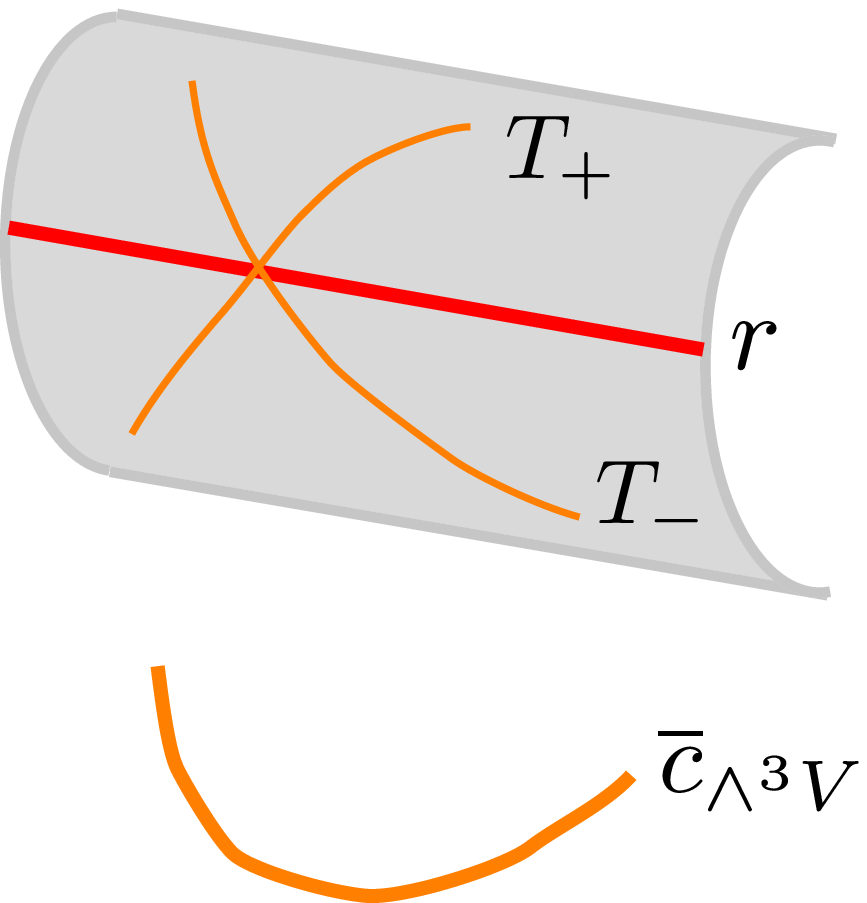} \\
  (i) & (ii) \\
 \end{tabular}
 \caption{\label{fig:rk6} The left panel (i) is a schematic picture 
of various matter curves in the zero section for cases with rank-6 
bundles. It shows how the curves intersect one another, and what kind 
of singularities they have. The right panel (ii) describes a geometry 
associated with the type (f) points, which arise in the analysis 
$\wedge^3 V$ bundles of rank-6 bundles $V$. 
Two irreducible components $T_+$ and $T_-$ of a curve $T$ in $C_V$ 
intersect when $T_\pm$ intersects the ramification divisor $r$ without 
being ramified over $\bar{c}_{\wedge^3 V}$. This is where the type (f) 
points are found.}
\end{center}
\end{figure}

In a local neighborhood of a triple point, $C_{\wedge^2 V} \subset Z$ 
consists of three irreducible components, one for $C_{(ij)}$, 
one for $C_{(kl)}$ and the other for $C_{(mn)}$. Intersection 
of any two of the three irreducible components are double-curve 
singularity of $C_{\wedge^2 V}$, and the triple points are 
where three double-curve singularities collide.
As this type of codimension-2 singularity inevitably appears on 
the zero section in the case of rank-6 bundles, we need to modify 
the argument that we presented in section~\ref{ssec:dbl-curve}.

Only a straightforward generalization is required, however. 
We choose 
\begin{equation}
 \widetilde{C}_{\wedge^2 V} = C_{(ij)} \coprod C_{(kl)} \coprod C_{(mn)}
\end{equation}
locally around any triple points. $\nu_{C_{\wedge^2 V}}$ is 
defined around this codimension-2 singularity by 
\begin{equation}
 \nu_{C_{ij}} \coprod \nu_{C_{kl}} \coprod \nu_{C_{mn}}: 
  C_{(ij)} \coprod C_{(kl)} \coprod C_{(mn)} \rightarrow 
  C_{(ij)} \cup C_{(kl)} \cup C_{(mn)} = C_{\wedge^2 V}.
\end{equation}
By repeating almost the same argument as in 
section~\ref{ssec:dbl-curve}, one can see that 
i) ${\cal N}_{\wedge^2 V} = \nu_{C_{\wedge^2 V} * } 
\widetilde{\cal N}_{\wedge^2 V}$ exists, ii) 
(\ref{eq:asPushForward}) is satisfied as a sheaf 
of ${\cal O}_Z$ module, and iii) $\widetilde{\cal N}_{\wedge^2 V}$
on $\widetilde{C}_{\wedge^2 V}$ is a locally free rank-1 sheaf.
Thus, (\ref{eq:F4V2}) can be used for this case as well. 
The covering matter curve $\tilde{\bar{c}}_{\wedge^2 V}$ is 
defined as $\tilde{\bar{c}}_{\wedge^2 V} : = 
\nu_{C_{\wedge^2 V}}^{-1} (\bar{c}_{\wedge^2 V})$ 
as before,\footnote{\label{fn:normalization}
In all the cases that we considered 
in this article, $\widetilde{C}_{\rho(V)}$ corresponds to 
normalization of $C_{\rho(V)}$. ``Normalization'' is a jargon 
in algebraic geometry which means a normal variety that is associated 
with an original algebraic variety. The covering matter curve 
$\tilde{\bar{c}}_{\rho(V)}$ is defined as the inverse image 
of the matter curve $\bar{c}_{\rho(V)}$ in 
$\nu_{C_{\rho(V)}}: \widetilde{C}_{\rho(V)} \rightarrow C_{\rho(V)}$.
Reference~\cite{BMRW} introduced a curve $D'/\tau$ as a normalization 
of the matter curve $\bar{c}_{\wedge^2 V}$. The two 
curves $\tilde{\bar{c}}_{\wedge^2 V}$ and $D'/\tau$ are the same 
for most of the cases, 
because the covering matter curve $\tilde{\bar{c}}_{\wedge^2 V}$
is obtained by resolving double points in the rank-5 case, 
and by resolving triple points in the rank-6 case. 
In the rank-4 case for $\rho(V) = \wedge^2 V$ and in 
the rank-6 case for $\rho(V) = \wedge^3 V$, however, these two 
definitions are not the same. The matter curves 
$\bar{c}_{\wedge^2 V}$ for the rank-4 case and 
$\bar{c}_{\wedge^3 V}$ for the rank-6 case are smooth, 
and do not need normalization. \\
The other definition of $D'/\tau$ in \cite{BMRW} can also be 
read out from the notation itself, a quotient of $D'$ by $\tau$.
In all the cases we considered in this article, the covering 
matter curve for $\rho(V) = \wedge^2 V$ agrees with $D'/\tau$ 
in this definition.  
} 
and each triple point is resolved into three points 
in $\tilde{\bar{c}}_{\wedge^2 V}$, one in 
$C_{(ij)}$, one in $C_{(kl)}$ and the other in $C_{(mn)}$. 

The classification of the $D$--$r$ intersection goes exactly 
the same as in the case of a rank-5 bundle $V$. There are 
type (a), (b) and (c) $D$--$r$ intersection points, and only 
the type (c) points contribute to $\tilde{\pi}_{D*}(r|_D - R)/2$ 
in (\ref{eq:F4V2}).\footnote{
A footnote added in version 4: 
The type (c) points are characterized as follows. A pair of points 
$p_i$ and $p_j$ on $C_V$ are on $D$, if and only if their coordinates 
on the elliptic fiber are $(x_*, y_*)$ and $(x_*, -y_*)$, with 
$x_* = - a_3/a_5 = B/A$, and $y_*^2 = x_*^3 + f_0 x_* + g_0$.
$D$ is ramified over $\bar{c}_{\wedge^2 V}$ if 
\begin{equation}
\frac{ds}{dx} = \left( (a_2 + 2 a_4 x_* + 3 a_6 x_*^2 + a_5 y_*) 
   + \frac{dy}{dx}(a_3 + a_5 x_*) \right) = 0. \nonumber 
\end{equation}
The second term $(a_3 + a_5 x_*)$ vanishes, and this condition 
is equivalent to 
\begin{equation}
R^{(6)} := 
 \left(a_2 - 2 a_4 \frac{a_3}{a_5} + 3 a_6 \frac{a_3^2}{a_5^2}\right)^2 
 + a_5^2 \left(\frac{a_3^3}{a_5^3} + f_0 \frac{a_3}{a_5} - g_0 \right) =
 0. \nonumber 
\end{equation}
It should be noted that $a_5 \neq 0$ is assumed. 
Although $\pi_{D*} (r|_D - R)$ is not expected to leave 
contributions at $a_5 = 0$ locus of $\bar{c}_{\wedge^2 V}$, 
$R^{(6)}|_{\bar{c}_{\wedge^2 V}} = 0$ has non-zero fake 
contributions at type (a) points (where $a_5 = 0$), when 
it is applied naively to the $a_5 = 0$ locus. 
$R^{(5)}|_{\bar{c}_{\wedge^2 V}}$ has a pole of order 2 at 
every type (a) point, and the number of true type (c) points 
on the $a_5 \neq 0$ locus of $\bar{c}_{\wedge^2 V}$ is 
\begin{equation}
 {\rm deg} R^{(6)}|_{\bar{c}_{\wedge^2 V}} + 2 \# (a) 
 = \bar{c}_{\wedge^2 V} \cdot (4K_{B_2} + 2 \eta) + 2 D \cdot \sigma,
\nonumber 
\end{equation}
which is exactly the expected number of the type (c) points 
in (\ref{eq:nbr-bc}).}
\begin{equation}
 \widetilde{\cal F}_{\wedge^2 V} = 
  {\cal O}_{\tilde{\bar{c}}_{\wedge^2 V}}
        \left (\tilde{\imath}^*_{\wedge^2 V} K_{B_2} 
                + \frac{1}{2}\tilde{b}^{(c)} + 
                \tilde{\pi}_{D*} \gamma|_D\right),
\end{equation}
the same as in (\ref{eq:Ftil4V2-rk5}).
${\rm deg} \; \tilde{b}^{(c)}$ is given by (\ref{eq:nbr-bc}), now 
with $N=6$.
\begin{table}[tb]
\begin{center}
  \begin{tabular}{c||c|c|c|c|c|c|c|c}
$\eta$ & $\bar{c}_V$ & \#(a) & \#(d) & \#(e) &
   $g(\tilde{\bar{c}}_{\wedge^2 V})$ & \#(c) &
   $g(\tilde{\bar{c}}_{\wedge^3 V \pm})$ & \#(f) \\
\hline
$12 D_b + 19 D_f$ & $D_f$ & 2 & 16 & 32 & 261 & 680 & 270 & 678 \\
$13 D_b + 18 D_f$ & $D_b$ & 0 & 5 & 27 & 234 & 618 & 243 & 618 \\
$13 D_b + 19 D_f$ & $D_{b'}$ & 4 & 27 & 37 & 288 & 742 & 297 & 738 \\
  \end{tabular}
\caption{\label{tab:rk6-example} Some examples of geometry that 
result from rank-6 bundle compactification: We chose $F_1$ 
(Hirzebruch surface) as the base manifold $B_2$ as before. 
See the caption of \ref{tab:rk5-example} for the definition of 
the divisors $D_b$, $D_f$ and $D_{b'}$ of $B_2 = F_1$.
Three examples are chosen for a divisor $\eta$ on $B_2$, 
so that the matter curve $\bar{c}_V \in |6K_{B_2} + \eta|$ 
is effective, and $|\eta|$ is base-point free, conditions 
in \cite{OPP}. Genus of the covering matter curves, numbers 
of various types of intersection points of those matter curves 
and degree of some divisors are calculated and shown for the three 
examples. }
\end{center}
\end{table}

The covering matter curve $\tilde{\bar{c}}_{\wedge^2 V}$ has 
\begin{eqnarray}
 2g(\tilde{\bar{c}}_{\wedge^2 V}) - 2 & = &
 {\rm deg} \; K_{\tilde{\bar{c}}_{\wedge^2 V}}, \nonumber \\
 & = & {\rm deg} \; K_{\bar{c}_{\wedge^2 V}} - 6 \times \# (\rm{ e}), \\
 & = & (15 K_{B_2} + 4 \eta) \cdot (16 K_{B_2} + 4 \eta)
    - 6 (5K_{B_2} + \eta) \cdot (3K_{B_2} + \eta), \\
 & = & 150 K_{B_2}^2 + 76 K_{B_2} \cdot \eta + 10 \eta^2.   
\end{eqnarray}
In the second equality, we have used the fact that the genus of a curve reduces by $3$ when 
a triple point is blown up; see \cite{Hartshorne, GH}.
On the other hand, 
\begin{eqnarray}
{\rm deg} \; \left( \tilde{\imath}^*_{\wedge^2 V} K_{B_2} 
                   + \frac{1}{2}\tilde{b}^{(c)} \right)
 & = & (15 K_{B_2} + 4 \eta) \cdot K_{B_2} + 
  \frac{1}{2} D \cdot (2\sigma + 2 K_{B_2} + \eta), \\
 & = & 75 K_{B_2}^2 + 38 K_{B_2} \cdot \eta + 5 \eta^2 
 = \frac{1}{2} K_{\tilde{\bar{c}}_{\wedge^2 V}}.
\end{eqnarray}
Thus, using the Riemann--Roch theorem, the net chirality 
from the bundle $\wedge^2 V$ is given by  
\begin{equation}
\chi(\wedge^2 V) = 
\chi(\tilde{\bar{c}}_{\wedge^2 V}; \widetilde{\cal F}_{\wedge^2 V}) 
= \int_{\tilde{\bar{c}}_{\wedge^2 V}} \tilde{\pi}_{D*} \gamma
= D \cdot \gamma.
\label{eq:chi-rk6-V2-F}
\end{equation}

After studying the direct images $R^1\pi_{Z*} \wedge^2 V$ one by one 
for $V$ of various ranks, we find the the net chirality from these 
bundles is given by the same expression, 
$\chi(\wedge^2 V) = D \cdot \gamma$. It will be clear that 
the rank-4 (\ref{eq:chi-rk4-V2-F}), rank-5 (\ref{eq:chi-rk5-V2-F}) 
and rank-6 (\ref{eq:chi-rk6-V2-F}) cases have this form of expression. 
In the rank-3 case, $\chi(\wedge^2 V) = \int_{\bar{c}_{\wedge^2 V }} 
- j^* \gamma + \pi_{C* \gamma} = D \cdot \gamma$, too.
Thus, it is tempting to guess that 
\begin{equation}
 \chi(\wedge^2 V) = D \cdot \gamma 
\label{eq:chi-rkN-V2-F}
\end{equation}
for any $\U(N)$ bundles given by spectral cover construction.

In order to show that this is really the case for general $N$, 
we need a better way to obtain  
${\rm deg} \; K_{\tilde{\bar{c}}_{\wedge^2 V}}$. 
The relation between ${\rm deg} \; K_{\bar{c}_{\wedge^2 V}}$ 
and ${\rm deg} \; K_{\tilde{\bar{c}}_{\wedge^2 V}}$ was different 
for all the different ${\rm rank} \; V$ we have considered.  
To avoid this ${\rm rank} \; V$ dependence, the following 
observation in \cite{BMRW} is useful: 
$\tilde{\pi}_D: D \rightarrow \tilde{\bar{c}}_{\wedge^2 V}$ 
is a degree-2 cover for any ${\rm rank} \; V = N$, and  
\begin{eqnarray}
 {\rm deg} \; K_{\tilde{\bar{c}}_{\wedge^2 V}} & = & 
\frac{1}{2}\left({\rm deg} \; K_{D} - {\rm deg}\; R\right), \\
  & = & \frac{1}{2}\left(D \cdot (2C_V - (\sigma + \sigma'))
  - {\rm deg} \; R \right).
\end{eqnarray}
Using a relation $R = (\sigma + \sigma')|_D$ \cite{Penn5}, 
this expression can be also rewritten as 
\begin{equation}
 {\rm deg} \; K_{\tilde{\bar{c}}_{\wedge^2 V}} = 
   D \cdot C_V - {\rm deg} \; R. 
\end{equation}
On the other hand, 
\begin{equation}
 {\rm deg} \; \left(\tilde{\imath}^*_{\wedge^2 V} K_{B_2} 
   + \frac{1}{2}\tilde{\pi}_{D*}\left(r|_D-R \right)\right) 
 = \frac{1}{2} \left(D \cdot (\pi_D^* K_{B_2} + r ) - 
   {\rm deg} \; R \right) 
 = \frac{1}{2} \left(D \cdot C_V - {\rm deg} \; R \right).
\end{equation}
Thus, we always have 
\begin{equation}
 {\rm deg}\; \left(\tilde{\imath}^*_{\wedge^2 V} K_{B_2} 
+ \frac{1}{2}\tilde{\pi}_{D*}(r|_D - R)\right) 
 = \frac{1}{2} {\rm deg} \; K_{\tilde{\bar{c}}_{\wedge^2 V}}.
\label{eq:related-rkN}
\end{equation}
Now, the chirality formula (\ref{eq:chi-rkN-V2-F}) follows 
from (\ref{eq:F4V2}) and the Riemann--Roch theorem for 
general ${\rm rank} \; V = N$.

The net chirality in the matter multiplets from bundle $V$
is given by (\ref{eq:chi-V-F}), and that in the matter multiplets 
from bundle $\wedge^2 V$ by (\ref{eq:chi-rkN-V2-F}). Although 
both are determined by one and the same $\gamma$, it is not 
obvious what kind of relations the net chiralities in the 
two sectors satisfy. We know, if we calculate $\chi(Z;V)$ 
and $\chi(Z;\wedge^2 V)$ by applying the Hirzebruch--Riemann--Roch
theorem on a Calabi--Yau 3-fold, that a relation (\ref{eq:conj-rel})
should hold between them. Thus, we will show how 
the relation (\ref{eq:conj-rel}) follows also from the 
chirality formulae (\ref{eq:chi-V-F}) and (\ref{eq:chi-rkN-V2-F}).

First, note that $\gamma$ can be decomposed into 
\begin{equation}
 \gamma = \gamma_0 + \pi_Z^* \omega, \qquad 
 \gamma_0 = \lambda (N \sigma - \eta + N K_{B_2})
\label{eq:gamma0}
\end{equation}
for some $\lambda$ and a 2-form $\omega$ on $B_2$.
Since $\pi_{C*}\gamma_0 = 0$ and $\pi_{C*} \pi_Z^* \omega = N \omega$, 
only the$\gamma_0$ part is allowed for $\SU(N)$ bundles \cite{FMW}.
For $\SU(N)$ bundles,  
\begin{equation}
 \chi(V) = \bar{c}_V \cdot \gamma_0 = - \lambda \eta \cdot (N K_{B_2} + \eta)
\end{equation}
from (\ref{eq:chi-V-F}), and 
\begin{eqnarray}
 \chi(\wedge^2 V) & = & D \cdot \gamma_0 \nonumber \\
& = &   \left[\sigma \cdot \left(N(N-1)K_{B_2} + 2(N-2)\eta\right) + 
  \eta \cdot (3K_{B_2}+\eta)\right] \cdot \lambda (N\sigma -\eta
 -N K_{B_2}),  \nonumber \\
 & = & \lambda \left(- \eta \cdot (N(N-1)K_{B_2} + 2(N-2)\eta)
                      + N \eta \cdot (3K_{B_2} + \eta)\right), \\
 & = & -\lambda \eta \cdot (N K_{B_2} + \eta) \times (N-4)
\end{eqnarray}
from (\ref{eq:chi-rkN-V2-F}). Thus, the expression 
(\ref{eq:chi-rkN-V2-F}) yields a result consistent with the relation 
$\chi(\wedge^2 V) = (N-4) \chi(V)$ in (\ref{eq:conj-rel}) 
for the case with $c_1(V)=0$. It is also easy to show 
(through a similar calculation) that 
\begin{equation}
 D \cdot \pi_Z^* \omega = (N-4) \times \bar{c}_V \cdot \omega 
  + (3K_{B_2} + \eta) \cdot (N \omega).
\end{equation}
Therefore, the chirality formula (\ref{eq:chi-rkN-V2-F}) for 
$\U(N)$ bundles always yields a result consistent 
with (\ref{eq:conj-rel}). 

\subsubsection{$\wedge^3 V$}

In the Heterotic string compactification with an $\SU(6)$ 
bundle $V$ in $E_8$, another species of chiral multiplets 
arise from the cohomology group $H^1(Z; \wedge^3 V)$. 
Thus, the direct image $R^1\pi_{Z*} \wedge^3 V$ is studied 
here, so that its F-theory description is obtained.

The matter curve $\bar{c}_{\wedge^3 V}$ is characterized by 
a condition that the defining equation of the spectral surface 
(\ref{eq:CV-rk6-defeq}) factorizes\footnote{The structure group 
of the spectral surface reduces from $\SU(6)$ to 
$\SU(3) \times \SU(3)$, and the commutant enhanced from 
$\SU(3) \times \SU(2)$ to $\SU(3) \times \SU(3)$; note that 
$E_8 \supset \SU(3) \times E_6$, and 
$E_6 \supset \SU(3) \times \SU(3) \times \SU(3)$, as one can see 
by removing one node from the extended Dynkin diagram of $E_6$.} 
locally as 
\begin{equation}
 s  = (A y + B x + C)(P y + Q x + R).
\end{equation}
Three points $p_i, p_j, p_k$ satisfying $(A y + B x + C) = 0$ 
satisfy $p_i \boxplus p_j \boxplus p_k = e_0$. 
After some calculations, one finds that this factorization 
condition is equivalent to 
\begin{equation}
 Q^{(6)} := a_6 (\tilde{a}_2^2 - 4 a_4 \tilde{a}_0) 
  + (\tilde{a}_0 a_5^2 - \tilde{a}_2 a_5 a_3 + a_4 a_3^2) = 0.
\label{eq:def-Q6}
\end{equation}
$Q^{(6)}$ is a section of ${\cal O}(10K_{B_2} + 3\eta)$, and 
$\bar{c}_{\wedge^3 V}$ belongs to a class $|10 K_{B_2} + 3\eta|$.
This curve passes through the type (d) $\bar{c}_V$--$\bar{c}_{\wedge^2 V}$
intersection points, and it intersects with $\bar{c}_{V}$ only at 
such points. This is because $Q^{(6)}|_{a_6 = 0} = P^{(5)}$. 
See Figure~\ref{fig:rk6}.

The prescription of \cite{Penn5} in determining the topological 
class of $C_{\rho(V)}$ is that 
\begin{equation}
 C_{\rho(V)} \sim ({\rm dim.} \rho(V)) \sigma + 
 2 T_{\rho(V)} \pi^*_Z \eta,
\end{equation}
where $T_{\rho(V)}$ is the Dynkin index of representation $\rho$.
Because 
\begin{equation}
 {\rm dim.} \wedge^3 V = \frac{N(N-1)(N-2)}{3!}, \qquad 
 T_{\wedge^3 V} = \frac{(N-2)(N-3)}{4}
\end{equation}
for $\wedge^3 V$ of a rank-$N$ bundle $V$, the naive expectation 
is that $\bar{c}_{\wedge^3 V} \sim C_{\wedge^3 V} \cdot \sigma
\sim (20 K_{B_2} + 6\eta)$. This is twice as much as the result 
that we have obtained. This is because $\bar{c}_{\wedge^3 V}$ 
is actually a double curve in $C_{\wedge^3 V}$, just like 
$\bar{c}_{\wedge^2 V}$ is in $C_{\wedge^2 V}$ of a rank-4 bundle $V$.
The three points $\{p_l, p_m, p_n\}$ satisfying $(P y + Q x + R) =0$
also satisfy $p_l \boxplus p_m \boxplus p_n = e_0$ simultaneously.

An idea was presented in section~\ref{sec:Idea} how to study 
$R^1\pi_{Z*} \wedge^2 V$. The same idea can be applied to 
$R^1\pi_{Z*} \wedge^3 V$ only with quite a natural generalization.
The treatment in section~\ref{sec:Idea} allows us to obtain 
a locally free rank-1 sheaf $\widetilde{\cal F}_{\wedge^3 V}$ 
on a covering matter curve $\tilde{\bar{c}}_{\wedge^3 V}$, if 
the two conditions are satisfied:
i) the Fourier--Mukai transform of $\wedge^3 V$ on $Z$ 
is represented as a pushforward (as in (\ref{eq:asPushForward}))
as a sheaf of ${\cal O}_Z$-module, and ii) ${\cal N}_{\wedge^3 V}$ 
on $C_{\wedge^3 V}$ is given by a pushforward of a locally free
rank-1 sheaf $\widetilde{\cal N}_{\wedge^3 V}$ on
$\widetilde{C}_{\wedge^3 V}$, a resolution of $C_{\wedge^3 V}$.
In the situation we have, the matter curve $\bar{c}_{\wedge^3 V}$ 
itself is a double curve in $C_{\wedge^3 V}$, but this double-curve 
singularity is resolved by blowing up $Z$ with a center along 
the double-curve singularity, just like in the appendix~\ref{ssec:pinch}
where $\wedge^2 V$ bundle for a rank-4 bundle $V$ was discussed.
We now have a covering curve $\tilde{\bar{c}}_{\wedge^2 V}$, which 
is a degree-2 cover of $\bar{c}_{\wedge^2 V}$. Furthermore,
since $[A:B:C] = [P:Q:R]$ can be realized only on a codimension-2 
locus in curve $\bar{c}_{\wedge^3 V}$, the degree-2 cover does not 
ramify for a generic choice of moduli parameters $a_{0,2,3,4,5,6}$.
The covering matter curve is a disjoint union 
of two copies of $\bar{c}_{\wedge^3 V}$:
\begin{equation}
 \tilde{\bar{c}}_{\wedge^3 V} = \tilde{\bar{c}}_{\wedge^3 V +} 
  \coprod \tilde{\bar{c}}_{\wedge^3 V -}.
\end{equation}
We have no reason to expect that singularities appear on these curves.
Therefore, no extra complication arises other than the original 
double-curve singularity, and we have shown in section~\ref{sec:Idea} 
how to deal with double-curve singularity; thus, the idea 
in section~\ref{sec:Idea} is now applicable to the analysis 
of $R^1 \pi_{Z*} \wedge^3 V$ for a rank-6 bundle $V$. 

Instead of a curve $D$ in $Y = \pi_{Z}^{-1}(\bar{c}_{\wedge^2 V})$, 
a curve $T$ in $Y = \pi_Z^{-1}(\bar{c}_{\wedge^3 V})$ is introduced. 
A triplet of points $\{p, p', p''\}$ in $C_V|_{E_b}$ 
($b \in \bar{c}_{\wedge^3 V}$) satisfying $p \boxplus p' \boxplus p'' 
= e_0$ sweeps a curve in $Y$, and that is 
the definition of $T$. 
$\pi_T = \pi_Z|_{T}: T \rightarrow \bar{c}_{\wedge^3 V}$ is not 
necessarily a degree-3 cover, but a projection to the covering curve 
$\tilde{\pi}_{\tilde{T}}: \tilde{T} \rightarrow 
\tilde{\bar{c}}_{\wedge^3 V}$ is a degree-3 cover. 
$\tilde{T}$ is a resolution of $T$ as we will explain it later.
For the case of a rank-6 bundle $V$, the three solutions 
of $(A y + B x + C) = 0$ [resp. of $(P y + Q x + R) = 0$] 
form $T_+$ part [resp. $T_-$ part] of $T = T_+ \cup T_-$, 
and $\tilde{T} = T_+ \coprod T_-$.
$T_\pm$ is mapped to $\tilde{\bar{c}}_{\wedge^3 V \pm}$ separately.

A locally free rank-1 sheaf $\widetilde{\cal F}_{\wedge^3 V}$ 
on $\tilde{\bar{c}}_{\wedge^3 V}$ is given by 
\begin{equation}
 \widetilde{\cal F}_{\wedge^3 V} = {\cal O}\left(
  \tilde{\imath}^*_{\wedge^3 V} K_{B_2} + 
 \tilde{\pi}_{\tilde{T}*} 
    \left( \frac{1}{2} (r|_{\tilde{T}} - R_{(T)}) + \gamma|_T \right)
                                           \right), 
\label{eq:F4V3}
\end{equation}
a straightforward generalization of the discussion that has led to 
(\ref{eq:F4V2}). A divisor $R_{(T)}$ is a ramification divisor 
of $\tilde{\pi}_{\tilde{T}}: \tilde{T} \rightarrow
\tilde{\bar{c}}_{\wedge^3 V}$, and hence 
$R_{(T)}:= K_{\tilde{T}} - \tilde{\pi}_{\tilde{T}}^*
K_{\tilde{\bar{c}}_{\wedge^3 V}}$. 

For the rank-6 case, the covering matter curve is a 
disjoint union of two curves, $\tilde{\bar{c}}_{\wedge^3 V \pm}$, 
and each curve has a locally free rank-1 sheaf 
\begin{equation}
 \widetilde{\cal F}_{\wedge^3 V \pm} = {\cal O}\left(
   \tilde{\imath}^*_{\wedge^3 V \pm} K_{B_2} + \tilde{\pi}_{T\pm *} \left(
     \frac{1}{2} (r|_{T_\pm} - R_{(T_\pm)}) + \gamma|_{T_\pm} \right)
     \right), 
\end{equation}
where 
$\tilde{\pi}_{T\pm} := \tilde{\pi}_{\tilde{T}}|_{T_\pm}$ 
maps $T_\pm$ to $\tilde{\bar{c}}_{\wedge^3 V \pm}$, 
$R_{(T_\pm)}$ their ramification divisors, and 
$r|_{T_{\pm}}$ a restriction on $T_{\pm}$ of a pullback of $r|_T$ 
to $\tilde{T}$. $\tilde{\imath}^*_{\wedge^3 V \pm}$ denotes pullback 
via either one of 
$\tilde{\imath}_{\wedge^3 V \pm}: = 
(i_{\wedge^3 V} \circ \nu_{\bar{c}_{\wedge^3
V}})|_{\tilde{\bar{c}}_{\wedge^2 V}}$.

The remaining task is to understand the divisor 
$\tilde{\pi}_{T_\pm *} (r|_{T_\pm} - R_{(T_\pm)})$ better.
Let us begin with examining $R_{(T_\pm)}$.
We first count the number of points where the projection 
$\tilde{\pi}_{\tilde{T}}$ is ramified.
From the definition of the ramification divisor, we have 
\begin{equation}
 {\rm deg} \; R_{(T)} = {\rm deg} \; K_{\tilde{T}} - 
  6 \times {\rm deg} \; K_{\bar{c}_{\wedge^3 V}};
\end{equation}
the second term needs a factor 6 because 
$\tilde{\pi}_{\tilde{T}}$ is a degree-3 cover of 
$\tilde{\bar{c}}_{\wedge^3 V}$, and the latter consists 
of two copies of $\bar{c}_{\wedge^3 V}$.
The second term is easy to calculate:
\begin{equation}
 {\rm deg} \; K_{\bar{c}_{\wedge^3 V \pm}} =
 {\rm deg} \; K_{\bar{c}_{\wedge^3 V}} = 
 (10 K_{B_2} + 3 \eta) \cdot (11 K_{B_2} + 3 \eta).
\end{equation}
Next, let us calculate ${\rm deg} \; K_T$. 
Since the curve $T$ collects all the points in $C_V \cdot Y$
(for a rank-6 bundle $V$), $T$ is topologically 
\begin{equation}
 T \sim C_V \cdot \pi_Z^{-1}(10 K_{B_2} + 3\eta) = 
 (6\sigma + \eta) \cdot (10 K_{B_2} + 3 \eta).
\label{eq:T-topology}
\end{equation}
Applying the adjunction formula to the curve $T$ in $C_V$ (or in $Y$), 
we have 
\begin{equation}
 {\rm deg} \; K_T = (10 K_{B_3} + 3\eta) \cdot (6 \sigma + \eta) \cdot 
 (6 \sigma + \eta + 10 K_{B_2})
 = (10 K_{B_2} + 3 \eta) \cdot 6(16 K_{B_2} + 5 \eta).
\end{equation}
The curve $T$ has two irreducible components, $T_+$ and $T_-$, 
and the two components intersect at some points. $\tilde{T}$
is obtained by resolving the double points formed by 
$T_+$ and $T_-$. As we will see later, 
\begin{equation}
 T_+ \cdot T_- = (10 K_{B_2} + 3\eta) \cdot (9K_{B_2} + 3 \eta).
\label{eq:TT}
\end{equation}
The genus (resp. ${\rm deg} K$) of a curve decreases by 1 (resp. 2) 
when a double point is blown up. Thus, 
\begin{equation}
 {\rm deg} \; K_{\tilde{T}} = {\rm deg} \; K_T - 2 T_+ \cdot T_-
  = (10 K_{B_2} + 3\eta) \cdot (78 K_{B_2} + 24 \eta).
\end{equation}
Combining all the information we have obtained, one finds that 
\begin{equation}
 {\rm deg} \; R_{(T)} = (10 K_{B_2} + 3 \eta) \cdot (12 K_{B_2} + 6 \eta). 
\end{equation}
${\rm deg} \; R_{(T)}$ above contains both ${\rm deg} \; R_{(T_+)}$ 
and ${\rm deg} \; R_{(T_-)}$:
\begin{eqnarray}
 {\rm deg} \; R_{(T)} & = & 
 {\rm deg} \; R_{(T_+)} + {\rm deg} \; R_{(T_-)}, \nonumber \\
 &  = & (10 K_{B_2} + 3 \eta)  \cdot (6K_{B_2} + 6 \eta_+) + 
 (10 K_{B_2} + 3 \eta)  \cdot (6K_{B_2} + 6 \eta_-),   
\end{eqnarray}
where $\eta_+ + \eta_- = \eta$. Although we are not presenting details, 
these intersection numbers can be understood as the number of points 
in $\bar{c}_{\wedge^3 V \pm}$ where $(A y + B x + C) = 0$ 
[resp. $(P y + Q x + R) = 0$] has a double root in their fiber, and 
hence $T_+$ (resp. $T_-$) ramifies. Since $a_0$ is a global section 
of ${\cal O}(\eta)$, $C$ and $R$ are sections 
(only along $\bar{c}_{\wedge^3 V}$) of ${\cal O}(\eta_+)$ and 
${\cal O}(\eta_-)$, respectively. Divisors $\eta_\pm$ are defined 
on $\bar{c}_{\wedge^3 V}$, and 
$\eta_+ + \eta_- = \eta|_{\bar{c}_{\wedge^3 V}}$.

Whenever curves $T_\pm$ in $C_V$ ramify over
$\tilde{\bar{c}}_{\wedge^3 V \pm}$, the spectral surface $C_V$ does 
the same over the zero section $\sigma$. Thus, such ramification points 
also contribute to ${\rm deg}\; r|_{T}$. 
The entire contribution to ${\rm deg} \; r|_T$ is given by 
\begin{eqnarray}
 {\rm deg} \; r|_T & = & T \cdot r = 
 (6 \sigma+ \eta) \cdot (10 K_{B_2} + 3\eta) \cdot 
 (6\sigma + \eta - K_{B_2}), \nonumber \\
 & = & (10 K_{B_2} + 3 \eta) \cdot (30 K_{B_2} + 12 \eta).
\end{eqnarray}
Assuming that $T$--$r$ intersection takes place with multiplicity 
1 at all the ramification points of $T_{\pm}$, we find that 
\begin{equation}
 {\rm deg} \; r|_{T} - {\rm deg} \; R_{(T)}
 = (10 K_{B_2} + 3\eta) \cdot (18 K_{B_2} + 6\eta)
\label{eq:fromTT}
\end{equation}
remains. This should be the contributions to ${\rm deg} \; r|_T$ 
that are not from the ramification points of $T_{\pm}$.

When $T_+$ intersects with the ramification divisor $r$ on $C_V$ 
at a point where $\tilde{\pi}_+: T_+ \rightarrow
\tilde{\bar{c}}_{\wedge^3 V+}$ is not ramified, $T_-$ also runs 
through the same point; see Figure~\ref{fig:rk6}.
Such points contribute to 
${\rm deg}\; r|_{T} ={\rm deg} \; r|_{T_\pm}$, but 
not to ${\rm deg} \; R_{(T_{\pm})}$.
Let us find out where in $\bar{c}_{\wedge^3 V}$ we should expect 
this to happen, and how many such points there are.
At such a point, $(A y + B x + C) = 0$ and $(P y + Q x + R) = 0$
share a same root, $(x_*,y_*)$. Thus, 
\begin{equation}
\left(\begin{array}{c}
 C \\ R
      \end{array}\right) = 
 - \left(\begin{array}{cc}
    A & B \\ P & Q
	 \end{array}\right)
   \left(\begin{array}{c}
    y_* \\ x_*
	 \end{array}\right), \; \; {\rm and~hence} \; \; 
 \left(\begin{array}{c}
  y_* \\ x_*
       \end{array}\right) = \frac{-1}{AQ - BP}
   \left(\begin{array}{cc}
    Q & -B \\ - P & A
	 \end{array}\right)
   \left(\begin{array}{c}
    C \\ R
	 \end{array}\right).
\end{equation}
Since $(x_*,y_*)$ expressed in terms of $A,B,C,P,Q,R$ should 
satisfy the Weierstrass equation of the elliptic curve, 
we have an equation constraining $A \sim R$.
Writing the equation explicitly, 
\begin{equation}
 S^{(6)}:=  - (AQ-BP)(BR-CQ)^2 + (CP-AR)^3 + f_0 (AQ-BP)^2 (CP-AR)
  + g_0 (AQ -BP)^3 = 0.
\label{eq:def-S6}
\end{equation}
$S^{(6)}$ is a section of 
${\cal O}(9K_{B_2}+3\eta)|_{\bar{c}_{\wedge^3 V}}$. 
Let us denote the divisor of the zero locus of $S^{(6)}$ as $b^{(f)}$.
Therefore, 
\begin{equation}
 {\rm deg} \; b^{(f)} = (10 K_{B_2}+3\eta) \cdot (9K_{B_2} + 3\eta).
\end{equation}
For each zero locus of $S^{(6)}$, $T_+$, $T_-$ and $r$ intersect 
in $C_V$ (as in Figure~\ref{fig:rk6}), and all these type (f) 
intersection points give rise to the contributions (\ref{eq:TT}) 
and (\ref{eq:fromTT}). All the contributions to ${\rm deg} \; r|_T$
and ${\rm deg}\; R_{(T)}$ are now understood, and we conclude that 
\begin{equation}
 \tilde{\pi}_{T_\pm *} (r|_{T_\pm} - R|_{(T_\pm)}) = \tilde{b}_\pm^{(f)},
\end{equation}
where $\tilde{b}_\pm^{(f)} := (\nu_{\bar{c}_{\wedge^3 V}}|
_{\tilde{\bar{c}}_{\wedge^3 V \pm}})^{-1}(b^{(f)})$.

Therefore, we are now ready to write down the line bundles 
on the covering matter curves:
\begin{equation}
 \widetilde{\cal F}_{\wedge^3 V \pm} = 
 {\cal O}\left( \tilde{\imath}^*_{\wedge^2 V \pm} K_{B_2} + \frac{1}{2} \tilde{b}_\pm^{(f)} 
  + \tilde{\pi}_{T_\pm *} \gamma|_{T_\pm}\right).
\label{eq:F4V3-rk6}
\end{equation}
${\cal F}_{\wedge^3 V}$ on the matter curve $\bar{c}_{\wedge^3 V}$
is given by a pushforward of the two line bundles 
$\widetilde{\cal F}_{\wedge^3 V \pm}$, and hence becomes 
a direct product of two line bundles.
Massless chiral multiplets are identified with 
\begin{equation}
 H^1(Z; \wedge^3 V) \cong 
 H^0(\tilde{\bar{c}}_{\wedge^3 V +}; \widetilde{\cal F}_{\wedge^3 V +})
 \oplus 
 H^0(\tilde{\bar{c}}_{\wedge^3 V -}; \widetilde{\cal F}_{\wedge^3 V -}).
\end{equation}

It is now straightforward to see that 
\begin{equation}
 {\rm deg} \; \left(i^* K_{B_2} + \frac{1}{2} \tilde{b}_\pm^{(f)} \right)
  = (10 K_{B_2} + 3\eta) \cdot \frac{1}{2}(2K_{B_2} + (9K_{B_2} +
  3\eta))
 = \frac{1}{2} {\rm deg} \; K_{\tilde{\bar{c}}_{\wedge^3 V \pm}}.
\end{equation}
Therefore, 
\begin{eqnarray}
 \chi(\wedge^3 V) & = & 
\chi(\tilde{\bar{c}}_{\wedge^3 V +}; \widetilde{\cal F}_{\wedge^3 V +})
+ 
\chi(\tilde{\bar{c}}_{\wedge^3 V -}; \widetilde{\cal F}_{\wedge^3 V -}), 
\nonumber \\
& = & \int_{\tilde{\bar{c}}_{\wedge^3 V +}} \tilde{\pi}_{T_+*} \gamma
  + \int_{\tilde{\bar{c}}_{\wedge^3 V -}} \tilde{\pi}_{T_-*} \gamma
 = T_+ \cdot \gamma + T_- \cdot \gamma 
 = \int_{\bar{c}_{\wedge^3 V}} \pi_{T*} \gamma.
\end{eqnarray}
For physics application that we mentioned in section~\ref{sec:review} 
(in Table~\ref{tab:Het-matter}), $\wedge^3 V$ bundle of a rank-6 
bundle is purely of $\SU(6)$ bundle $V$; even when a structure group 
of $V$ is chosen to be $\U(6) \subset \SO(12)$, the bundle $\wedge^3 V$
is neutral under the $\U(1)$ symmetry in the structure group. 
Thus, $\pi_{C*} \gamma = 0$ should be use for the calculation 
of chirality here, and hence $\chi(\wedge^3 V) = 0$. 
This should be the case, since coming out of the bundle $\wedge^3 V$
are chiral multiplets in the doublet representation of an unbroken 
$\SU(2)$ gauge group, and there is no well-defined chirality 
associate with this representation (or gauge group).
This serves as a consistency check, giving a confidence 
in the description of the bundles we have provided.

\section{From Heterotic String to F-theory}
\label{sec:Het2F}

The Heterotic string theory compactified on an elliptic fibered 
manifold $Z$ has a dual description in F-theory. 
The matter curves $\bar{c}_{\rho(V)}$, the support of 
$R^1\pi_{Z*} \rho(V)$ in the Heterotic theory description, 
correspond to intersection curves of 7-branes in F-theory. 
Sheaves on the matter curves, ${\cal F}_{\rho(V)}$, obtained 
in the Heterotic theory are also believed to be shared by 
the dual F-theory description.

In the previous section, a detailed description of ${\cal F}_{\rho(V)}$ 
was obtained in terms of spectral surface $C_V$ and $\gamma$. 
The geometric data, $C_V$ and $\gamma$, were  introduced to describe 
vector bundles on $Z$, and hence the description of ${\cal F}_{\rho(V)}$
in the previous section is still phrased in terms of data of 
Heterotic string compactification. We will take necessary steps 
in this section to translate the description of ${\cal F}_{\rho(V)}$ 
into F-theory language.

A dictionary for the translation already exists since 1990's. 
The holomorphic sections $a_r$ ($r = 0,2,3,\cdots, N$) become a part 
of complex structure moduli of an elliptic Calabi--Yau 4-fold 
for the dual F-theory compactification \cite{Vafa-Het-F,MV1, MV2, FMW, 
Het-F-4D, CD}, and 
$\gamma$ corresponds to four-form flux $G$ in F-theory \cite{CD}. 
We find, however, that the dictionary has to be refined in order 
to complete the translation, and that is what we do 
in sections~\ref{ssec:dP8} and \ref{ssec:gamma2G}. 
After the dictionary is completed, we will see in
section~\ref{ssec:codim-3} that some components of the divisors 
describing ${\cal F}_{\rho(V)}$ correspond to codimension-3 
singularities in F-theory geometry $X$.

\subsection{Describing Vector Bundles via $dP_8$ Fibration}
\label{ssec:dP8}

Reference \cite{FMW} explains how a del Pezzo surface $dP_r$ 
describes flat bundles on an elliptic curve, and Ref.~\cite{CD, DW} 
refined the correspondence between the moduli space of complex 
structure of $dP_8$ and data determining spectral surface of 
$\SU(N)$ bundles, but details are left to readers. 
In the first subsection of section~\ref{sec:Het2F}, we begin 
with filling the details that were not spelt out explicitly 
in the literature, so that ordinary physicists (like majority 
of the authors of this article) can understand.  

We denote a del Pezzo surface $dP_8$ as $S$. 
Its second cohomology group is generated by $L_0$
and $L_I$ $(I=1,2,\cdots,8)$, with their intersection form 
given by 
\begin{equation}
 L_0 \cdot L_0 = 1, \qquad L_0 \cdot L_I = 0, \qquad 
 L_I \cdot L_J = - \delta_{IJ} \qquad ({\rm for~}1 \leq I, J \leq 8).
\end{equation}
The anti-canonical divisor of $S$ is given by 
\begin{equation}
 x := -K_S = 3L_0 - \sum_{I=1}^8 L_I.
\end{equation}
General elements $E$ of the class $|x|$ is a curve of genus 1. 

The subsets of $H^2(S; \Z)$, $I_8$ and $R_8$, are defined as follows:
\begin{eqnarray}
 I_8 & := & \left\{ l \in H^2(S; \Z)|  l \cdot l = -1, \quad l \cdot x = 1
		\right\}, \\
 R_8 & := & \left\{ C \in H^2(S; \Z)| C \cdot C = -2, \quad C \cdot x = 0
                \right\}.
\end{eqnarray}
Elements of $I_8$ and $R_8$ are in one to one correspondence 
through $l = C + x$. $R_8 \otimes \Z$ is the subspace of 
$H^2(S ; \Z)$ orthogonal to $x$ in the intersection form, 
and it is known that the intersection form restricted on 
$H^2(S; \Z)^\perp$ is given by the Cartan matrix of $E_8$ 
Lie algebra multiplied by $(-1)$.
Elements of $R_8$ (and hence those of $I_8$) are in one 
to one correspondence with roots of $E_8$ Lie algebra.
\begin{equation}
 C_I = L_I - L_{I+1} \quad ({\rm for~}I=1,\cdots,7)\qquad {\rm and~} \quad 
 C_8 = L_0 - (L_1 + L_2 + L_3)
\label{eq:E8generator}
\end{equation}
can be chosen as the generators of $R_8$ (and of the root 
lattice).\footnote{Root lattices of $E_r$ ($r=6,7,8$) are generated 
by $C_I$ ($I=1,\cdots, (r-1)$) and $C_8$.} 
When a complex structure of $S$ is given (with an elliptic curve 
$E \in |x|$ embedded in $S$), a flat bundle on $E$ is given by \cite{FMW}
\begin{equation}
 {\cal O}({\rm div}\; C_\alpha|_E) \simeq {\cal O}(p_\alpha - e_0)
\label{eq:dP8-Ebdl}
\end{equation}
for a root $\alpha$ of $\mathfrak{e}_8$; here, 
$C_\alpha$ is an element of $R_8$ that corresponds to $\alpha$.
$p_\alpha$ is a point on $E$ given by $l_\alpha \cdot E$, and 
$e_0$ is the unique base point of $|x|$.

Spectral surface describes a bundle on an elliptic fibration 
$\pi_Z: Z\rightarrow B_2$ by specifying a set 
$\{ p_\alpha \}_{\alpha \in R_8}$ for each $E_b$ ($b \in B_2$). 
The same role can be played by a $dP_8$ fibration 
$\pi_U: U \rightarrow B_2$. $Z$ is identified with a subset 
of $U$, so that $\pi_U|_Z = \pi_Z$. Elliptic fiber 
$E_b = \pi^{-1}_Z(b)$ is a subset of a $dP_8$ surface 
$S_b := \pi^{-1}_U(b)$, and complex structure of $S_b$ 
determines a flat bundle on $E_b$ through (\ref{eq:dP8-Ebdl}).

\subsubsection{Two Descriptions of a $dP_8$ Surface}

A $dP_8$ surface $S$ can be described in two different ways, 
each of which has its own advantage.
\begin{itemize}
 \item [A:] $S$ is given by blowing up 8 points $p_{1,2,\cdots, 8}$ 
      in $\P^2$; $\pi: S \rightarrow \P^2$, 
 \item [B:] $S$ is a subvariety of $W\mathbb{P}^3_{1,1,2,3}$ given by an equation 
of homogeneous degree 6.
\end{itemize}
The description A is useful in capturing the 240 $(-1)$-lines 
of $I_8$, while it is easier to identify the elliptic curve 
$E$ in the description B.

The two descriptions are related as follows. 
In the description A, $L_0$ is a line of $\P^2$, and 
$L_I$ ($I = 1, \cdots, 8$) are exceptional curves, 
$L_I = \pi^{-1}(p_I)$ (set theoretic inverse image).
There are two independent global holomorphic sections 
in $H^0(S; {\cal O}(x))$; there are ${}_5 C_2 - 8 = 2$ 
degrees of freedom in a cubic form on $\P^2$ that have 
all $p_I$'s as zeros of order one. Let us take their 
generators as $F_0$ and $F_1$.
Similarly, there are ${}_8 C_2 - 8 \times 3 = 4$ degrees 
of freedom in $H^0(S; {\cal O}(2x))$, and we choose 
a generator $G$ so that $H^0(S; {\cal O}(2x))$ is generated 
by $G$, $F_0^2$, $F_0 F_1$ and $F_1^2$. 
Through a similar argument, one also finds that another 
generator---which we denote as $H$---is necessary for 
$H^0(S; {\cal O}(3x))$. 
A map $\Phi: S \rightarrow W\P^3_{1,1,2,3}$ is given by 
\begin{equation}
 \Phi: S \ni s \mapsto \left[Z':Z:X:Y\right] = 
 [F_0(s):F_1(s):G(s):H(s)] \in W\P^3_{1,1,2,3}, 
\label{eq:WP-embed}
\end{equation}
where $\left[Z':Z:X:Y\right]$ are homogeneous coordinates of 
$W\P^3_{1,1,2,3}$. $\Phi(S)$ does not occupy the entire 
weighted projective space $W\P^3_{1,1,2,3}$. 
Because 
\begin{equation}
 \dim_{\C} H^0(S; {\cal O}(6x)) = {}_{20} C_2 - 8 \times \frac{6 \times 7}{2}
 = 22,
\end{equation}
there must be an algebraic relation among 23 monomials 
of homogeneous degree 6 made out of $F_0$, $F_1$, $G$ and $H$:
\begin{equation}
 H^2 + (c_0 F_0 + c_1 F_1) H G + \cdots = 0.    
\end{equation}
Thus, the image $\Phi(S)$ is mapped in a subspace of 
$W\P^3_{1,1,2,3}$ given by an equation obtained by replacing 
$F_0, F_1, G$ and $H$ in the relation above by homogeneous 
coordinates $Z', Z, X$ and $Y$:
\begin{equation}
 Y^2 + (c_0 Z' + c_1 Z)Y X + \cdots = 0.
\end{equation}

Complex structure moduli of a $dP_8$ are described by eight 
complex parameters. In the description A, 16 complex numbers 
are needed to specify 8 points in $\P^2$, but there is 
redundancy of $PGL_3 \C$, which is of dimension 8. 
Thus, the dimension of the moduli space is 8. 
One arrives at the same conclusion in the description B. 
The defining equation of a $dP_8$ surface in $W\P^3_{1,1,2,3}$ 
can be cast into Weierstrass form 
\begin{equation}
 Y^2 = X^3 + F^{(4)}(Z',Z) X + G^{(6)}(Z',Z)
\label{eq:dP8-Weierstrass}
\end{equation}
by redefining $Y$ and $X$. $F^{(4)}$ and $G^{(6)}$ are
homogeneous function of $Z$ and $Z'$ and are of 
degree 4 and 6, respectively. Thus, they are described 
by $5 + 7 =12$ complex numbers. Since the $GL_2 \C$ 
coordinate transformation of $(Z,Z')$ can be still used 
to reduce the freedom, there are 8 moduli parameters left. 
Those eight moduli correspond to those of $E_8$ flat 
bundles on an elliptic curve $E$. 

\subsubsection{$\SU(5)$ Bundle in Description A}

For physics application, it is often more interesting 
to think of a bundle with smaller structure group, 
because the commutant of the structure group is left 
unbroken and can be seen in low-energy physics.
We are definitely interested in such situations for 
phenomenological applications. Smaller structure 
group corresponds to a restricted moduli space. 
We take SU(5)$_{\rm bdl}$ $(\subset E_8)$ structure group 
as an example; this is certainly the most motivated 
case in phenomenology.

In the description A, the $\SU(5)_{\rm bdl}$ structure group of 
a flat bundle on $E$ corresponds to choosing 
four points $p_{A+1}$ ($A = 1, \cdots, 4$) infinitesimally near 
$p_{A}$, i.e., on $\P^1$ that is obtained by blowing up 
$p_A$. Because these four points are chosen within $\P^1$'s and not 
from the entire $\P^2$, the dimension of the moduli space
is reduced  by four, leaving $8-4 = 4$ moduli.
This agrees with our expectation coming from 
$\dim_\C \P^4 = 4$, the dimension of the moduli space of 
flat $\SU(5)_{\rm bdl}$ bundles on $E$. 
Among the generators of $R_8$, $C_A = L_A - L_{A+1}$
for $A=1,2,3,4$ are $\P^1$ obtained right after blowing up 
the point $p_A$, and are effective curves. 
Their intersection form is the $(-1) \times$
Cartan matrix of $\mathfrak{su}(5)_{\rm GUT}$.
When the $\P^1$ cycles obtained by the first four blow-ups 
are of zero size, then $S$ develops an $A_4$ type 
singularity. $dP_8$ surface contains an element 
of $I_8$ 
\begin{equation}
 l_0 := 3L_0 - 2L_1 - L_2 - L_3 - L_4 - L_6 - L_7 - L_8,
\end{equation}
which is\footnote{Curves of the form 
$mL_0 - \sum_I d_I L_I$ are interpreted as zero locus 
of a homogeneous function of degree $m$ on $\P^2$ that 
has $p_I$ as a zero of order $d_I$ \cite{Hartshorne, Manin}.} 
a cubic curve in $\P^2$ that has a double point at $p_1$.
Intersection diagram of 
$l_0$ (or $C_0 := l_0 - x = - L_1 + L_5$) and 
$C_A$ ($A=1,2,3,4$) forms the extended Dynkin diagram of $A_4$.
Since 
\begin{equation}
 l_0 + \sum_{A=1}^4 C_A = x, 
\label{eq:sing-fiber}
\end{equation}
one of generators of $H^0(S; {\cal O}(x))$, $F_1$, can be chosen 
so that its zero locus becomes an effective divisor 
$l_0 + (C_1 + \cdots + C_4)$.
Effective divisors $C_A$'s do not intersect with a generic 
element $E \in |x|$, and hence the vector bundles on $E$ 
are trivial for the roots generated by $C_A$'s.
Thus, the $\mathfrak{su}(5)_{\rm GUT}$ algebra generated by $C_A$'s 
($A=1,2,3,4$) is the commutant of the SU(5)$_{\rm bdl}$ structure group
of vector bundles on $E$.

\begin{figure}[t]
\begin{center}
\begin{tabular}{cc}
\includegraphics[width=.45\linewidth]{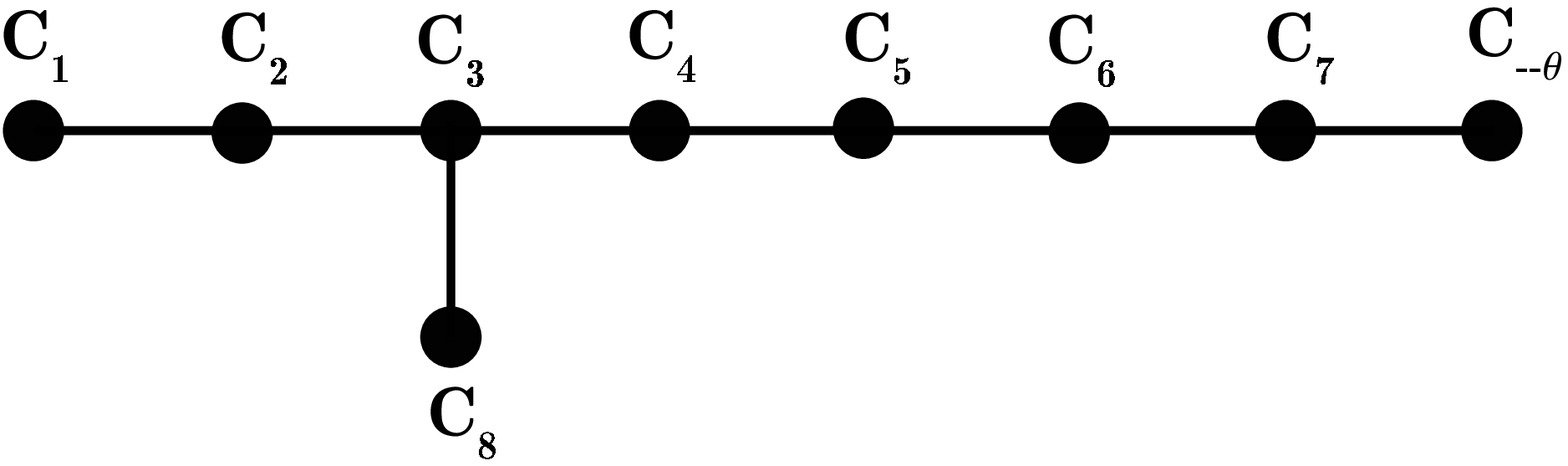}  &
\includegraphics[width=.45\linewidth]{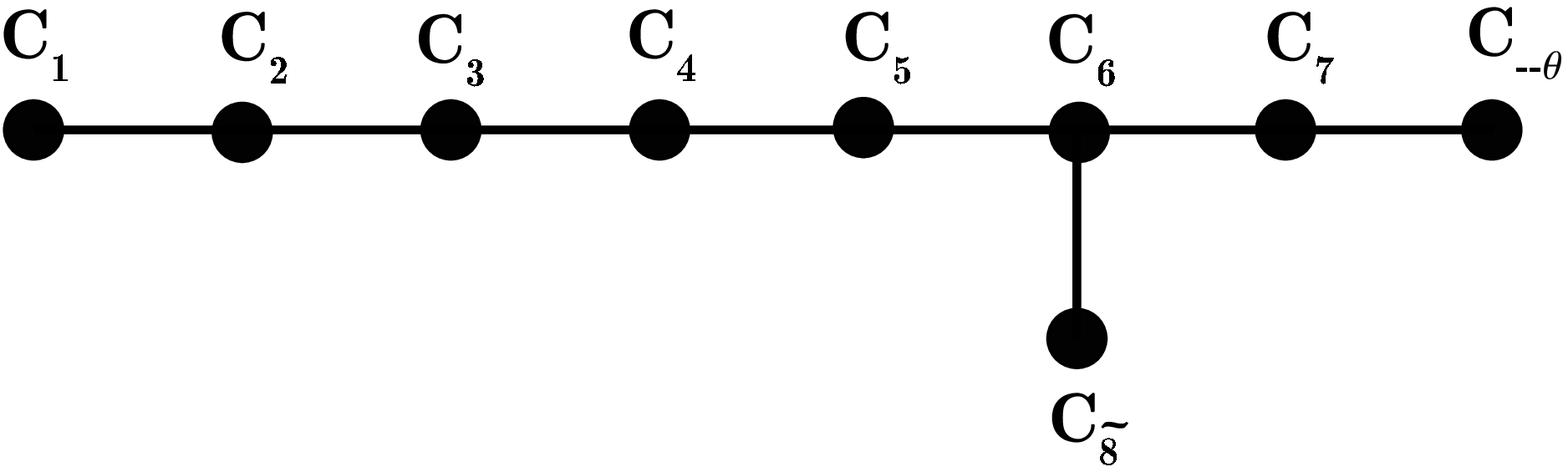} \\
(i) & (ii)
\end{tabular}
 \caption{\label{fig:e8dynkin} Extended Dynkin diagrams of $E_8$.
The two diagrams correspond to two different choices of Weyl chamber. 
By removing the node $C_5$ from the diagram (ii), one finds 
$\mathfrak{su}(5)_{\rm GUT}+\mathfrak{su}(5)_{\rm bdl}$ subalgebra 
generated by $C_{A}$ ($A=1,2,3,4$) and $C_{\tilde{8}}$, $C_{6,7}$ 
and $C_{-\theta}$.} 
\end{center}
\end{figure}

The Lie algebra of $\mathfrak{e}_8$ has 240 roots, and those 
of $\mathfrak{su}(5)_{\rm GUT}$ account for only 20 in the first summand 
of the irreducible decomposition
\begin{equation}
 {\bf 248} \rightarrow ({\bf 24},{\bf 1})\oplus ({\bf 1},{\bf 24})
   \oplus \left[({\bf 10},{\bf 5}) \oplus (\bar{\bf 5},{\bf 10})\right] 
+ {\rm h.c.};
\end{equation}
here, $(R,R')$ denotes an irreducible component that is 
in $R$ representation of $\mathfrak{su}(5)_{\rm GUT}$, and 
in $R'$ of $\mathfrak{su}(5)_{\rm bdl}$.
A group of roots in the $({\bf 10}, {\bf 5})$ representation of 
$\mathfrak{su}(5)_{\rm GUT} + \mathfrak{su}(5)_{\rm bdl}$ is given by 
\begin{equation}
 C^{ab;p} = \left(\begin{array}{l}
	     C^{ab;6^\flat} \\ C^{ab;p=6,7,8} \\ C^{ab;8^\sharp}
		  \end{array}\right)
  := L_a + L_b + \left(\begin{array}{l}
  L_0 - (L_1 + \cdots + L_5) \\ 
  -L_0 + L_p \\
  2L_0 - (L_1 + \cdots + L_8 )
		\end{array}\right), \qquad (a \neq b);
\label{eq:line-V}
\end{equation}
indices $1 \leq a,b \leq 5$ label five weights of the 
$\SU(5)_{\rm GUT}$ fundamental representation, and 
$p = \{ 6^\flat, 6,7,8,8^\sharp \}$ (on the left hand side 
of (\ref{eq:line-V})) five weights of the $\SU(5)_{\rm bdl}$ 
fundamental representation. It is clear that $C^{ab:p}$ form 
a ${\bf 10}$ representation of $\mathfrak{su}(5)_{\rm GUT}$. 
One can also see that they are in the fundamental representation 
of $\mathfrak{su}(5)_{\rm bdl}$; the structure group is generated by 
four simple roots, $C_{\tilde{8}}$, $C_6$, $C_7$ and $C_{-\theta}$, 
where 
\begin{eqnarray}
 C_{\tilde{8}} & = & C_1 + 2 C_2 + 3 C_3 + 2 C_4 + C_5 + 2 C_8,
  \nonumber \\
  & = & 2L_0 - (L_1 + L_2 + L_3 + L_4 + L_5 + L_6), \\
 C_{-\theta} & = & - (2 C_1 + 4 C_2 + 6 C_3 + 5 C_4 + 4 C_5 + 3 C_6 + 2 C_7
  + 3 C_8), \nonumber \\
 & = & -3L_0 + (L_1 + \cdots + L_8) + L_8.
\end{eqnarray}
$C_{-\theta}$ corresponds to the minimal root of $E_8$ 
when $C_I$ ($I=1,\cdots 8$) are chosen as a set of positive 
simple roots, and $C_{\tilde{8}}$ is determined so that 
$C_{1,2,\cdots,7}$, $C_{-\theta}$ and $C_{\tilde{8}}$ form 
an extended Dynkin diagram of $E_8$ in Figure~\ref{fig:e8dynkin}~(ii).
Through a straightforward calculation, one can see that 
the five weights in (\ref{eq:line-V}) for a given $(a,b)$
are obtained from $C^{ab;6^\flat}$ by applying $-C_{\tilde{8}}$, 
$-C_{6,7}$ and $-C_{-\theta}$ successively:
\begin{equation}
 C^{ab;6} = C^{ab;6^\flat} - C_{\tilde{8}}, \quad
 C^{ab;p+1} = C^{ab;p} - C_{I=p} \quad (p=6,7), \quad
 C^{ab;8^\sharp} = C^{ab;8} - C_{-\theta}.
\end{equation}
Lines in $I_8$ that corresponds to those $10 \times 5$ roots 
are given by $l^{ab;p} = C^{ab;p} + x$.  Those in the 
$(\overline{\bf 10}, \bar{\bf 5})$ representation 
are given by just multiplying $(-1)$ to $C^{ab;p}$. 

Roots in the $(\bar{\bf 5},{\bf 10})$ representation are 
given by 
\begin{equation}
 C_{a;}^{\; \; \; pq} = - L_a + \left(\begin{array}{lll}
	0 & L_q & 3L_0 - (L_1 + \cdots + L_8) \\
	 & - 2L_0+(L_1 + \cdots +L_5) +(L_p+L_q) & L_0  -
	     (L_6 + L_7 + L_8) + L_p \\
        & & 0   \end{array}\right),
\label{eq:line-X}
\end{equation}
where $a = 1,\cdots , 5$ is the $\SU(5)_{\rm GUT}$ index, and 
$p,q \in \{ 6^\flat, 6,7,8,8^\sharp\}$ ($p, q \in \{6,7,8\}$ on the
right hand side). The $5 \times 5$ matrix on the right-hand side 
forms a ${\bf 10}$ representation of $\mathfrak{su}(5)_{\rm bdl}$.
$5 \times 10$ lines are given by 
$l_{a;}^{\;\;\; pq} = C_{a;}^{\;\; \; pq} + x$, and 
those for the $({\bf 5},\overline{\bf 10})$ representation 
are by $-C_{a;}^{\;\;\; pq} + x$.

\subsubsection{$\SU(5)$ Bundle in Description B}

Reference~\cite{CD, DW} proposed how to describe $\SU(5)$
bundles on an elliptic fibration in terms of $dP_8$ fibration 
in description B. (See also \cite{KMV-BM}.)
First, the authors of $\cite{CD, DW}$ take  
\begin{eqnarray}
 Y^2 & = & X^3 + f_0 Z^4 X + g_0 Z^6 \nonumber \\
     & & \qquad \qquad
  + Z' \left(a_0 Z^5 + a_2 Z^3 X + a_3 Z^2 Y + a_4 Z X^2 + a_5 X Y
       \right)
\label{eq:dP8fiber} 
\end{eqnarray}
as the equation determining a del Pezzo surface $S$ in 
$W\P^3_{1,1,2,3}$. For a del Pezzo surface 
$S_b = \pi^{-1}_U(b)$ ($b \in B_2$),  spectral data 
$a_{0,2,\cdots,5}(b)$ are used in the equation above. 
We consider that this is a very non-trivial discovery, 
and we will take time in the following to examine 
the geometry given by this equation until this idea sinks in.

The $Z'=0$ locus is an elliptic curve given by 
a Weierstrass equation $y^2 = x^3 + f_0 x + g_0$, where 
$(x,y) = (X/Z^2, Y/Z^3)$, which is $E_b$.
The $Z=0$ ($z_f = 0$) locus is an elliptic curve given by
\begin{equation}
 y^2 - x^3 - a_5 xy = 
 \left(y - \frac{a_5}{2}x \right)^2 
 - x^2 \left(x + \frac{a_5^2}{4}\right) = 0,
\end{equation}
where now $(x,y) = (X/Z^{'2}, Y/Z^{'3})$. 
Thus, this curve has a double point at $(x,y) = (0,0)$.
Moreover, one can see by examining the equation (\ref{eq:dP8fiber})
around $[Z':Z:X:Y] = [1:0:0:0]$ that this point is 
an $A_4$ singularity, and the $Z=0$ locus consists of 
five irreducible components with their intersection form 
that of the extended Dynkin diagram of $A_4$.
In the description A of del Pezzo surfaces that correspond 
to $\SU(5)$ bundles and unbroken $\mathfrak{su}(5)_{\rm GUT}$ 
symmetry, a curve given by $F_1 = 0$ also has exactly the same
property. The map (\ref{eq:WP-embed}) identifies the two curves 
$Z=0$ and $F_1 = 0$ that have the same property.
Thus, it is likely that the way $\SU(5)$ bundles are formulated 
in the description B here is the same as the one in the description A. 

Among the 240 lines in $I_8$, $10 \times 5$ of them form a group 
$l^{ab;p} = C^{ab;p} + x$, but for a given 
$p \in \{6^\flat,6,7,8,8^\sharp\}$, the ten lines are different 
only by $C_{A=1,2,3,4}$ that are buried in the $A_4$ singularity, 
and they do not look different anywhere else in $S$. Thus, 
those lines can be treated as five sets of lines 
$l^p := l^{ab;p}$ mod $C_{A=1,2,3,4}$.
Reference~\cite{CD, DW} provided how to describe those five ``lines'' 
in the description B: they are the zero locus of 
\begin{equation}
 FL := a_0 Z^5 + a_2 Z^3 X + a_3 Z^2 Y + a_4 Z X^2 + a_5 X Y.
\label{eq:5lines}
\end{equation}

Suppose that a flat $\SU(5)$ bundle on an elliptic curve $E_b$ 
is given by a spectral surface (\ref{eq:CVeq-rk5}) at $b\in B_2$. 
The spectral surface determines five points $\{p_i\}$. 
Let us denote the coordinates of $p_i$ as $(x_i,y_i)$. 
Then, a map  
\begin{equation}
 l^i: \P^1 \ni [Z':Z] \mapsto [Z':Z:x_i Z^2: y_i Z^3] \in W\P^3_{1,1,2,3}
\label{eq:intoFL}
\end{equation}
is defined. The image $l^i(\P^1)$ falls into the zero locus of 
$FL$, and also satisfies the equation (\ref{eq:dP8fiber}). 
Thus, each one of $l^i(\P^1)$'s $(i=1,\cdots,5)$ 
becomes a line in $S$, and is an irreducible component
of the zero locus of $FL$. Because $FL = 0$ admits only five 
solutions for a given $[Z':Z]$, those five irreducible 
components are all in the zero locus of $FL$.
Those lines intersect a general element 
$E \in |x|$ just once, since such elements are in one to one
correspondence with $\P^1$ parametrized by $[Z':Z]$. The five 
lines $l^i(\P^1)$ intersect $E_b$ at the $Z'=0$ locus at 
$[1:x_i:y_i]$, which is $p_i$ itself \cite{CD, DW}. 

All those five lines pass through the $A_4$ singularity point. 
Once the $A_4$ singularity is blown up, then one can see explicitly 
that the five lines remain distinct irreducible components.

When $FL$ is pulled back to $S$ itself by $\Phi$ 
in (\ref{eq:WP-embed}), $\Phi^*(FL)$ is a global section 
of ${\cal O}(5x)$. $\Phi^*(FL)$ should be factorized into 
five irreducible pieces, because the $FL=0$ locus consists 
of five irreducible components in the description B.
Because the five lines $l^p$ in the description A
satisfy 
\begin{equation}
 \sum_{p=6^\flat}^{8^\sharp} l^p \equiv 
 \sum_{p=6^\flat}^{8^\sharp} \left(C^{ab;p} + x \right)
 \equiv 5x 
\label{eq:sum25x}
\end{equation}
mod $C_{A=1,2,3,4}$, $l^p$'s can be the irreducible 
zero loci of $\Phi^*(FL) \in H^0(S;{\cal O}(5x))$, and 
hence $l^i(\P^1)$'s in the description B.

\subsubsection{Enhancement of Singularity}

For a structure group smaller than $\SU(5)_{\rm bdl}$, 
the description B using $W\P^3_{1,1,2,3}$ is more convenient.
One only needs to turn off $a_5$ to obtain an $\SU(4)$ bundle, 
and $a_4$ to an $\SU(3)$ bundle.

Let us consider a limit $a_5 \rightarrow 0$. 
It can be regarded as a limit $B_2 \ni b \rightarrow \bar{c}_V$.
Then, $FL$ is factorized:
\begin{equation}
 FL \rightarrow Z \left(a_0 Z^4 + a_2 Z^2 X + a_3 Z Y + a_4 X^2 \right).
\label{eq:FL}
\end{equation}
Four lines among five still scan over $\P^1$ parametrized by $[Z':Z]$, 
but one of the five is absorbed in the $Z=0$ locus.
Suppose that the absorbed line is 
$l^{p=6^\flat} \equiv (L_0 - (L_1 + L_2 + L_3))+ x = C_8 + x$ 
mod $C_{A=1,2,3,4}$. 
This process adds $C_8$ to the set of roots whose bundle 
in (\ref{eq:dP8-Ebdl}) is trivial. 
The unbroken symmetry group is enhanced from $\SU(5)_{\rm GUT}$ to 
$\SO(10)$, because the intersection form 
of $C_{1,2,3,4}$ and $C_8$ is that of $\SO(10)$. 

If $a_4$ is further set to zero, another line is absorbed to the 
$Z=0$ locus. When the absorbed line is $l^{p=6}$
then another 2-cycle (and corresponding root)
\begin{eqnarray}
 l^{p=6} & \equiv & - L_0 + (L_1 + L_2) + L_6 + x 
 \quad {\rm mod~}C_{A=1,2,3,4}, \\
 & \equiv & - C_5 + x \quad {\rm mod~} C_{1,2,3,4,8}
\end{eqnarray}
joins the unbroken symmetry group, which is now $E_6$.
 
Similar process of symmetry (singularity) enhancement is observed 
when one of lines $l^{pq} := l_{a;}^{\;\;\; pq}$ mod $C_{A=1,2,3,4}$ 
is absorbed in the $Z=0$ locus. 
If $l^{6^\flat 6} \equiv (-L_5 + L_6) + x$ is absorbed, $-C_5$ is now 
buried in the $Z=0$ locus, and the intersection form of 
$C_{1,2,3,4,5}$ becomes the $(-1) \times$ Cartan matrix of $\SU(6)$.

\subsection{Chirality from Four-Form Fluxes}
\label{ssec:gamma2G}

\subsubsection{From $dP_8$ to $dP_9$}

A $dP_8$ surface $S$ containing an elliptic curve $E$ 
determines a flat bundle on $E$, and a $dP_8$ fibration 
$\pi_U: U \rightarrow B_2$ is able to play the same role 
as the spectral surface.
Spectral data $a_{r}$ in (\ref{eq:dP8fiber}) are promoted 
to sections of ${\cal O}(r K_{B_2} + \eta)$, and 
the homogeneous coordinates $[Z':Z:X:Y]$ of (\ref{eq:dP8fiber})
should now be regarded as sections of 
\begin{equation}
 {\cal O}(J_H - \eta) \otimes {\cal L}_H^6, \qquad 
 {\cal O}(J_H), \qquad 
 {\cal O}(2J_H) \otimes {\cal L}_H^2, \qquad 
 {\cal O}(3J_H) \otimes {\cal L}_H^3, 
\end{equation}
respectively. 

Dual F-theory geometry is given by a $dP_9$ fibration\footnote{
The stable degeneration limit of $K3$-fibration in F-theory 
corresponds to the situation in Heterotic theory where the 
volume of the fiber $T^2$ is sufficiently large relatively 
to $\alpha'$.} \cite{FMW} on $B_2$, 
$\pi_W: W \rightarrow B_2$, rather than this $dP_8$ fibration. 
$dP_9$ (fiber) is obtained by blowing up 
$e_0 = \left[ 0:0:c^3:c^2\right]$ ($c \neq 0$).

Such correspondence between the Heterotic and F-theories is 
rather a well-known story. The process of blowing up $dP_8$ 
to obtain $dP_9$ is well understood, and no new problems should be 
posed. Nevertheless, we will carefully follow this process, 
in order to make our presentation pedagogical, and also not 
to make a mistake.

In order to blow up a $dP_8$ surface $S$ (to obtain a strict 
transform of $S$), we begin with blowing up the ambient space 
$W\P^3_{1,1,2,3}$. Blowing-up of $W\P^3_{1,1,2,3}$ is a
$W\P^2_{1,2,3}$-fibration over $\P^1$. Two patches cover 
the new ambient space; the base $\P^1$ is covered by 
$z_f \neq \infty$ patch and $z_f \neq 0$ patch, and 
so is the entire ambient space. $(z_f,[Z':X:Y])$ 
(resp. $(z'_f,[Z:X:Y])$) is the coordinate set in the 
$z_f \neq \infty$ patch (resp. $z_f \neq 0$ patch).
The map to $W\P^3_{1,1,2,3}$ is given by $Z = Z' z_f$ from the 
$z_f \neq \infty$ patch and by $Z' = Z z'_f$ from the 
$z_f \neq 0$ patch. $z'_f = 1/z_f$.
The exceptional locus $\sigma$ that is mapped to the center 
of blow-up $e_0 = [0:0:c^2:c^3] \in W\P^3_{1,1,2,3}$ is 
given by $({}^\forall z_f,[0:c^2:c^3])$ in the 
$z_f \neq \infty$ patch (resp. $({}^\forall z'_f, [0:c^2:c^3])$ 
in the $z_f \neq 0$ patch).
The defining equation of the blown-up $dP_9$ surface is 
given in the new ambient space by \cite{KMV-BM}
\begin{eqnarray}
 y^2 & = & x^3 + z_f^4 f_0 x + z_f^6 g_0 
 + (a_0 z_f^5 + a_2 z_f^3 x + a_3 z_f^2 y + a_4 z_f x^2 + a_5 y x), 
  \label{eq:dP9fiber0}\\
 y^2 & = & x^3 + f_0 x + g_0 
 + z'_f(a_0 + a_2 x + a_3 y + a_4 x^2 + a_5 y x);   
  \label{eq:dP9fiberoo}
\end{eqnarray}
the first one is in the $z_f \neq \infty$ patch, and 
the second one in the $z_f \neq 0$ patch. 
Inhomogeneous coordinates $(y,x)$ correspond to 
$(Y/Z^{'3}, X/Z^{'2})$ and $(Y/Z^3, X/Z^2)$ in the 
two patches, and hence they are sections of 
${\cal O}_{\P^1}(3)$ and ${\cal O}_{\P^1}(2)$, 
respectively.
A del Pezzo surface $dP_9$ obtained this way is an 
elliptic fibration on $\P^1$. The exceptional locus 
of this blow up $\sigma$ passes through the infinity 
points, $(y,x) = (\infty,\infty)$.

Geometry of $dP_9$ fibration $\pi_W: W\rightarrow B_2$
is now given by the same data $a_{0,2,3,4,5}$ that described the 
vector bundles. It is now straightforward to cast the 
equation into the Weierstrass form 
\begin{equation}
 y^2 = x^3 + f x + g,
\label{eq:dP9-Weierstrass}
\end{equation}
where we now use the $z_f \neq \infty$ patch. 
After a redefinition of the coordinates $(x,y)$, 
$f$ and $g$ in the $z_f \neq \infty$ patch are given in 
$z_f = Z/Z'$ expansion as 
\begin{equation}
 f  :=  \sum_{i=0}^4 z_f^{4-i} f_i, \qquad
 g  :=  \sum_{i=0}^6 z_f^{6-i} g_i, 
\end{equation}
\begin{eqnarray}
 f_0 & = & f_0, \label{eq:dict-f0} \\
 f_1 & = & a_2, \\
 f_2 & = & -\frac{1}{3}a_4^2 + \frac{1}{2}a_5 a_3, \\
 f_3 & = & -\frac{1}{6} a_5^2 a_4, \\
 f_4 & = & - \frac{1}{48} a_5^4, \label{eq:dict-f4}
\end{eqnarray}
and 
\begin{eqnarray}
 g_0 & = & g_0, \label{eq:dict-g0} \\
 g_1 & = & a_0 - \frac{1}{3} a_4 f_0, \label{eq:dict-g1}\\
 g_2 & = & \frac{1}{4} a_3^2 - \frac{1}{3}a_4 a_2 - \frac{1}{12}a_5^2
  f_0, \label{eq:dict-g2}\\
 g_3 & = & \frac{2}{27} a_4^3 - \frac{1}{6}a_5 a_4 a_3 
         - \frac{1}{12} a_5^2 a_2, \\
 g_4 & = & a_5^2 \left(\frac{1}{18} a_4^2 - \frac{1}{24} a_5 a_3
		 \right),\\
 g_5 & = & \frac{1}{72} a_5^4 a_4, \\
 g_6 & = & \frac{1}{864} a_5^6. \label{eq:dict-g6}
\end{eqnarray}
The overall rescaling redundancy of the spectral data 
$[a_0:a_2:a_3:a_4:a_5]$ corresponds to the rescaling redefinition 
of the coordinate $z_f$. Apart from this rescaling, all the 
coefficients are determined. This precise dictionary proves 
very powerful later in translating the Heterotic theory description 
of the sheaves ${\cal F}_{\rho(V)}$ into F-theory language.

Now, suppose that a $dP_9$ surface $S'$ is a blow up of 
a $dP_8$ surface $S$:
\begin{equation}
 \pi: S' \rightarrow S.
\end{equation}
The second cohomology group of $S$ is generated by 
$C_I$'s ($I=1, \cdots, 8$) of (\ref{eq:E8generator}) and 
the anti-canonical divisor $x_8 = x$ of $S$, while that of 
$S'$ by $\pi^*(C_I)$, $\sigma$ and the anti-canonical divisor 
$x_9 = x$ of $S'$. Note that the anti-canonical divisors of $S$ and 
$S'$ are related via 
$\pi^{*}(x_8) = \pi^{-1}(x_8) + \sigma = x_9 + \sigma$, since 
$x$ passes through the base point $e_0$.
Note also that a topological relation 
\begin{equation}
 \pi^*(l) \sim \pi^*(C) + x_9 + \sigma
\end{equation}
holds for a pair of $l \in I_8$ and $C \in R_8$.
Intersection form among the 2-cycles of $S'$ is given by 
\begin{equation}
 (-C_{E_8}) \oplus \left(\begin{array}{cc}
		 0 & 1 \\ 1 & -1
		      \end{array}\right),
\end{equation}
where $\pi^*(C_I)$'s are the basis of the $(-C_{E_8})$ part 
and $(x_9,\sigma)$ of the latter $2 \times 2$ matrix.
$C_{E_8}$ means the Cartan matrix of $E_8$.

\subsubsection{Four-Form Fluxes}
\label{sssec:4-form}

In the Heterotic string theory description, matter 
multiplets are characterized in terms of spectral surfaces
and line bundles on them. All these pieces of information 
are associated with the fiber elliptic curve, which is now 
found in the $z'_f= 0$ ($z_f = \infty$) locus of 
the $dP_9$ fibration. On the other hand, in the F-theory 
description, non-Abelian gauge field of the unbroken 
symmetry group are localized within the locus of 
enhanced singularity, which is found in 
$z_f = 0$ locus. Chiral matter multiplets are also supposed 
to be at the $z_f = 0$ locus. So, how are these two 
descriptions related?

The spectral surface $C_V$ in the Heterotic description only 
determines $N$ points for an $\SU(N)$ bundle in a 
given elliptic fiber (which is at $z_f = \infty$), but 
each point corresponds to a line $l^p$ belonging to $I_8$. 
The $N$ lines specified by (\ref{eq:5lines}, \ref{eq:intoFL}) cover 
all the region of the base $\P^1$, including $z_f = Z/Z' = \infty$
and $z_f = 0$. Thus, in a description using $dP_8$ fibration 
(and $dP_9$ fibration), the information of spectral surface is 
not particularly localized at either end of the elliptic fibration 
over $\P^1$. In fact, the data $a_{0, 2,\cdots, 5}$ specifying 
the spectral surface controls the entire geometry of 
the del Pezzo fibration in 
(\ref{eq:dP8fiber}, \ref{eq:dP9fiber0}, \ref{eq:dP9fiberoo}). 

More important in generating {\it chiral} matter spectrum in 
low-energy physics is the line bundle ${\cal N}_V$ on $C_V$, 
or to be more precise, $\gamma$ determining $c_1({\cal N}_V)$ 
through (\ref{eq:defgamma}).
Reference~\cite{CD} introduced four-form flux $G^{(4)}_H$ 
in a description using $dP_8$ fibration, so that it 
plays the role of $\gamma$ in the Heterotic theory. 
$C_V$ is regarded locally as $N$ copies of a local patch 
of $B_2$, and each copy corresponds to a point $p_p$ 
for one of $p \in \{6^\flat,6,7,8,8^\sharp\}$ 
(in case of an $\SU(5)$ bundle) sweeping over $B_2$.
$\gamma$ on $C_V$ is locally described by two forms 
on the each one of those copies. 
Suppose that a four-form flux $G_H^{(4)}$ is given 
in a $dP_8$-fibration $\pi_U:U \rightarrow B_2$. 
Then, $\gamma$ on the copy of $p_p$, $\gamma_p$, 
is given by \cite{CD}
\begin{equation}
 \gamma_p = \int_{l^p} i_{l^p}^* G^{(4)}_H.
 \label{eq:gamma-Het-l}
\end{equation}
Because of this correspondence between $\gamma$ and $G^{(4)}_H$, 
only topological aspects of $G^{(4)}_H$ in $dP_8$ matter.

When considering  $\SU(N)$ vector bundle, $\gamma$ on $C_V$ has a constraint. 
The vanishing of the first Chern class 
$c_1(V) = 0$ means that 
\begin{equation}
 \pi_{C*} \gamma = 0. 
\label{eq:traceless}
\end{equation}
This implies that the integration of $G^{(4)}_H$ over the five 
lines specified by (\ref{eq:5lines}) should vanish. 
Because of the topological relation (\ref{eq:sum25x}) satisfied 
by the five lines, the condition above is equivalent to 
\begin{equation}
 \int_{5x} G^{(4)}_H = 0, \qquad {\rm and~hence~} 
 \int_{x_8} G^{(4)}_H = 0.
\label{eq:G4-cond-Het}
\end{equation}
Here, we assume that only $\SU(5)_{\rm GUT}$ preserving fluxes 
are introduced in the $dP_8$ fibration.
Because of these constraints, $G^{(4)}_H$ can be expressed as\footnote{
There may or may not be an issue along the ramification locus of the 
spectral surface $C_V$. We do not address this issue in this article.} 
\begin{equation}
 G^{(4)}_H \equiv \sum_{P=\tilde{8},6,7,-\theta} C_P \otimes \pi_Z^*
  \omega^P, 
\label{eq:GH-rk5}
\end{equation}
where $\omega^P$'s are 2-forms on $B_2$, and $C_P$'s are
2-cycles---Poincar\'e dual of 2-forms---in $S = dP_8$. 
Fluxes proportional to $x_8$ should not be introduced. 

We should be clear what we mean by (\ref{eq:GH-rk5}). 
Four-form $G^{(4)}_H$ is classified by $H^4(U; \Z)$, 
where $\pi_U: U \rightarrow B_2$ is a $dP_8$-fibration.
Using Leray spectral sequence, one finds that $H^4(U; \Z)$
has a filtration structure:
\begin{equation}
 H^4(U; \Z) = F_0 \supset F_2 \supset F_4 \supset \{ 0 \},
\label{eq:H4-U-filter}
\end{equation}
with 
\begin{equation}
 F_4 \cong H^4(B_2; R^0 \pi_{U*} \Z), \quad 
 F_2/F_4 \cong H^2(B_2; R^2 \pi_{U*} \Z), \quad 
 F_0/F_2 \cong H^0(B_2; R^4 \pi_{U*} \Z).
\end{equation}
$G^{(4)}_H$ in (\ref{eq:GH-rk5}) is understood as an element of 
$F_2/F_4$ modulo $F_4 = H^4(B_2; \Z)$, and $C_P$ as local 
generators of $R^2 \pi_{U*} \Z$. Although the Poincare dual 
2-forms of $C_P$'s are well-defined in $H^2(U; \Z)$ only modulo 
$H^2(B_2; \Z)$, this ambiguity does not appear in (\ref{eq:GH-rk5}) 
because $G^{(4)}_H$ is given in (\ref{eq:GH-rk5}) only modulo $H^4(B_2; \Z)$.
Since $x_8$ is a cycle in the fiber direction, differential forms 
on $B_2$ is trivial when pulled back to $x_8$, and hence 
(\ref{eq:G4-cond-Het}) cannot determine the $F_4$ part. 
Because of the same reason, however, (\ref{eq:gamma-Het-l}) 
does not depend on the $F_4$ part either. 
Therefore, in describing vector bundles in Heterotic theory, 
it is sufficient to have a four-form $G^{(4)}_H$ in $F_2/F_4$, 
and leave the ambiguity in $F_4$ unfixed.

By using this explicit expression of $G^{(4)}_H$ and the 
intersection form
\begin{equation}
 C^{ab;p=6^\flat,6,7,8,8^\sharp} \cdot C_{P=\tilde{8},6,7,-\theta} 
 = \left( \begin{array}{cccc}
    -1 & & & \\
    1 & -1 & & \\
     & 1 & -1 & \\
     & & 1 & -1 \\
     & & & 1 \\
	  \end{array}\right),
\end{equation}
one can see explicitly that \cite{CD}
\begin{equation}
 \pi_{C*}(\gamma \cdot \gamma) = \sum_p \gamma_p \wedge \gamma_p
  = \sum_{P,Q} C_{A_4 \; PQ} \omega^P \wedge \omega^Q  
  = - \pi_{U*} G^{(4)}_H \wedge G^{(4)}_H.
\end{equation}
Although $G^{(4)}_H$ in (\ref{eq:GH-rk5}) has the ambiguity 
$F_4 = H^4(B_2; \Z)$, $G^{(4)}_H \wedge G^{(4)}_H$ does not depend 
on the ambiguity.
The same is true for other $\SU(N)$ bundles with $N < 5$; 
$\omega^{P=\tilde{8}}$ is set to zero for $\SU(4)$ bundles, 
and $\omega^{P=6}=0$ is further imposed for $\SU(3)$ bundles.

In the F-theory compactification, there is totally an independent 
condition for 4-form flux $G^{(4)}_F$ on a Calabi--Yau 
4-fold compactification: the (2,2) part of the four-form flux 
has to be primitive in order to preserve ${\cal N} = 1$ 
supersymmetry. Reference~\cite{CD} observed that the condition 
(\ref{eq:traceless}) and the primitiveness condition 
\begin{equation}
 J \wedge G^{(4)}_F = 0
\label{eq:primitive}
\end{equation}
are quite ``similar,'' and certainly they are. 
On the other hand, $H^2(dP_9;\Z)$ is larger than $H^2(dP_8; \Z)$
by rank one. Thus, with only one constraint on $H^2(dP_8; \Z)$ 
and one for $H^2(dP_9; \Z)$, there should be no one-to-one 
correspondence between Heterotic and F-theory vacua. This 
gap has to be filled in order to complete the dictionary 
of the Heterotic--F theory duality.

The primitiveness condition (\ref{eq:primitive}) involves 
a two-form $J$ on $W$ and a four-form $G^{(4)}_F$ on $W$, 
where $\pi_W:W \rightarrow B_2$ is a $dP_9$ fibration.
$H^4(W; \Q)$, in which $G^{(4)}_F$ takes its value, has 
a filtration structure just like in (\ref{eq:H4-U-filter}):
\begin{equation}
 H^4(W; \Q) = F_0 \supset F_2 \supset F_4,
\end{equation}
with 
\begin{equation}
 F_4 \cong H^4(B_2; \Q), \quad 
 F_2/F_4 \cong H^2(B_2; R^2 \pi_{W*} \Q), \quad 
 F_0/F_2 \cong H^0(B_2; R^4 \pi_{W*} \Q);
\end{equation}
notations $F_{0,2,4}$ are recycled here, as we expect little 
confusion. $\Z$ in (\ref{eq:H4-U-filter}) is replaced by 
$\Q$ here, because the four-form flux in F-theory is 
not necessarily quantized as integral value \cite{WittenFlux}.
It is known that the four-form flux in $F_0$ has its two legs 
in the $T^2$-fiber directions of the $dP_9$, and results in 
non SO(3,1) Lorentz symmetric vacuum \cite{DRS}. Thus, we only consider 
$G^{(4)}_F$ that belongs to $F_2$ in the following. 
$G^{(4)}_F$ being an element of $F_2$ is not a sufficient condition 
for the $\SO(3,1)$ Lorentz symmetry; we will elaborate on it later.

Similarly, the K\"{a}hler form $J$ takes its value in 
$H^2(W; \R)$, and this cohomology group also has a filtration 
structure: 
\begin{equation}
 H^2(W; \R) = E_0 \supset E_2, \quad E_2 \cong H^2(B_2 ; \R), \quad 
  E_0/E_2 \cong H^0(B_2; R^2 \pi_{W*} \R).
\end{equation}
Thus, the K\"{a}hler form $J$ is written as 
\begin{equation}
 J = \pi_{W}^* J_{B_2} + t_2 J_0; 
\end{equation}
projection of $J$ into $E_0/E_2$ specifies a 2-form on $dP_9$, 
and $J_0$ is a representative of the class specified by 
the 2-form on $dP_9$. $t_2 \geq 0$ is a parameter.

In the Heterotic--F theory duality, moduli space is shared 
by the two theories, but one of the two theories provides 
a better description of some part of the moduli space, and 
the other of some other parts. The description in the Heterotic theory 
(without stringy excitations taken into account in calculations)
becomes unreliable either when the Heterotic theory dilaton 
expectation value is large, or when the volume of the $T^2$ fiber 
becomes comparable to $\alpha'$. In the first case, 
the base $\P^1$ manifold of $S'=dP_9$ has a large volume. 
Thus, whenever F-theory provides a better description, 
the volume of the base $\P^1$ of $dP_9$ is larger than 
that of the $T^2$ fiber. Therefore, in the F-theory limit, 
we can take it that the K\"{a}hler form on $dP_9$ specified by 
$J$ or $J_0$ has a dominant contribution only from the 
the $\P^1$ base of $dP_9$, not from the $T^2$ fiber. 
Thus, $J_0$ (or $J$) regarded as a 2-form on $dP_9$ is a 
Poincar\'e dual of $x_9$.

The filtration structure of $J$ and $G^{(4)}_F$ 
makes the analysis of 
the primitiveness condition (\ref{eq:primitive}) easier.
The condition (\ref{eq:primitive}) takes its value in 
$H^6(W; \R)$, and this group also has a filtration structure 
\begin{equation}
 H^6(W; \R) = G_2 \supset G_4 \supset \{ 0 \}, 
\end{equation}
with 
\begin{equation}
 G_4 \cong H^4(B_2; R^2 \pi_{W*} \R), \qquad 
 G_2/G_4 \cong H^2(B_2; R^4 \pi_{W*} \R).
\end{equation}
We begin with the primitiveness condition in $G_2/G_4$, and 
we will come back later to the condition in the $G_4$ part.
The $G_2/G_4$ part of the primitiveness condition receives 
contributions only from the wedge product of 
the $E_0/E_2$ part and the $F_2/F_4$ part, 
and we find that
%
\begin{equation}
 t_2 x_9 \cdot G^{(4)}_F 
 \equiv 0
\label{eq:G4-cond-F}
\end{equation}
mod $G_4$. 
%

The primitiveness condition (\ref{eq:G4-cond-F}) allows 
two types of local expressions for the four-form flux:
\begin{eqnarray}
 G^{(4)}_{F;\gamma} & \equiv & \sum_{I=1}^8 C_I \otimes \omega^I, 
\label{eq:GF-C}\\
 G^{(4)}_{F;9} & \equiv & x_9 \otimes \omega^{I=9},
\label{eq:GF-x9}
\end{eqnarray}
here, we abuse the notation, and denote $\pi^*(C_I)$ of 
$H^2(dP_9; \Z)$ as $C_I$, because the intersection 
form of $\pi^*(C_I)$'s are the same as those of $C_I$'s 
in $H^2(dP_8; \Z)$. The four-form flux $G^{(4)}_H$ 
(\ref{eq:GH-rk5}) in the Heterotic theory description 
can be mapped into the first type of $G^{(4)}_F$:
\begin{equation}
 G_{F;\gamma}^{(4)} \equiv \pi^* G^{(4)}_H;
\label{eq:FGfromGHet}
\end{equation}
everything is in modulo $F_4 = H^4(B_2; \Q)$ here.
We understand that the four-form flux $G_\gamma$ in \cite{DW}
belongs to this class modulo $F_4 = H^4(B_2; \Q)$.

A little more attention has to paid in interpreting the other 
contribution (\ref{eq:GF-x9}). F-theory dual of a Heterotic 
compactification involves a Calabi--Yau 4-fold that is a 
$K3$-fibration on a base 2-fold. Although the $K3$ fiber 
becomes two $dP_9$ surfaces in the stable degeneration limit, 
$K3$ fiber, rather than two $dP_9$'s, is better in understanding 
this aspect. As explained clearly in \cite{DRS}, out of 22 two-cycles 
of a $K3$ fiber, $2 \times 8 = 16$ two-cycles correspond to 
the $C_I$'s in two $dP_9$'s. Four-form fluxes associated with these 
two-cycles, like (\ref{eq:GF-C}), satisfy the primitiveness condtion 
in the $G_2/G_4$ part. Fluxes associated with the zero section 
of the elliptic fibered $K3$ (like $\sigma$ of $dP_9$) and 
with the $T^2$-fiber class (like $x_9$ of $dP_9$), on the other hand, 
do not either satisfy the primitiveness condition or preserve 
the $\SO(3,1)$ Lorentz symmetry. Thus, such fluxes should not be
introduced. Two other (1,1) two-cycles remain, and 
the four-form fluxes associated with these two two-cycles 
as well as the (2,0) and (0,2) two-cycles of $K3$ fiber correspond 
to the three-form fluxes of the Type IIB string theory \cite{DRS}.
Therefore, when the three-form fluxes are set to zero, 
\begin{eqnarray}
 \pi_{C*} \left(\gamma \wedge \gamma \right) 
& = & - \pi_{U*} \left(G^{(4)}_H \wedge G^{(4)}_H \right), \nonumber \\
& = & - \pi_{W*} \left(G^{(4)}_{F;\gamma} \wedge G^{(4)}_{F;\gamma}
		 \right)
 = - \pi_{W*} \left(G^{(4)}_{F} \wedge G^{(4)}_{F}
		 \right).
\label{eq:gg-GG}
\end{eqnarray}
Once again, the $F_4 = H^4(B_2; \Z)$ ambiguity in $G^{(4)}_F$ does not 
matter to the relation (\ref{eq:gg-GG}).

The correspondence between the number of $M5$-branes in the Heterotic 
theory and the number of 3-branes in F-theory is one of the 
most important clues of the Heterotic--F theory duality.
The number of $M5$-branes wrapped on the elliptic fiber 
is given by \cite{FMW}
\begin{equation}
 n_5 = \int_{B_2} \left(c_2(TZ) - c_2(V_1)|_{\gamma_1= 0} 
  - c_2(V_2)|_{\gamma_2 = 0}\right) + \frac{1}{2} \gamma_1^2 
  + \frac{1}{2}\gamma_2^2,
\end{equation}
where $V_i$ and $\gamma_i$ ($i=1,2$) are vector bundle 
and discrete twisting data in (\ref{eq:defgamma}), respectively, 
in the visible ($i=1$) and hidden ($i=2$) sector. 
The number of 3-branes in F-theory is given by \cite{DM} 
\begin{equation}
 n_3 = \frac{\chi(X)}{24} - \sum_{i=1,2} 
   \frac{1}{2} G_{Fi}^{(4)} \wedge G_{Fi}^{(4)},
\end{equation}
where $F^{(3)} \wedge H^{(3)}$ contribution from the three form 
fluxes of the Type IIB string theory are set to zero.
The equality between the first terms in $n_5$ and $n_3$
was proved in \cite{FMW, AC1, AC}. 
The equality (\ref{eq:gg-GG}) was basically shown in \cite{CD}.
When the $F^{(3)} \wedge H^{(3)}$ contribution is turned on, 
the $n_3$ in F-theory will be different from the original $n_5$ 
in a Heterotic compactification (that is no longer a dual).
%
%
What we did so far in section~\ref{sssec:4-form} is basically 
to collect references (mainly \cite{CD, DRS}) and tell a combined 
story. 

The extra degree of freedom in the four-form flux $G^{(4)}_F$ 
(the 3-form flux $F^{(3)}$ and $H^{(3)}$ in the Type IIB language)
brings about another issue. In the Heterotic theory description, 
$\gamma$ on $C_V$ has an alternative expression 
\begin{equation}
 \gamma_p = \int_{C^p} G^{(4)}_H,
\label{eq:gamma-Het-C}
\end{equation}
where $C^p := l^p - x_8 = C^{ab;p}$ mod $C_{A=1,2,3,4}$, 
because the difference from (\ref{eq:gamma-Het-l}), 
$x_8 \cdot G^{(4)}_H$, vanishes.
In F-theory, however, the two natural guesses 
\begin{eqnarray}
 \gamma_p & = & \int_{\pi^*(l^p)} G^{(4)}_F, 
 \label{eq:gamma-F-A}\\
 \gamma_p & = & \int_{\pi^*(C^p)} G^{(4)}_F
 \label{eq:gamma-F-B}
\end{eqnarray}
are not necessarily the same. 
The meaning of (\ref{eq:gamma-F-A}) is not even well-defined, 
because $\pi^*(l_p)$'s are well-defined two-cycles in $dP_9$, but 
their meaning has not been specified in $K3$. Depending on how 
the $\pi^*(l_p)$'s are defined in $K3$, (\ref{eq:gamma-F-A}) may 
or may not depend on the three-form fluxes $F^{(3)}$ (and $H^{(3)}$).
$\pi^*(C_p)$ on the other hand, are naturally identified with one 
of two sets of two-cycles of $K3$ whose intersection form is 
$(-1) \times$ Cartan matrix of $E_8$. Since those two-cycles have 
vanishing intersection numbers with the two-cycles to which 
the three-form fluxes are associated with (see \cite{DRS}), 
(\ref{eq:gamma-F-B}) does not depend on the choice of 
the extra discrete degrees of freedom in F-theory, 
and is the same as (\ref{eq:gamma-Het-l}, \ref{eq:gamma-Het-C}).
Thus, we adopt (\ref{eq:gamma-F-B}) in translating $\gamma$ in 
Heterotic theory into F-theory language.
Note that $\gamma_p$'s defined by (\ref{eq:gamma-F-B}) 
(and those by (\ref{eq:gamma-F-A})) do not depend on the $F_4$ part 
of $G^{(4)}_F$.

\subsubsection{Line Bundles on Discriminant Locus of F-theory}

Here is a side remark on $\U(N)$ bundles and line bundles. 
In section~\ref{sec:Examples}, we studied vector bundles 
whose structure group is $\U(N)$, as well as $\SU(N)$ bundles. 
$\U(N)$ bundles appear in phenomenological applications
of Heterotic string compactification in a form 
$V \cong U_N \oplus U_M$, where $U_N$ and $U_M$ are bundles 
with structure group $\U(N)$ and $\U(M)$, respectively, and 
$\U(N) \times \U(M) \subset \SU(N+M) \subset E_8$.
If the bundles $U_N$ and $U_M$ are given by spectral cover 
construction on an elliptic fibered Calabi--Yau 3-fold, 
then ``the structure group of the spectral surface'' is 
$\SU(N) \times \SU(M)$, and the geometry of dual F-theory 
description has a locus of enhanced singularity 
that corresponds to the commutant of $\SU(N) \times \SU(M)$
in $E_8$. The unbroken symmetry is smaller than the 
commutant, because the structure group of the vector bundle $V$
is $\U(N) \times \U(M)$, not $\SU(N) \times \SU(M)$. In 
the dual F-theory description, this symmetry breaking is 
given by a line bundle on the locus of singularity. 
The four-form flux determines the line bundle. 

Let us see this correspondence between the Heterotic and 
F-theory descriptions more explicitly. We use 
a $V \cong U_3 \oplus U_2$ bundle with the structure group 
$\U(3) \times \U(2) \subset \SU(5)_{\rm bdl}$. Much the same story 
follows for a bundle $V \cong U_4 \oplus U_1$ with the 
structure group $\U(4) \times \U(1) \subset \SU(5)_{\rm bdl}$.
In the spectral cover construction of the bundles $\U_3$ and $\U_2$, 
two two-forms $\gamma_3$ and $\gamma_2$ are used. Since the structure 
group is $\U(3) \times \U(2)$, the condition (\ref{eq:traceless})
does not have to be imposed separately for $\gamma_3$ and $\gamma_2$; 
only $\pi_{C*}(\gamma_3 + \gamma_2) = 0$ is required.
Because of the overall traceless condition, 
$\gamma_3+\gamma_2$ corresponds to a four-form flux of the form 
\begin{equation}
 G^{(4)}_F \equiv \sum_{P=\tilde{8},6,7,-\theta} C_P \otimes \omega^P
\end{equation}
mod $F_4 = H^4(B_2; \Z)$ in F-theory description. 
Four-form fluxes that correspond to the traceless parts 
of $\gamma_3$ and $\gamma_2$ are proportional to $C_{P=7,-\theta}$
and $C_{P=\tilde{8}}$, respectively, and 
preserve the $\SU(6)$ symmetry generated by $C_{A=1,2,3,4,5}$.
However, a four-form flux of the form\footnote{Coefficients of 
$C_{P=\tilde{8},6,7,-\theta}$ are determined so that the linear
combination has vanishing intersection number with the $\SU(3)$ 
generators $C_{P=7,-\theta}$ and with the $\SU(2)$ generator 
$C_{\tilde{8}}$. The same logic is behind the choice of 
the linear combination coefficients in (\ref{eq:GF4Y}).} 
\begin{equation}
 G_F^{(4)} \equiv \frac{1}{6}(3C_{\tilde{8}} + 6 C_6 + 4 C_7 + 2 C_{-\theta})
  \otimes \pi_{C*} \gamma_3
  \equiv \frac{1}{6}(-(L_1 + \cdots + L_5) + 5 L_6) \otimes 
  \pi_{C*} \gamma_3
\label{eq:GF4q7}
\end{equation}
breaks the $\SU(6)$ symmetry.

In section~\ref{ssec:rk2}, we studied the direct images of 
$\wedge^2 U_2$ of a rank-2 bundle $U_2$. Matter multiplets from 
$(\wedge^2 U_2)^{\pm 1}$ are described in terms of a line bundle 
$E^{\pm 1} = {\cal O}_{B_2}(\pm \pi_{C*} \gamma_2)$ on $B_2$. 
Remembering that the matter multiplets from $(\wedge^2 U_2)^{\pm 1}$ 
correspond to two-cycles 
$\pm C_{a;}^{\;\;\; pq = 6^\flat 6} = \pm (-L_a + L_6)$, 
one can see that the four-form flux $G^{(4)}_F$ in (\ref{eq:GF4q7}) 
solely reproduces the divisor of the line bundle $E^{\pm 1}$:
\begin{equation}
 \pm C_{a;}^{\;\;\; 6^\flat 6} \cdot G^{(4)}_F = \mp \pi_{C*} \gamma_3
 = \pm \pi_{C*} \gamma_2.
\label{eq:bdl4Hu-F}
\end{equation} 
Note that we use the dictionary (\ref{eq:gamma-F-B}) here.

The dictionary (\ref{eq:gamma-F-B}) determines the gauge field 
that matter field feels in a very natural way. In F/M-theory, 
matter multiplets that correspond to the roots 
$\pm C_{a;}^{\;\;\; pq}$ (or any other roots) are fluctuations of 
an $M2$-brane wrapped on these two-cycles. The three-form field 
$C^{(3)}$ of M-theory is integrated over a two-cycle to become 
a gauge field for the matter field associated with the two-cycle. 
Degrees of freedom that appear in low-energy physics come from 
$M2$-branes wrapped on collapsed two-cycles. In the case of 
$\U(3) \times \U(2)$ bundle, the two-cycles 
$\pm C_{a;}^{\;\; 6^\flat 6}$ ($a=1,\cdots,5$) are also collapsed 
everywhere along the  $z_f = 0$ locus isomorphic to $B_2$, 
a locus of $A_5$ singularity.
Thus, the fields from those two-cycles propagate over the 
entire $A_5$ singularity locus, and are under the influence of 
the gauge field $\pm \int_{C_{a;}^{\;\;\; 6^\flat 6}}C^{(3)}$ everywhere.

Here, bundles for roots in adjoint representation of $\SU(6)$ 
were obtained directly, without considering a bundle 
in the fundamental representation on the locus of $A_5$ singularity.
Since we used (\ref{eq:bdl4Hu-F}), the bundles are well-defined 
as long as the relevant part of the four-form flux $G^{(4)}_F$
is integral. It is true that the four-form flux of F-theory has to be
shifted from integral-valued quantization by $c_2(TX)/2$
\cite{WittenFlux}, 
where $X$ is an elliptic fibered Calabi--Yau 4-fold $\pi_X: X \rightarrow B_3$
for F-theory compactification.
But at least for cases with a Heterotic dual, i.e., $B_3$ is 
a $\P^1$-fibration over $B_2$, $c_2(TX)$ can be calculated. 
Relevant to the issue above is a component of $c_2(TX)$ 
that is 2-form on $B_2$ and 2-form on the K3-fiber.
Explicit calculation of $c_2(TX)$ reveals\footnote{
$c_2(TX) = 24 \sigma \cdot J_0 - 
(12 (K_{B_2} + t) \cdot \sigma + 46 K_{B_2} \cdot J_0) + \cdots$.
Here, $t := 6K_{B_2} + \eta$, and $J_0$ is a divisor that 
corresponds to $z_f = 0$. Ellipses stand for a four-form that 
comes from $B_2$. $J_0$ here is the same as $J_0$ in \cite{DW} 
modulo $H^2(B_2)$. Since the coefficient of $\sigma \cdot J_0$ is even, 
the four-form flux $G^{(4)}_F$ can be chosen within $F_2$.}  
that the terms proportional to $\sigma$ and $J_0$ are even. 
Therefore, $G^{(4)}_F$ is quantized integrally for these components, 
and this is sufficient in guaranteeing that the bundles for fields 
in the roots of $E_8$ are well-defined.

It is also possible to turn on a line bundle in the $\U(1)_Y$
direction within $\SU(5)_{\rm GUT}$, so that the $\SU(5)_{\rm GUT}$
symmetry is broken down to local $\SU(3)_C \times \SU(2)_L$
and possibly global $\U(1)_Y$ symmetry, as in \cite{Blumenhagen}. 
If the line bundle in the $\U(1)_Y$ direction is trivial in the fiber 
direction in the Heterotic string compactification, then its 
F-theory dual exists, and a four-form flux 
\begin{equation}
 \Delta G^{(4)}_F \propto (2 C_1 + 4 C_2 + 6 C_3 + 3 C_4) \otimes \omega_Y 
\label{eq:GF4Y}
\end{equation}
(modulo $H^4(B_2)$) turns on a $\U(1)_Y$ bundle on a locus of $A_4$ 
singularity in the F-theory dual description. 
Because the $\SU(3)_C$ and $\SU(2)_L$
gauge interactions come from the same worldvolume 
($A_4$ singularity locus) in F-theory, unification of the 
gauge coupling constants at the Kaluza--Klein scale is maintained.
The doublet--triplet splitting problem in the Higgs sector 
can also be solved with a $\U(1)_Y$ line bundles, because 
the spectrum in the doublet part and triplet part of 
$\SU(5)_{\rm GUT}$-${\bf 5}+\bar{\bf 5}$ representations are 
different in the presence of a line bundle. Of course, 
a flat bundle in the $\U(1)_Y$ direction on an $A_4$ singularity 
locus can also maintain gauge coupling unification and 
solve the doublet--triplet splitting problem \cite{Witten-Wilson}, 
if the $A_4$ singularity locus has a non-trivial fundamental group.

We have yet to study the $G_4$ part of the primitiveness condition 
(\ref{eq:G4-cond-F}). As long as $G^{(4)}_F$ belongs to $F_2$ and 
is further expressed as (\ref{eq:GF-C})\footnote{We assume here 
that the $F^{(3)}$ and $H^{(3)}$ components of the four-form 
flux $G^{(4)}_F$ vanish.} in $F_2/F_4$, 
then $J \wedge G^{(4)}_F$ vanish in $G_2/G_4$. 
Thus, $J \wedge G^{(4)}_F$ is contained in $G_4$.
The four-form flux is primitive if and only if the condition 
in $G_4$ is satisfied.
Reference \cite{DW} chose a representative $G_\gamma \in F_2$ 
from a class in $F_2/F_4$ specified by (\ref{eq:GF-C},
\ref{eq:FGfromGHet}).
We understand that the calculations in \cite{DW} mean that 
\begin{equation}
J \wedge (G_\gamma (+ G^{(4)}_{F;B_2})) = 
 t_2 x_9 \otimes (\bar{c}_V \cdot \gamma (+ G^{(4)}_{F;B_2})) 
 + C_I \otimes (\omega^I \wedge J_{B_2}).
\label{eq:FI-Dterm-F}
\end{equation}
Here, $G^{(4)}_{F;B_2}$ means an element of $F_4 = H^4(B_2; \Q)$, 
and $(+G^{(4)}_{F;B_2})$ is added on the left-hand side, so that 
we can see how $J \wedge G^{(4)}_F$ changes when the representative 
is chosen differently. Here, we follow the way the authors of \cite{DW} 
specify in separating $t_2 J_0$ from $\pi_{W}^* J_{B_2}$.
Because $x_9$ and $C^I$'s remain mutually independent over the 
entire base 2-fold $B_2$, the first and second term should vanish 
separately in order for $G^{(4)}_F$ to be primitive (and 
the ${\cal N} = 1$ supersymmetry is preserved).

The $\bar{c}_V \cdot \gamma$ term is the non-primitive contribution 
in \cite{DW}. This contribution, however, might be cancelled 
by choosing a representative in $F_2$ differently (put another 
way, by exploiting the ambiguity in $G^{(4)}_{F;B_2} \in F_4$). 
If the projective cylinder map \cite{CD} formulated in $dP_{9}$ fibration 
is used instead of the cylinder map in the determining a representative, 
then the ``$x_9$ component'' on the right-hand side of \eqref{eq:FI-Dterm-F} 
vanishes, and at the same time, this new $G_{\gamma}$ is odd under the involution 
flipping the elliptic fiber, implying that the $SO(3,1)$ Lorentz symmetry may be restored.
Furthermore, in the entire $K3$-fibered Calabi--Yau 4-fold $X$ 
(rather than in one of $dP_9$-fibred 4-fold $W$'s), there is 
another contribution proportional to the $T^2$-fiber class $x_9$
from the hidden sector as well. It is more appropriate to study 
the primitiveness condition in the ``$x_9$ component'' in $G_4$
using the entire $K3$-fibred geometry. Contributions from 
$F^{(3)}$ and $H^{(3)}$ may or may not mix into the business. 
On the other hand, we need to make sure that we obtain a low-energy 
effective theory with $\SO(3,1)$ Lorentz symmetry, which means that 
a representative from a class in $F_2/F_4$ cannot be chosen arbitrarily. 
Although the specific choice mentioned above in the $dP_{9}$ fibration seems to be 
consistent with the $SO(3,1)$ Lorentz symmtery, things should be re-considered 
carefully in terms of $K3$-fibration once more. 
There is also a quantization condition 
on $G^{(4)}_{F;B_2}$ (possibly shifted by half integral value, 
depending on $c_2(TX)/2$). 
Therefore, we find that it is prematured to conclude that 
the four-form flux cannot (or can) be chosen primitive, although 
there is a non-primitive contribution pointed out by \cite{DW}.

Reference \cite{DW} showed that the second term vanishes 
when $C_V$ is irreducible. Although $C_I$'s are independent 
generators of $R^2\pi_{W*} \Z$ locally in $B_2$, they are not 
globally over $B_2$ for the $C_P's$ in the generator of 
the structure group of $C_V$. There is only one independent 
condition for them, and the tracelessness condition 
(\ref{eq:traceless}) guarantees that the condition is satisfied.

When the spectral surface $C_V$ is reducible, for example, when 
the degree-5 cover $C_V$ consists of irreducible degree-3 and 
degree-2 covers $C_{U_3}$ and $C_{U_2}$, then the second term 
of the right-hand side of (\ref{eq:FI-Dterm-F}) consists of 
three independent components: 
one for the $\SU(3)$ part, $C_{P=7,-\theta}$, 
one for the $\SU(2)$ part, $C_{\tilde{8}}$ and 
one for the U(1) part $\propto (3C_{\tilde{8}}+6C_6+4C_7+2C_{-\theta})$.
The primitiveness condition is satisfied in the first two 
components, which is not more than a special case of \cite{DW}.
If a $U_3 \oplus U_2$ bundle is chosen semi-stable in the Heterotic 
compactification, then the primitiveness condition is satisfied 
for the last component as well in F-theory. If the U(1) part 
has non-vanishing Fayet--Iliopoulos parameter, 
$\xi \propto \int_{B_2} \pi_{C*} \gamma_3 \wedge J_{B_2}$, then 
the U(1) symmetry in F-theory description also has a non-vanishing 
Fayet--Iliopoulos parameter 
$\xi_C \propto (C \cdot C_J) \int_{B_2} \omega^J \wedge J_{B_2}$, 
with $C$ the two-cycle that appear in (\ref{eq:GF4q7}).
This Fayet--Iliopoulos parameter is an F-theory generalization of 
an expression in \cite{FI-IIB} in Type IIB orientifold compactification.
A related subject is discussed in \cite{DRS, DW, BHV}.

If a line bundle in the $\U(1)_Y$ direction is introduced by 
the four-form flux in (\ref{eq:GF4Y}), the Fayet--Iliopoulos 
parameter may not vanish in general. In order to find a model 
of the real world, the four-form flux $\Delta G^{(4)}_F$ and 
$J_{B_2}$ should be chosen so that the Fayet--Iliopoulos parameter 
vanishes. This is an F-theory translation of a condition 
in \cite{Blumenhagen}.

\subsubsection{Chirality in $\rho(V) = V$}

The matter curve $\bar{c}_V$ in $B_2$ in the Heterotic 
string description is where a line $l=C+x_8$ 
in (\ref{eq:line-V}) is absorbed in the $Z=0$ locus of $dP_8$ fibration. 
In the $dP_9$ fibration for the F-theory description, 
$\pi^*(C)$ (often simply denoted in this article as $C$)
is in the $z_f = 0$ locus, and moreover, shrinks to zero size.
$C$ is a two-cycle isomorphic to $\P^1$, and have 
self intersection number $-2$. Singularity of $dP_9$ fibration 
is enhanced along the matter curve because of the extra collapsed 
two-cycle.

Chiral matter multiplets are localized along the matter curve, 
and they are identified with the global holomorphic sections 
of sheaves ${\cal F}_V$ on $\bar{c}_V$. The sheaves (\ref{eq:F4V}) 
are calculated in the Heterotic string compactification.
All the topological quantities associated with these sheaves 
should be the same in F-theory dual description, or otherwise, 
that is not dual. Non-topological aspects of the sheaves may 
be subject to corrections; we will discuss this issue 
in section~\ref{ssec:codim-3} later.

One of the components of the divisors determining the sheaves 
(\ref{eq:F4V}) is $j^*\gamma$, $\gamma$ pulled back onto the 
matter curve $\bar{c}_V$ from $C_V$. In order to have a description 
of matter multiplets entirely in terms of F-theory, we need 
a translation. It is well-known that the matter curves $\bar{c}_V$ 
are identified with loci of codimension-2 singularity (loci of 
enhanced singularity) in F-theory geometry, and the moduli space 
controlling the locus of the codimension-2 singularity in F-theory 
is the same as that of the spectral surface in Heterotic theory 
description. $j^* \gamma$ corresponds to (\ref{eq:gamma-F-B}) 
in F-theory. Now $j^* \gamma = \int_{C} i^* G^{(4)}_F$ 
is the field strength tensor of a gauge field obtained by 
integrating the 3-form field $C^{(3)}$ over the collapsed 
two-cycle $C$ an $M2$-brane is wrapped on.

Now the chirality formula (\ref{eq:chi-V-F}) can be rewritten 
in a more F-theory fashion:
\begin{equation}
 \chi(V) = \int_{\bar{c}_V} j^* \gamma = 
 \int_{C \times \bar{c}_V} G^{(4)}_F.
\label{eq:chi-V-F-G}
\end{equation}
This expression would be the most natural expectation 
for the chirality formula, even before passing through 
all the calculation of direct images and translation between 
the Heterotic string--F-theory duality; $\chi(V)$ is obtained 
by counting the number of vortices that the gauge field 
$\int_C C^{(3)}$ creates. In this sense, this is a beautiful 
result, but not surprising. 
The true benefit of all these processes starting from the Heterotic 
theory is to know that codimension-3 singularities of F-theory 
do not give rise to extra contributions to the chirality formula.
It would be difficult to say something about the codimension-3
singularities of F-theory without a better (and fundamental) 
formulation of F-theory itself or using duality with the Heterotic 
string theory.

The chirality formula above generalizes a corresponding formula 
in the Type IIB string theory. In the Type IIB string theory, 
$\SU(5)_{\rm GUT}$ vector multiplets can be realized by 
wrapping five D7-branes on a holomorphic four-cycle $\Sigma$, 
and chiral multiplets in the ${\bf 10}$ representation 
of $\SU(5)_{\rm GUT}$ are localized on a curve $\bar{c}_{\bf 10}$ 
that is the intersection of $\Sigma$ and an O7-plane.
The chirality formula in the Type IIB string theory is 
given\footnote{The $B$-field has to be chosen half integral, 
if $c_1(T\Sigma)$ is not even (Freed--Witten anomaly) \cite{FW}.
But, the field strength for the Type IIB open strings in the rank-2 
anti-symmetric representation receives a contribution $2B$, 
and the vector bundle for these fields are well-defined.} 
by \cite{BH,WY,TW1}
\begin{equation}
 \chi({\bf 10}) = 
 2 \int_{\bar{c}_{\bf 10}} \left(\frac{F}{2\pi} - \frac{B}{(2\pi)^2\alpha'}\right).
\end{equation}
The net chirality can also be expressed in terms of K-theory 
pairings \cite{MM} of D7-brane charge and O7-brane charge in a Calabi--Yau 
3-fold for the Type IIB orientifold compactification \cite{orientifold-pair}. 
But the expression in terms only of local geometry along the 
intersection curve allows straightforward generalization in F-theory.

Among the five two-cycles $C^{ab;p}$ for 
$p \in \{6^\flat, 6,7,8,8^\sharp\}$ in (\ref{eq:line-V}) 
in $dP_8$ fibration, four are linearly independent. 
(Here, we ignore the difference in the choice of $ab$.) 
In the language of spectral surface, only one point in the 
spectra surface intersects the zero section along the 
matter curve generically.\footnote{Two points among 
$\{p_i\}_{i=1,\cdots,N}$ in an elliptic fiber $E_b$ are 
on the zero section only for special isolated points 
on the matter curves. Such exceptional points 
are the subject of section~\ref{ssec:codim-3}, and 
we will ignore this issue here.}
Thus, only one out of five is absorbed in the $Z=0$ locus 
at a generic point on $\bar{c}_V$. In F-theory language, 
this means that only on two-cycle $C$ that corresponds to 
(\ref{eq:line-V}) collapses along the matter curve $\bar{c}_V$.

It is possible only in a local patch of $B_2$ to individually 
trace the five points $\{p_i\}_{i=1,\cdots,5}$ of the spectral surface, 
or five two-cycles $C^p$ ($p=\{6^\flat,6,7,8,8^\sharp\}$).
Globally on $B_2$, those five objects have to be glued 
together between two adjacent patches by the Weyl group 
$\mathfrak{S}_5$ of $A_4$. A system of five two-cycles glued 
by $\mathfrak{S}_5$ along $B_2$ is a part of $R^2\pi_{W*} \Z$ 
introduced in \cite{CD}. It will often be the case 
(though we do not have a proof) that there is only one topological 
four-cycle coming out of $H^2(B_2; R^2\pi_{W*} \Z)$, one given by 
$C \times \bar{c}_V$, where $C$ is now the collapsed two-cycle along 
the curve $\bar{c}_V$. The net chirality is given by the topological 
number of the four-form flux $G^{(4)}_F$ on this four-cycle.
Only one topological number matters. This will be in one to one 
correspondence with the parameter $\lambda$ in (\ref{eq:gamma0}).

\subsubsection{Chirality in $\rho(V) = \wedge^2 V$}

We are now ready to study the sheaf ${\cal F}_{\wedge^2 V}$ 
on the matter curve $\bar{c}_{\wedge^2 V}$, or 
$\widetilde{\cal F}_{\wedge^2 V}$ on the covering curves 
$\tilde{\bar{c}}_{\wedge^2 V}$. 
As we have learnt in section~\ref{sec:Examples}, 
divisors of the line bundles $\widetilde{\cal F}_{\wedge^2 V}$
always contain $\tilde{\pi}_{D*} \gamma$. 
Let us study what this contribution means in the F-theory language.

$\tilde{\pi}_D: D \rightarrow \tilde{\bar{c}}_{\wedge^2 V}$ 
is a degree-2 cover, allocating two points $\{p_i, p_j\}$ to  
a point in $\tilde{\bar{c}}_{\wedge^2 V}$ so that 
$p_i \boxplus p_j = e_0$. Let us denote the lines in $I_8$ 
for those two points (in the fundamental representation of 
$\SU(5)_{\rm bdl}$) as $l^p$ and $l^q$ 
($p,q \in \{6^\flat,6,7,8,8^\sharp\}$); here, we consider those 
lines modulo $C_A$ ($A=1,2,3,4$) for the unbroken $\SU(5)_{\rm GUT}$ 
symmetry. Now
\begin{equation}
 \tilde{\pi}_{D*} \gamma = \int_{l^p} G^{(4)}_H + \int_{l^q} G^{(4)}_H 
  = \int_{C^p + C^q} G^{(4)}_H 
  = \int_{C^{pq}} G^{(4)}_H. 
\end{equation}
Here, a topological relation $C^p + C^q \equiv C^{pq}$ 
(mod $C_A$ ($A=1,2,3,4$)) between the 2-cycles in (\ref{eq:line-V})
and (\ref{eq:line-X}) was used in the last equality. 

There is a uniqueness problem in translating the Heterotic theory result 
of $\tilde{\pi}_{D*}\gamma$ into F-theory language, as we encountered 
in translating $j^* \gamma$ into (\ref{eq:gamma-F-A}) or (\ref{eq:gamma-F-B}).
We adopt 
\begin{equation}
 \tilde{\pi}_{D*} \gamma = \int_{\pi^* (C^{pq})} G^{(4)}_F,
\label{eq:gamma-D-F}
\end{equation}
in the same spirit as we chose (\ref{eq:gamma-F-B}) for $j^* \gamma$.
This is the field strength of a gauge field obtained by integrating 
the 3-form field $C^{(3)}$ over the two-cycle 
$C_{a;}^{\;\;\; pq}$---a gauge field an $M2$-brane wrapped on 
the collapsed two-cycle $C_{a;}^{\;\;\;pq}$ is coupled to.
As long as we adopt this rule of translation, the flux quanta 
associated with the non-$E_8$ part of the two-cycles in $K3$-fiber 
do not have an influence on the net chirality, or even on $\gamma$ 
that describes a line bundle in F-theory. 

It is interesting to note that the notion of the covering curve 
$\tilde{\bar{c}}_{\wedge^2 V}$ we introduced in sections~\ref{sec:Idea}
and \ref{sec:Examples} is not only for mathematical convenience. 
An $M2$-brane wrapped on a cycle $\pi^*(C^{pq})$ 
propagates on the covering matter curve $\tilde{\bar{c}}_{\wedge^2 V}$, 
not on the matter curve $\bar{c}_{\wedge^2 V}$, because the each point 
of the covering mater curve is in one to one correspondence with the 
collapsed two-cycle.
 
The chirality formula in this (pair of) irreducible representation(s) 
follows immediately:  
\begin{equation}
 \chi(\wedge^2 V) = \int_{\tilde{\bar{c}}_{\wedge^2 V}} 
  \tilde{\pi}_{D*} \gamma = 
 \int_{C^{pq} \times \tilde{\bar{c}}_{\wedge^2 V}} G^{(4)}_F.
\label{eq:chi-V2-F-G}
\end{equation}
This is quite a natural result, once again. But all the hard work 
in section~\ref{sec:Examples} that has led to this conclusion 
tells us that we do not need to add an extra contributions associated 
with codimension-3 singularities of F-theory; it was the part hardly 
accessible with limited intuition in F-theory, yet our study using 
the Heterotic--F theory duality shows that (\ref{eq:chi-V2-F-G})
is indeed fine.

This expression is an F-theory generalization of the Type IIB 
chirality formula in a corresponding system. 
Here, we imagine a Type IIB set up where five D7-branes are wrapped 
on a holomorphic four-cycle $\Sigma_{\bf 5}$ of a Calabi--Yau 3-fold, 
and another D7-brane on another four-cycle $\Sigma_{\bf 1}$. 
Topological U(1) gauge field configuration $F_{5}$ and $F_{1}$ 
is assumed on the both four-cycles, $\Sigma_{\bf 5}$ and 
$\Sigma_{\bf 1}$, respectively. Then, the net chirality in the
$\SU(5)_{\rm GUT}$-$\bar{\bf 5}$ representation is given by \cite{WY}:
\begin{equation}
 \# (\bar{{\bf 5}},{\bf 1}^+) - \# ({\bf 5},{\bf 1}^-) = 
\int_{\Sigma_{\bf 5} \cdot \Sigma_{\bf 1}} 
  i^* \left(\frac{F_1}{2\pi}\right) - i^* \left(\frac{F_5}{2\pi}\right).
\label{eq:IIB-chiralityD7D7'}
\end{equation}
This expression, written only in terms of local geometry around 
the D7--D7 intersection curve, is equivalent to the one 
in \cite{quintic} given by pairing of D-brane charge 
vectors in K-theory \cite{MM, CY, charge}.
The F-theory formula (\ref{eq:chi-V2-F-G}) is the most natural 
generalization of the local formula of the Type IIB string 
theory (\ref{eq:IIB-chiralityD7D7'}).

\subsubsection{Chirality in $\rho(V) = \wedge^3 V$}

It is now straightforward to provide an F-theory interpretation 
for the $\tilde{\pi}_{T \pm *} \gamma|_{T_\pm}$ contribution 
to the sheaves $\widetilde{\cal F}_{\wedge^3 V \pm}$ in 
(\ref{eq:F4V3-rk6}). In the Heterotic theory description, 
\begin{equation}
 \tilde{\pi}_{T_\pm *} \gamma = \int_{l^p + l^q + l^r} G^{(4)}_H
  = \int_{C^p + C^q + C^r} G^{(4)}_H
  = \int_{C^{pqr}} G^{(4)}_H,
\end{equation}
where $C^{pqr}$'s are now two-cycles that correspond to 
the roots in the $(\wedge^3 V,{\bf 1}, {\bf 2})$ of the 
group $\SU(6) \times \SU(3) \times \SU(2) \subset E_8$.
In F-theory, this is replaced by $\int_{\pi^*(C^{pqr})} G^{(4)}_F$.

In the $\SU(6)$-bundle compactification of the Heterotic string 
theory, there are two types of massless chiral multiplets 
in the $({\bf 1}, {\bf 2})$ representation of the unbroken 
symmetry group $\SU (3) \times \SU (2)$. One group of multiplets is  
$H^0(\bar{c}_{\wedge^3 V}; \nu_* \widetilde{\cal F}_{\wedge^3 V+})$,
and the other  
$H^0(\bar{c}_{\wedge^3 V}; \nu_* \widetilde{\cal F}_{\wedge^3 V-}) \\
 \simeq 
[H^1(\bar{c}_{\wedge^3 V}; \widetilde{\cal F}_{\wedge^3 V+})]^\times$. 
Thus, a net chirality can be defined in the $\SU(2)$-doublet sector 
as the difference between the degrees of freedom of the two groups.
It is 
\begin{equation}
 \chi(\wedge^3 V)_+ := 
 \chi(\bar{c}_{\wedge^3 V}; \nu_* \widetilde{\cal F}_{\wedge^3 V+})
  = T_+ \cdot \gamma = 
 - \chi(\bar{c}_{\wedge^3 V}; \nu_* \widetilde{\cal F}_{\wedge^3 V-}).
\end{equation}
In F-theory, this chirality is given by 
\begin{equation}
 \chi(\wedge^3 V)_+ = \int_{C^{pqr} \times \bar{c}_{\wedge^3 V}} G^{(4)}_F.
\end{equation}
%

\subsection{Codimension-3 Singularities in F-theory Geometry}
\label{ssec:codim-3}

There are many aspects in low-energy physics that do not depend 
only on the net chirality in each representation. 
One will be surely interested in whether the two Higgs doublets 
of the Minimal Supersymmetric Standard Model can be vector-like 
in nature. If there are light vector-like $\SU(5)_{\rm GUT}$-charged 
multiplets, they may serve as messenger sector of gauge mediated 
supersymmetry breaking, for example. For these purposes, 
we need to know both $h^0(\bar{c}_{\wedge^2 V}; {\cal F}_{\wedge^2 V})$
and $h^1(\bar{c}_{\wedge^2 V}; {\cal F}_{\wedge^2 V})$ separately, 
not just the difference between these two numbers. 
Even if there are no vector-like pairs of multiplets in low energy, 
heavy vector-like states can make some qualitative differences 
in physics observed in low energy (e.g. \cite{KNW}). 
However, $d = {\rm deg} \; c_1(\widetilde{\cal F}_{\rho(V)})$ 
alone cannot determine both $h^0$ and $h^1$, 
if $0 \leq d \leq 2g - 2$, where $g$ is the genus of 
$\tilde{\bar{c}}_{\rho(V)}$ \cite{Penn5}. 
Since $g$ is generically large for the curve
$\tilde{\bar{c}}_{\wedge^2 V}$ (see Tables~\ref{tab:rk5-example}, 
\ref{tab:rk6-example}), and $d$ is not different very much from $g-1$ 
(c.f. (\ref{eq:related-rkN})), $d$ is quite likely to be in the 
window above, indeed.
More numerical information such as values of Yukawa couplings 
of quarks and leptons depend on more detail of the divisors specifying 
the sheaves on the matter curves; Yukawa couplings may depend on 
the values of global holomorphic sections at intersection points 
of matter curves \cite{BHV}, and just $d$ is clearly not enough 
information in determining the values of the sections.
 
In order to obtain all this information, one needs to use 
all the information of the line bundles $\tilde{\cal F}_{\rho(V)}$ 
on the matter curves, or of the divisors that determine them.
We have clarified how the divisors proportional to $\gamma$ 
originate in F-theory geometry. It is now time to do the rest.
All the divisors of $\tilde{\cal F}_{\rho(V)}$'s contain 
a pullback of the canonical divisor of the base manifold $B_2$.
Since the Heterotic and F-theory share the same base 2-fold $B_2$, 
$K_{B_2}$ is well-defined in F-theory as well.
Thus, we study the rest of the components of the divisors 
on the (covering) matter curves in this subsection.
We will see that most of the divisors that we identified in 
section~\ref{sec:Examples} in Heterotic theory compactification 
correspond to codimension-3 singularities in F-theory geometry.

Geometry of $dP_9$ fibration $\pi_W: W \rightarrow B_2$ is specified 
by equations (\ref{eq:dP9fiber0}, \ref{eq:dP9fiberoo}), and 
(singular) geometry along $z_f = 0$ locus is described better 
by (\ref{eq:dP9fiber0}). $a_r$ in this equation are global holomorphic 
sections of ${\cal O}(r K_{B_2} + \eta)$ on $B_2$, and 
they originally described the spectral surfaces. 
Parameters $a^{\rm Tate}_i$ that appeared in \cite{6authors} are related 
to these $a^{\rm SS}_r$ (SS is a short hand notation of spectral surface)
are related through 
\begin{equation}
 a^{\rm Tate}_{6 - i} = (-1)^i a^{\rm SS}_i z_f^{5 - i}, 
\end{equation}
and the property 
\begin{equation}
 {\rm ord} \; a^{\rm Tate}_{r} = r-1
\end{equation}
required for geometry with $A_4$ singularity is reproduced 
from the correspondence above. 
Note in the Heterotic--F theory dictionary in 
(\ref{eq:dict-f0}--\ref{eq:dict-g6}), however, that 
all the coefficients are already fixed except the 
rescaling of the coordinate $z_f$.
Codimension-3 singularities of F-theory geometry 
were studied in detail in \cite{AC}. (See also \cite{Candelas}.)
We will use the 
precisely determined dictionary (\ref{eq:dict-f0}--\ref{eq:dict-g6}) 
instead and do the same calculation over again in the following.

\subsubsection{Geometry with a Locus of $E_6$ Singularity}

If the sections $a_4$ and $a_5$ vanish, then the $dP_9$-fibered 
geometry develops a locus of $E_6$-type singularity at $z_f = 0$:
\begin{equation}
 y^{2} = x^{3} + g_2 z_f^4.
\end{equation}
The discriminant of the elliptic fibration is given by 
\begin{equation}
 \Delta = z_f^8 \left(\frac{27}{16}a_3^4  + 
  \frac{z_f}{2}(8 a_2^3 + 27 a_3^2 a_0) + 
  z_f^2 (27 a_0^2 + \cdots) + \cdots \right).
\label{eq:Det-E6}
\end{equation}

The sheaves ${\cal F}_V$ and ${\cal F}_{\wedge^2 V}$ on the matter curve
$\bar{c}_V$ involve a divisor $b^{(a)} = j^* r$. In order to obtain 
a description of these sheaves in F-theory, we would not want 
these divisors to be expressed in terms of the ramification divisor 
$r$, which is rather closely associated with geometry of vector bundles 
in Heterotic theory. In section~\ref{ssec:rk3}, 
\begin{equation}
 b^{(a)} := {\rm div} \; a_2
\end{equation}
was the definition of the divisor on $\bar{c}_V$.
Since the defining equation of $dP_9$-fibered geometry 
of F-theory uses the same data $a_0$, $a_2$ and $a_3$, 
we know where the support of $b^{(a)}$ is in F-theory 
geometry as well. 

The codimension-1 $z_f = 0$ locus in the base 3-fold is now 
a zero of the discriminant $\Delta$ of order $z_f^8$. 
The matter curve $\bar{c}_V$ is a codimension-2 locus in the 
3-fold and $\Delta \sim {\cal O}(z_f^9)$ there. Singularity 
is enhanced from $E_6$ to $E_7$ there. Because the coefficient 
of $z_f^9$ term is $4 a_2^3$ when it is evaluated on the matter 
curve $a_3 = 0$, $b^{(a)}$ is actually the codimension-3 locus 
in the 3-fold where $\Delta \sim {\cal O}(z_f^{10})$. 
Singularity is now enhanced to $E_8$. 
Thus, the divisor $b^{(a)}$ on the matter curve can be 
defined as the codimension-3 singularity of the $dP_9$-fibered 
geometry of F-theory.

We are now able to describe the sheaves (\ref{eq:F4V})
and (\ref{eq:F4V2-rk3}) entirely in terms of geometric 
object in F-theory. We lack an explanation for why 
the coefficient of the divisor $b^{(a)}$ is $1/2$, 
nothing else. Of course we know that it has to be $1/2$, 
because otherwise, 
\begin{equation}
 {\rm deg} \; \left(K_{B_2} + c_a b^{(a)} \right)
 = \frac{1}{2} {\rm deg} \; K_{\bar{c}_V} + 
   \left( c_a - \frac{1}{2}\right) {\rm deg}\; b^{(a)},
\label{eq:inconsistency}
\end{equation}
and a consistency relation 
${\cal F}_V \cong K_{\bar{c}_V}\otimes {\cal F}_{V^\times}^{-1}$ 
no longer holds for $c_a \neq 1/2$. 
We believe that there must be an explanation for $1/2$ 
in terms of local geometry around the codimension-3 
singularity within F-theory itself, not just from a global 
consistency above, but we do not have one; the coefficient was 
determined through the Heterotic--F theory duality, instead. 
For practical purposes such as model building, one can 
just use the coefficient $1/2$, and there is nothing wrong.
A local explanation of $1/2$ in F-theory itself remains 
an (academic but quite interesting) open problem for the
future.\footnote{(note in v.4) A clear answer is now given 
to this problem; see section 5 of \cite{Hayashi-2}.}

\subsubsection{Geometry with a Locus of $E_5 = D_5$ Singularity}

If only the global holomorphic section $a_5$ vanishes, 
and $a_{0,2,3,4}$ are generic, then we have a locus of 
$D_5$ singularity.
\begin{equation}
 y^{2} = x^{3} + f_2 z_f^2 x + g_3 z_f^3.
\end{equation}
The discriminant is given by 
\begin{equation}
 \Delta = z_f^7 \left(a_4^{3} a_3^2 + 
   z_f \left(\frac{27}{16} a_3^4 - \frac{9}{2} a_3^2 a_2 a_4
             - a_4^2 (a_2^2 - 4 a_0 a_4) \right)
   + {\cal O}(z_f^2)\right).
\label{eq:Det-E5}
\end{equation}
Singularity is $D_5$ along the codimension-1 $z_f = 0$ locus, and 
$\Delta \sim {\cal O}(z_f^7)$. 
Along codimension-2 locus $a_4 = 0$ ($\bar{c}_V$) and $a_3 = 0$ 
($\bar{c}_{\wedge^2 V}$), $\Delta \sim {\cal O}(z_f^8)$, and 
the singularity is enhanced to $E_6$ and $D_6$, respectively.

The description of ${\cal F}_V$ involves a divisor $j^* r$
on $\bar{c}_V$, and that of $\tilde{\cal F}_{\wedge^2 V}$ 
a divisor $\tilde{b}^{(c)}$ on $\tilde{\bar{c}}_{\wedge^2 V}$ 
[resp. $b^{(c)}$ on $\bar{c}_{\wedge^2 V}$]. 
Thus, let us think of characterizing those divisors 
in terms of F-theory geometry.

Because the argument around (\ref{eq:cv-r}) is valid 
independent of rank $N$ of the vector bundles 
in Heterotic theory compactification, the relation 
(\ref{eq:jr=ba}) holds for any $N$. Here, the definition 
of $b^{(a)}$ is now 
\begin{equation}
 b^{(a)} := {\rm div} \; a_{N-1}
\label{eq:def-ba}
\end{equation}
on the matter curve $\bar{c}_V$ ($a_N=0$).
The $E_6$ singularity along the matter curve $a_4 = 0$
is enhanced to $E_7$ at the codimension-3 singularity, 
$a_3 = 0$, and this is where we find the divisor $b^{(a)}$. 
Thus, this can be used as the F-theory characterization of 
the divisor $b^{(a)}$. At such codimension-3 singularities, 
the order of the discriminant $\Delta$ is enhanced, an extra 
two-cycle collapses, and sometimes, multiple two-cycles 
exhibit a monodromy around a codimension-3 singularity. 
Such nature of F-theory geometry may be able to account 
for the coefficient $1/2$ of the divisor $b^{(a)}$, but 
we do not have a clear answer for this problem, apart 
from the global consistency condition we mentioned after
(\ref{eq:inconsistency}).

The divisor $\tilde{b}^{(c)}$ is where the covering 
matter curve $\tilde{\bar{c}}_{\wedge^2 V}$ is ramified 
over the matter curve $\bar{c}_{\wedge^2 V}$, and the 
branched locus was characterized as the zero locus of 
$R^{(4)}$; see (\ref{eq:def-R4}) and (\ref{eq:def-bc-rk4}).
Since the coefficient of the $z_f^8$ term of the 
discriminant is 
\begin{equation}
 a_4^2 R^{(4)}  + {\cal O}(a_3^2), 
\end{equation}
it is at $b^{(c)}$ that this coefficient vanishes,\footnote{
It was reported in \cite{AC} that the coefficient of the 
${\cal O}(z_f^8)$ term vanishes on $\bar{c}_{\wedge^2 V}$ 
when 
\begin{equation}
R \propto - \frac{3}{4}f_1^2 + 2 g_1 h + 3 f_0 h^2 = 0.
\end{equation}
There is a loose Heterotic--F-theory correspondence between 
sections of a common line bundle: 
\begin{eqnarray}
f_1, a_2 \in \Gamma(B_2; {\cal O}(2K_{B_2} + \eta)), & &
g_1, a_0 \in \Gamma(B_2; {\cal O}(\eta)),   \\
h, a_4 \in \Gamma(B_2; {\cal O}(4K_{B_2} + \eta)), & & 
f_0 \in \Gamma(B_2; {\cal O}(-4 K_{B_2})).
\end{eqnarray}
It is only with the precise Heterotic--F dictionary 
(\ref{eq:dict-f0}--\ref{eq:dict-g6}), however, that one can 
find (or even discuss) an agreement between the divisors 
of $R^1\pi_{Z*} \wedge^2 V$ in the Heterotic 
theory and the codimension-3 singularities in F-theory.
Note that $f_0$ responsible for the complex structure of the elliptic 
fiber enters also in (\ref{eq:dict-g1}, \ref{eq:dict-g2}).}  
and $\Delta \sim {\cal O}(z_f^9)$.

It is interesting to note that there is a contribution 
$\tilde{b}^{(c)}$ to the divisor of 
$\tilde{\cal F}_{\wedge^2 V}$, but from the other codimension-3 
singularity such as $b^{(a)}$ which also defines a divisor on
$\bar{c}_{\wedge^2 V}$. On the other hand, the divisor $b^{(a)}$ 
contributes to ${\cal F}_V$ with a coefficient $1/2$.
We regret that we only have the results, and do not have 
a local explanation for these interesting phenomena, 
apart from the global consistency conditions such as 
$\tilde{\cal F}_{\wedge^2 V} \cong K_{\tilde{\bar{c}}_{\wedge^2 V}}
\otimes \tilde{\cal F}_{\wedge^2 V}^{-1}$ (or equivalently 
(\ref{eq:relate-rk4})).  

\subsubsection{Geometry with a Locus of $E_4=A_4$ Singularity}

For fully generic choice of $a_{0,2,3,4,5}$, a locus of 
$A_4$ singularity exists in the $z_f = 0$ locus. 
The singularity is at $(x,y) = (0,0)$ in the coordinate used 
in (\ref{eq:dP9fiber0}), and $(x,y)=(a_5^2/12,0)$ in the coordinates 
for the Weierstrass-form equation in (\ref{eq:dP9-Weierstrass}).
The discriminant around the $A_4$ singularity locus is given by 
\begin{eqnarray}
 \Delta & = & z_f^5 \left( \frac{1}{16}a_5^4 P^{(5)} 
   + \frac{z_f}{16} a_5^2 
    \left(12 a_4 P^{(5)} - a_5^2 R^{(5)}\right) \right. \nonumber \\
  & & \qquad \qquad   \left. +z_f^2 \left(a_3^2 a_4^{3} 
      + {\cal O}(a_5)\right) 
      + {\cal O}(z_f^3) \right).
\label{eq:Det-E4}
\end{eqnarray}
The matter curve $\bar{c}_V$ is given by $a_5 = 0$, and
$\bar{c}_{\wedge^2 V}$ by $P^{(5)} = 0$, where $P^{(5)}$ is defined 
in (\ref{eq:5bar-curve-eq}).

The sheaf ${\cal F}_V$ on the matter curve $\bar{c}_V$ involves 
a divisor $j^* r = b^{(a)}$, which is identified with the locus of 
$a_5 = a_4 = 0$, just like we argued for the case with $D_5$ 
singularity. This locus corresponds to the type (a) intersection 
points of $\bar{c}_V$ and $\bar{c}_{\wedge^2 V}$ 
(see Figure~\ref{fig:rk5}). Singularity is enhanced here, and 
$\Delta \sim {\cal O}(z_f^8)$.

The sheaf $\tilde{\cal F}_{\wedge^2 V}$ on the covering matter curve 
$\tilde{\bar{c}}_{\wedge^2 V}$ involves $b^{(c)}$.
Among the type (c) ramification points, we have identified the 
locus of type (c1) points, and their positions on $\bar{c}_{\wedge^2 V}$ 
was specified by the zero locus of $R^{(5)}$ defined in (\ref{eq:R5-def}). 
One can see from (\ref{eq:Det-E4}) that this is exactly the place 
in F-theory geometry where singularity is enhanced, and 
the discriminant becomes $\Delta \sim {\cal O}(z_f^7)$ from 
$\Delta \sim {\cal O}(z_f^6)$ on generic points on $\bar{c}_{\wedge^2 V}$.

Although the divisors of the sheaves ${\cal F}_V$ and 
${\cal F}_{\wedge^2 V}$ correspond to codimension-3 singularities 
of F-theory geometry, however, not all those singularities 
contribute to the divisors, just like we have already seen in the 
case of $D_5$ singularity locus. For example, along the matter curve 
$\bar{c}_V$, $\Delta \sim {\cal O}(z_f^8)$ at type (d) intersection 
points of $\bar{c}_V \cdot \bar{c}_{\wedge^2 V}$ as well 
($a_3 = 0$ as well as $a_5 = 0$), but there is not contribution 
to the divisor of ${\cal F}_V$ there.
Similarly, there is no contribution at the type (a) intersection 
points to ${\cal F}_{\wedge^2 V}$, although $\Delta \sim {\cal O}(z_f^7)$
at $a_5 =0$ along $\bar{c}_{\wedge^2 V}$. We do not have an explanation 
which codimension-3 singularities contribute by how much, apart 
from the global consistency conditions 
${\cal F}_V \cong K_{\bar{c}_V} \otimes {\cal F}_{V^\times}^{-1}$ and 
$\tilde{\cal F}_{\wedge^2 V} \cong K_{\tilde{\bar{c}}_{\wedge^2 V}}
\otimes \tilde{\cal F}_{\wedge^2 V}^{-1}$ (or equivalently 
(\ref{eq:relate-rk5})).

\section{Describing Matter Multiplets in F-theory}

F-theory is compactified on an elliptic Calabi--Yau 4-fold 
\begin{equation}
 \pi_X : X \rightarrow B_3
\label{eq:XB3}
\end{equation}
in order to obtain low-energy effective theory with 
${\cal N} = 1$ supersymmetry.
Suppose that the elliptic fibration is given by 
\begin{equation}
 y^2 = x^3 + f x + g,
\end{equation}
where $f$ and $g$ are global holomorphic sections of 
line bundles ${\cal L}_F^{\otimes 4}$ and ${\cal L}^{\otimes 6}_F$, 
respectively. Calabi--Yau condition of $X$ requires that 
${\cal L}_F \simeq {\cal O}(-K_{B_3})$.

The discriminant locus of the elliptic fibration is given by 
\begin{equation}
 {\rm div} \; \Delta = -12 K_{B_3}.
\end{equation}
Suppose that the discriminant locus has an irreducible component 
$S$ with multiplicity $c$:
\begin{equation}
 {\rm div} \; \Delta = c \Sigma + \cdots.
\end{equation}
$c = 5$ when $X$ has $A_4$ singularity along $\Sigma$, 
$c = 7$ for $D_5$ singularity and $c=8$ for $E_6$ singularity.
A topological class of divisor $\eta$ on $\Sigma$ is defined 
by \cite{Rajesh}
\begin{equation}
 N_{\Sigma|B_3} = {\cal O}_\Sigma(6 K_\Sigma +\eta),
\end{equation}
where $K_\Sigma$ is the canonical divisor of $\Sigma$, 
and $N_{\Sigma|B_3}$ is the normal bundle of $\Sigma \hookrightarrow B_3$.
Normal coordinate of $\Sigma$ in $B_3$, $z_f$, is a section 
of the normal bundle.

Global holomorphic sections on $B_3$, $f$ and $g$,
can be expressed around $\Sigma$ by expansion in the normal 
coordinate $z_f$. 
\begin{eqnarray}
 f & = & \sum_{i=0} z_f^{i} f_{4-i}, \\
 g & = & \sum_{i=0} z_f^i g_{6-i}. 
\end{eqnarray}
Because 
\begin{equation}
 -K_{B_3}|_\Sigma = - K_\Sigma + N_{\Sigma|B_3} = 5K_\Sigma + \eta,
\end{equation}
$f_{4-i}$ and $g_{6-i}$ are holomorphic sections of the following:
\begin{eqnarray}
 f_{4-i} & \in & \Gamma(\Sigma; {\cal O}(4(5K_\Sigma + \eta)) \otimes
  N_{\Sigma|B_3}^{-i}) = 
    \Gamma(\Sigma; {\cal O}(20-6i) K_\Sigma + (4-i)\eta), \\
 g_{6-i} & \in & \Gamma(\Sigma; {\cal O}(6(5K_\Sigma + \eta)) 
   \otimes N_{\Sigma|B_3}^{-i}) 
  = \Gamma(\Sigma; {\cal O}(6(5-i)K_\Sigma + (6-i) \eta)).
\end{eqnarray}
In order to preserve $A_4$, $D_5$ or $E_6$ singularity along $\Sigma$, 
there should exist global holomorphic sections 
\begin{equation}
 a_r \in \Gamma(\Sigma; {\cal O}(rK_\Sigma + \eta)), \qquad r=0,2,3,4,5,
\end{equation}
so that $f_{4-i}$'s and $g_{6-i}$'s are given globally 
on $\Sigma$ as in (\ref{eq:dict-f0}--\ref{eq:dict-g6}). 
Note that $a_r$'s and the divisor $\eta$ are characterized only 
in terms of geometry of $X$ around $\Sigma$, and one does not 
need to refer to a dual description in the Heterotic string 
theory, or even to assume that $B_3$ is a $\P^1$ fibration 
over $\Sigma$ and a Heterotic dual exists.

Matter curves are determined by the sections $a_{0,2,\cdots, 5}$ 
on $\Sigma$, and various divisors on the matter curves by 
locus of codimension-3 singularities. Three-form background 
configuration influences the sheaves ${\cal F}_{\rho(V)}$ 
only through its behavior on the collapsed two-cycles along 
$\Sigma$ or matter curves on it. Thus, the sheaves 
${\cal F}_{\rho(V)}$ are described only in terms of local geometry 
around $\Sigma$.

Since only local information of $X$ around $\Sigma$ is involved, 
descriptions of ${\cal F}_{\rho(V)}$ still hold true, as long 
as local geometry remains the same. In particular, the same 
description of the sheaves can be used even when the Heterotic 
dual does not exist. It is true that the calculation of 
$R^1 \pi_{Z*} \rho(V)$ and hence $\tilde{\cal F}_{\rho(V)}$ that 
we carried out is reliable only in the region of the moduli space 
where the volume of $T^2$ is reasonably larger than $\alpha'$ 
(but the stability of the bundle given by spectral cover construction 
is guaranteed only when the size of $T^2$ fiber is smaller 
than the typical size of the base manifold $B_2$).
Such large $T^2$ region of the moduli space corresponds to 
the stable degeneration limit of a K3-fiber into two $dP_9$-fibration, 
one for the visible sector $E_8$, and the other for the hidden sector 
$E_8$. As the volume of $T^2$ becomes comparable to $\alpha'$, 
spectral surface parameters $[a_0:a_2:a_3:a_4:a_5] \in \P^4$ 
describing four Wilson lines in the $T^2$ fiber directions 
become a part of the $\SO(18,2)/\SO(18) \times \SO(2)$ Narain moduli of 
the $T^2$ compactification of the Heterotic string theory, 
and should be treated in a way mixed up with the K\"{a}hler and 
complex moduli parameters of $T^2$. Although field theory calculation 
in the Heterotic theory is unreliable and the meaning of $a_{0,2,3,4,5}$
is not clear, we can rather take (\ref{eq:dict-f0}--\ref{eq:dict-g6}) 
as the definition of $a_{0,\cdots,5}$ in such region of the moduli
space. An idea that the divisors specifying the sheaves 
${\cal F}_{\rho(V)}$ are associated with codimension-3 singularities 
seems so natural (at least to the authors) that we speculate 
that the relations between the sheaves and the codimension-3 
singularities persist without a correction in the entire region 
of the F-theory moduli space.
We obtain ${\cal F}_{\rho(V)}$ for generic configuration of 
F-theory geometry $X$ in this way. We will be more explicit 
for specific cases later on in this section.

Since the description of ${\cal F}_{\rho(V)}$ determined as above 
relies only on local geometry of $X$ along $\Sigma$, 
such a set up may allow for local model building of particle physics; 
``local'' in the sense that the geometry 
in the other parts of $X$ does not matter (very much) 
to particle physics of the visible sector. 
That will be an F-theory version of \cite{Madrid} in the Type IIB string, 
\cite{Uranga} in Type IIA and \cite{Witten-G2} in $G_2$ holonomy 
compactification of eleven-dimensional supergravity.

\subsection{On an $E_6$-Singularity Locus}
\label{ssec:concl-E6}

The Calabi--Yau 4-fold $X$ develops a locus of $E_6$ singularity 
along $\Sigma$, when $a_4$ and $a_5$ are set to zero. 
The discriminant of the elliptic fibration (\ref{eq:XB3}) 
is given by (\ref{eq:Det-E6}) around $\Sigma$. 
Zero locus of $a_3$ defines a matter curve\footnote{
When we use representations of unbroken symmetries 
(like ${\bf 27}$ of $E_6$) as subscripts, instead of representations 
of structure groups, we will do so by using parenthesis, 
like $({\bf 27})$.} 
$\bar{c}_{({\bf 27})}$, along which the singularity in the 
directions transverse to $\Sigma$ becomes $E_7$. 
There is an extra collapsed two-cycle along the matter 
curve, so that the intersection form of the $\P^1$'s 
becomes $E_7$. The matter curve $\bar{c}_{({\bf 27})}$
belongs to a topological class $|3K_\Sigma + \eta|$.
The singularity is enhanced even to $E_8$ at some 
special points on the matter curve, determined by 
the condition $a_2 = 0$. Collection of these points 
define a divisor $b^{(a)}$ on $\bar{c}_{({\bf 27})}$ as 
in (\ref{eq:def-ba}).

Suppose that a vector bundle ${\cal E}$ is turned 
on the $E_6$ singular locus $\Sigma$. Then, the 
unbroken symmetry at low energy is a subgroup 
$H$ of $E_6$ that commutes with the structure group 
of ${\cal E}$. First group of chiral multiplets 
in low energy effective theory arises from the 
entire bulk of $\Sigma$. By generalizing $B_2$ in 
(\ref{eq:bulk-matter-1-Het}, \ref{eq:bulk-matter-2-Het}) 
to a general $E_6$ discriminant locus $\Sigma$ of F-theory, 
the matter multiplets are in \cite{DW, BHV}
\begin{equation}
 H^1(\Sigma; {\rm adj.}({\cal E})) \oplus 
 H^0(\Sigma; {\cal O}(K_\Sigma) \otimes {\rm adj.} ({\cal E})).
\label{eq:chiral-78}
\end{equation}
References~\cite{DW, BHV} build an intrinsic formulation of 
F-theory itself and explain why the latter cohomology group is 
for a bundle involving $K_\Sigma$, rather than $N_{\Sigma |B_3}$.
Calculation in Heterotic dual also concludes that $K_{\Sigma}$ 
should be used, rather than $N_{\Sigma |B_3}$, regardless of 
whether the discriminant locus $\Sigma$ has codimension-2 (and -3)
loci of enhanced singularity or not \cite{DW}.
The Heterotic theory calculation in section~\ref{sec:trivial} 
suggests, however, that not all the generators of the cohomology group 
$H^0(\Sigma; {\cal O}(K_\Sigma) \otimes {\rm adj.} ({\cal E}))$ are 
massless in fact. Only the kernel and cokernel 
of the map (\ref{eq:d2}) remain massless there, and 
it may be that similar phenomenon exists in F-theory.
The formula for the net chirality 
itself does not depend on this subtlety, however, and 
a generalization of (\ref{eq:E-chi-K}) gives 
\begin{equation}
 \chi(R_H) = - \int_\Sigma c_1(T\Sigma) \wedge c_1(\rho({\cal E})),
\end{equation}
just like in \cite{TW1,BHV}; here, we assume that the structure 
group of ${\cal E}$ is a proper subgroup of $E_6$ and its commutant 
is $H$, and the net chirality is considered for a pair of 
Hermitian conjugate pair of irreducible representations 
$(\rho({\cal E}),R_H) + (\rho({\cal E})^\times, R_H^\times)$. 

The second group of chiral multiplets are localized 
on the matter curve $\bar{c}_{({\bf 27})}$.
\begin{eqnarray}
&&  H^0\left(\bar{c}_{({\bf 27})}; 
  {\cal O}\left(i^* K_\Sigma + \frac{1}{2}b^{(a)} \right) \otimes 
  {\cal L}_{G} \otimes \rho_{({\bf 27})}({\cal E}) \right), 
  \label{eq:chiral-27}\\
&& H^0\left(\bar{c}_{({\bf 27})}; 
  {\cal O}\left(i^* K_\Sigma + \frac{1}{2}b^{(a)} \right) \otimes 
  {\cal L}_{G}^{-1} \otimes \rho_{(\overline{\bf 27})}({\cal E})
      \right). 
  \label{eq:chiral-27bar}
\end{eqnarray}
where ${\cal L}_{G}$ is a line bundle on $\bar{c}_{({\bf 27})}$
determined by a gauge field obtained by integrating the 
3-form field on the vanishing 2-cycle along $\bar{c}_{({\bf 27})}$. 
See (\ref{eq:gamma-F-B}). The formula for the net chirality 
is given by 
$\chi = \int_{\bar{c}_{({\bf 27})}} c_1({\cal L}_G \otimes 
\rho_{({\bf 27})}({\cal E}))$, or simply by (\ref{eq:chi-V-F-G}) 
in the absence of the bundle ${\cal E}$ on the locus of $E_6$ 
singularity.

\subsection{On a $D_5=E_5$-Singularity Locus}

If $a_5$ is set to zero and $a_{0,2,3,4}$ do not 
vanish, then a locus of $E_5 = D_5$ singularity develops 
along $\Sigma$. The discriminant $\Delta$ is given by 
(\ref{eq:Det-E5}) around $\Sigma$. 

As for the chiral matter multiplets arising from the entire 
worldvolume of the $D_5$ singularity locus $\Sigma$, 
everything stated in the second paragraph of 
section~(\ref{ssec:concl-E6}) holds true, after 
replacing $E_6$ by $\SO(10)$, and interpret ${\cal E}$ 
as a bundle in $\SO(10)$.

The zero locus of $a_4$ defines a matter curve $\bar{c}_{({\bf 16})}$, 
and singularity of $X$ in the direction transverse to $\Sigma$ 
becomes $E_6$. An extra two-cycle is collapsed along 
this matter curve, so that the intersection form becomes $E_6$.
The singularity becomes $E_7$ on special points on 
$\bar{c}_{({\bf 16})}$, specified by $a_3 = 0$. These points 
define a divisor $b^{(a)}$ as in (\ref{eq:def-ba}). 
Chiral multiplets localized on the matter curve are 
\begin{eqnarray}
&& H^0\left(\bar{c}_{({\bf 16})}; 
   {\cal O}\left(i^* K_\Sigma + \frac{1}{2} b^{(a)}\right) \otimes 
   {\cal L}_{G} \otimes \rho_{({\bf 16})} ({\cal E})\right), \\
&& H^0\left(\bar{c}_{({\bf 16})}; 
   {\cal O}\left(i^* K_\Sigma + \frac{1}{2} b^{(a)}\right) \otimes 
   {\cal L}^{-1}_{G} \otimes 
   \rho_{(\overline{\bf 16})} ({\cal E})\right), 
\end{eqnarray}
where ${\cal L}_{G}$ is a line bundle determined by a 2-form 
on $\bar{c}_{({\bf 16})}$ which is obtained by integrating 
the four-form field strength $G^{(4)}_F$ over the two-cycle 
collapsed along $\bar{c}_{({\bf 16})}$. The net chirality is 
given by (\ref{eq:chi-V-F-G}), if ${\cal E}$ is trivial.

Another group of chiral multiplets arises from another matter 
curve $\bar{c}_{({\bf vec})}$, which is given by zero locus of $a_3$.
Singularity of $X$ becomes $D_6$ along this curve. 
There are two two-cycles collapsing along this curve. 
We are already familiar with this phenomenon in the Type IIB 
string theory. When a D7-brane intersects a stack of 
D7-branes and an O7-plane that forms an $\SO(10)$ symmetry, 
an orientifold mirror D7-brane always intersects the stack of 
7-branes at the same intersection curve. Two different kinds of 
open strings become massless simultaneously on this curve. 
Codimension-3 singularities along this curve are $b^{(a)}$ 
that we have already mentioned, and zero locus of 
$R^{(4)} = a_2^2 - 4 a_0 a_4$.

The two collapsed two-cycles turns into one another, when 
they are traced around one of the codimension-3 singularities 
at a zero of $R^{(4)}$. Therefore, it is convenient to think of a covering 
curve $\tilde{\bar{c}}_{({\bf vec})}$ that traces the collapsed two-cycles.
$\tilde{\bar{c}}_{({\bf vec})}$ is a degree-2 cover of $\bar{c}_{({\bf vec})}$, 
and ramifies at the zero locus of $R^{(4)}$. Divisor of the branch 
points on $\bar{c}_{({\bf vec})}$ is denoted by $b^{(c)}$, and 
that of ramification points on $\tilde{\bar{c}}_{({\bf vec})}$ by 
$\tilde{b}^{(c)}$. This degree-two cover 
$\nu_{\bar{c}_{({\bf vec})}}: \tilde{\bar{c}}_{({\bf vec})} \rightarrow 
\bar{c}_{({\bf vec})}$ has branch cuts whose number is given 
by the half of (\ref{eq:nbr-bc-rk4}).
The genus of the covering curve is given by (\ref{eq:genus-formula-kr4}).
Chiral multiplets on this matter curves are 
\begin{equation}
 H^0\left(\tilde{\bar{c}}_{({\bf vec})}; 
  {\cal O}\left( i^* K_\Sigma + \tilde{b}^{(c)}\right) \otimes 
  {\cal L}_{G} \otimes \rho_{({\bf vec})}({\cal E})\right).
\label{eq:chiral-vect}
\end{equation}
${\cal L}_{G}$ is defined by field strength given 
by (\ref{eq:gamma-D-F}). Since only one collapsed two-cycle 
is associated with each point in the covering matter curve, it is
well-defined as a line bundle on the covering curve. 
The covering matter curve is where $M$2-brane propagates, and 
is more appropriate object in describing this group of 
matter multiplets than the ordinary matter curve $\bar{c}_{({\bf vec})}$. 
One could also describe the same chiral multiplets, though, 
as global holomorphic sections of a rank-2 vector bundles 
on $\bar{c}_{({\bf vec})}$ obtained by pushing forward the line 
bundle in (\ref{eq:chiral-vect}) by $\nu_{\bar{c}_{({\bf vec})}}$.

The two different matter curves $\bar{c}_{({\bf 16})}$ and 
$\bar{c}_{({\bf vec})}$ intersect at codimension-3 singular loci 
$b^{(a)}$. There are $(4K_\Sigma + \eta) \cdot (3K_\Sigma + \eta)$ such 
points. This is where Yukawa couplings 
\begin{equation}
 \Delta W_{(a)} = {\bf 16} \; {\bf 16} \; {\bf 10}
\end{equation}
can be generated \cite{BHV, DW}. Simple algebraic relation 
among the collapsed two-cycles 
there---$C^p + C^q = C^{pq}$--allows $M$2-branes to reconnect.

\subsection{On an $A_4=E_4$-Singularity Locus}

When all $a_{0,2,3,4,5}$ are allowed to be non-zero, 
$S$ is an $A_4 = E_4$ singular locus. The discriminant 
of the elliptic fibration (\ref{eq:XB3}) is given locally 
around $\Sigma$ by (\ref{eq:Det-E4}). 
See (\ref{eq:5bar-curve-eq}) and (\ref{eq:R5-def})
for the definitions of $P^{(5)}$ and $R^{(5)}$, respectively.
No arguments on chiral multiplets from the bulk of $\Sigma$ have 
to be changed from the cases with $D_5$ or $E_6$ singularities.
A bundle ${\cal E}$ may be turned on in the $\U(1)_Y$ 
direction to break the $\SU(5)_{\rm GUT}$ symmetry 
of unified theories. The four-form flux to be used 
in this case is (\ref{eq:GF4Y}), where $\omega_Y$ is a two-form 
on the $A_4$ singularity locus $\Sigma$, and $C_{A=1,2,3,4}$ 
are collapsed four two-cycles forming a basis whose intersection 
form is $(- C_{A_4})$.

A matter curve $\bar{c}_{({\bf 10})}$ is the zero locus 
of $a_5$, along which the $A_4$ singularity on a generic 
point of $\Sigma$ is enhanced to $D_5$. There are two groups of 
codimension-3 singularities on $\bar{c}_{({\bf 10})}$. 
One is where $a_4$ also vanishes (type (a) intersection points), 
and the other is where $a_3$ does (type (d) intersection points). 
Those points define divisors $b^{(a)}$ and $b^{(d)}$ 
on $\bar{c}_{({\bf 10})}$. See Figure~\ref{fig:rk5}.
Chiral multiplets on $\bar{c}_{({\bf 10})}$ are 
\begin{eqnarray}
&& H^0\left(\bar{c}_{({\bf 10})}; 
   {\cal O}\left(i^* K_\Sigma + \frac{1}{2} b^{(a)}\right) \otimes 
   {\cal L}_{G} \otimes \rho_{({\bf 10})} ({\cal E})\right), \\
&& H^0\left(\bar{c}_{({\bf 10})}; 
   {\cal O}\left(i^* K_\Sigma + \frac{1}{2} b^{(a)}\right) \otimes 
   {\cal L}^{-1}_{G} \otimes 
   \rho_{(\overline{\bf 10})} ({\cal E})\right).  
\end{eqnarray}
The line bundle ${\cal L}_G$ is given by (\ref{eq:gamma-F-B}), and 
the net chirality by (\ref{eq:chi-V-F-G}).

Another matter curve $\bar{c}_{(\bar{\bf 5})}$ is the 
zero locus of $P^{(5)}$, along which the singularity 
is enhanced to $A_5$. Codimension-3 singularities along 
$\bar{c}_{\bar{\bf 5}}$ are $b^{(a)}$, $b^{(d)}$, and 
$b^{(c)} = {\rm div} R^{(5)} + b^{(a)}$.
The matter curve $\bar{c}_{(\bar{\bf 5})}$ forms a double point 
singularity at $b^{(d)}$, and it is convenient to discuss 
its blow up, the covering matter curve 
$\tilde{\bar{c}}_{(\bar{\bf 5})}$. 
This covering matter curve is where $M$2-brane propagates, 
and not the matter curve $\bar{c}_{(\bar{\bf 5})}$, because 
each point in the covering matter curve is in one-to-one 
correspondence with the collapsed two-cycle along the 
matter curve.
Chiral multiplets are global holomorphic sections 
of line bundles on the covering matter curve, 
\begin{eqnarray}
&& H^0\left(\tilde{\bar{c}}_{(\bar{\bf 5})}; 
   {\cal O}\left(i^* K_\Sigma + \frac{1}{2} b^{(c)} \right) \otimes 
   {\cal L}_{G} \otimes \rho_{(\bar{\bf 5})} ({\cal E})\right), \\
&& H^0\left(\tilde{\bar{c}}_{(\bar{\bf 5})}; 
   {\cal O}\left(i^* K_\Sigma + \frac{1}{2}b^{(c)} \right) \otimes 
   {\cal L}^{-1}_{G} \otimes 
   \rho_{({\bf 5})} ({\cal E})\right).    
\end{eqnarray}
\addtocounter{equation}{1}
See Figure~\ref{fig:rk5} for the rough sketch of the geometry 
of the two matter curves and the variety of their intersection 
points. Table~\ref{tab:rk5-example} shows sets of geometric 
data such as genus of the covering matter curve 
$\tilde{\bar{c}}_{({\bf 5}}$ and the number of various types of 
codimension-3 singularities for a few examples of $\eta$. 

The topological relation among the collapsing cycles, 
\begin{equation}
C^{ab;p} + C^{cd;q} = \epsilon^{abcde} C_{e;}^{\;\;\; pq},  
\end{equation}
allows a reconnection of $M2$-branes wrapped on the relevant 
two-cycles and Yukawa couplings of the form (up-type like)
\begin{equation}
 \Delta W_{(a)} = {\bf 10}^{ab} \; {\bf 10}^{cd} \; {\bf 5}^e
 \epsilon_{abcde}
\end{equation}
may be generated. These types of Yukawa couplings are generated 
at type (a) intersection points of 
$\bar{c}_{({\bf 10})} \cdot \bar{c}_{(\bar{\bf 5})}$, because, 
all the two-cycles $C^p$, $C^q$ and $C^{pq}$ mode $C_A$ ($A=1,2,3,4$)
collapse to zero size there. Singularity is enhanced from $A_4$ 
to $E_6$ at each type (a) intersection point \cite{TW1, BHV}.

Another relation 
\begin{equation}
 C_{a;}^{\;\;\;pq} + C_{b;}^{\;\;\; rs} + \epsilon_{pqrst} C^{ab;t} = 0
\end{equation}
allows a different kind of reconnection of $M2$-branes, and hence 
Yukawa couplings (down-type like)
\begin{equation}
 \Delta W_{(d)} = \bar{\bf 5}_a \; {\bf 10}^{ab} \; \bar{\bf 5}_b
\end{equation}
may be generated. This type of reconnection is possible at 
the type (d) intersection points, because all the two-cycles 
$C^{t}$, $C^{pq}$ and $C^{rs}$ can collapse to zero size 
simultaneously there. 
Singularity is enhanced from $A_4$ to $D_6$ there. 
Three branches of matter curves intersect at each type (d) 
intersection point (see Figure~\ref{fig:rk5}). 
Local geometry around this type (d) intersection point 
allows a Type IIB interpretation. This is where 
a stack of five D7-branes, an O7-plane, one D7-brane and 
its orientifold mirror image intersect simultaneously.

It is interesting that the up-type and down-type 
Yukawa couplings are associated with different kinds of 
the intersection points of the two matter curves 
$\bar{c}_{({\bf 10})}$ and $\bar{c}_{(\bar{\bf 5})}$. 
This is an important observation in an attempt to understand 
Yukawa couplings of quarks and leptons.

\subsection{On an $A_2+A_1=E_3$-Singularity Locus }

The $\SU(6)$ bundle compactification of the Heterotic theory can be 
used to study various properties of F-theory vacua with a locus 
of $A_2 + A_1 = E_3$ singularity. There are three matter curves 
on the $E_3$ singularity locus $\Sigma$; $\bar{c}_{(Q)}$, where the 
singularity is enhanced to $A_4 = E_4$, 
$\bar{c}_{(\bar{U})}$ where the symmetry is enhanced to
$\SU(4) \times \SU(2)$, and finally $\bar{c}_{(L)}$, where 
the enhanced symmetry is $\SU(3) \times \SU(3)$.
See Figure~\ref{fig:rk6} for how those curves intersect one 
another. Table~\ref{tab:rk6-example} shows numerical data 
of the geometry of those curves for a few examples. 
Although the analysis in section~\ref{ssec:rk6} relies on 
field theory approximation of the Heterotic string theory, 
qualitative nature of the intersection of those curves 
are believed to be the same in dual F-theory vacua. 

In the Heterotic theory language, the matter curve $\bar{c}_{(Q)}$ 
is given by $a_6 = 0$. Chiral multiplets $Q$ and $Q^c$ in the 
$(\bar{\bf 3}, {\bf 2})$ and $({\bf 3}, {\bf 2})$ representation 
of the $\SU(3) \times \SU(2)$ unbroken symmetry group are localized 
on this curve, and they are identified with the independent 
generators of the cohomology groups, 
\begin{eqnarray}
 & & H^0\left(\bar{c}_{(Q)}; {\cal O}\left(i^* K_{\Sigma} + \frac{1}{2} b^{(a)}
    \right)\otimes {\cal L}_G\right), \label{eq:chiral-Q}\\
 & & H^0\left(\bar{c}_{(Q)}; {\cal O}\left(i^* K_{\Sigma} + \frac{1}{2} b^{(a)}
    \right)\otimes {\cal L}_G^{-1} \right); 
\end{eqnarray}
the divisor $b^{(a)}$ is defined by (\ref{eq:def-ba}) with $N=6$, 
and corresponds to the type (a) intersection points 
in Figure~\ref{fig:rk6}. 

The matter curve $\bar{c}_{(\bar{U})}$ is defined by
(\ref{eq:cV2-eq-rk6}). Chiral multiplets $\bar{U}$ and $\bar{U}^c$ 
in the $({\bf 3}, {\bf 1})$ and $(\bar{\bf 3}, {\bf 1})$ 
representations correspond to 
\begin{eqnarray}
 & &  H^0\left(\tilde{\bar{c}}_{(\bar{U})}; {\cal O}\left(i^* K_{\Sigma}
  + \frac{1}{2}\tilde{b}^{(c)}\right)\otimes {\cal L}_G \right), 
   \label{eq:chiral-Ubar}\\
 & &  H^0\left(\tilde{\bar{c}}_{(\bar{U})}; {\cal O}\left(i^* K_{\Sigma}
  + \frac{1}{2}\tilde{b}^{(c)}\right)\otimes {\cal L}_G^{-1}
	 \right);  
\end{eqnarray}
here, the covering matter curve $\tilde{\bar{c}}_{(\bar{U})}$ is obtained 
by blowing up and resolving triple points of the curve 
$\bar{c}_{(\bar{U})}$, the type (e) points in Figure~\ref{fig:rk6}. 
The divisor $b^{(c)}$ is given by ${\rm div} R^{(6)} + 2 b^{(a)}$.

The last group of chiral multiplets, denoted by $L$, come from 
\begin{equation}
 H^0\left(\bar{c}_{(L)}; {\cal O}\left( i^* K_\Sigma + \frac{1}{2} b^{(f)}\right)
    \otimes {\cal L}_G \right) \oplus 
 H^0 \left(\bar{c}_{(L)}; {\cal O}\left( i^* K_\Sigma + \frac{1}{2} b^{(f)}\right)
    \otimes {\cal L}_G^{-1} \right). \label{eq:chiral-L}
\end{equation}
These matter multiplets are localized on the curve $\bar{c}_{(L)}$. 
Its defining equation is given by the zero locus of (\ref{eq:def-Q6}).
The divisor $b^{(f)}$ is the zero locus of (\ref{eq:def-S6}).

The duality map (\ref{eq:dict-f0}--\ref{eq:dict-g6}) between the moduli 
parameters of the Heterotic and F-theories was established 
only for bundles with a rank $N \leq 5$. Thus, the divisors 
$b^{(c)}$ and $b^{(f)}$ are still characterized in terms of 
the data $a_{0,2,3,4,5,6}$ describing the vector bundle $V$.
Thus, we have not seen for the rank-6 bundle compactification 
of F-theory that those divisors correspond to codimension-3
singularities in F-theory.

There are three types of the way matter curves intersect, 
as we see in Figure~\ref{fig:rk6}. 
Topological relations of collapsed two-cycles are all different 
for those different kinds of intersection points. 
Thus, the interactions generated at the codimension-3 singularities 
are different for different types of singularities.
At type (a) intersection points, 
\begin{equation}
 \Delta W_{(a)} = Q^{a\alpha} Q^{b\beta} \bar{U}^c_{ab} 
   \epsilon_{\alpha\beta}
\end{equation}
may be generated, where $a,b$ are $\SU(3)$ indices and 
$\alpha, \beta$ $\SU(2)$ indices. 
At the type (e) points, we may have 
\begin{equation}
 \Delta W_{(e)} = \bar{U}_a \bar{U}_b \bar{U}_c \epsilon^{abc}.
\end{equation}
The other type of three point couplings is 
\begin{equation}
 \Delta W_{(d)} = Q^{a\alpha} \bar{U}_a L_\alpha, 
\end{equation}
and this type of interactions may be generated at the 
type (d) points. The enhanced singularity at the type (d) 
points is $A_5$, and this interaction is what we expect 
when six D7-branes are separated into three, two and 
one coincident D7-branes intersecting one another \cite{Madrid}.

Suppose that the low-energy spectrum of chiral multiplets 
consists of the minimal anomaly free choice. That is, 
the cohomology groups (\ref{eq:chiral-Q}), (\ref{eq:chiral-Ubar}) 
and (\ref{eq:chiral-L}) have one, two and one independent generators, 
respectively, and all other cohomology groups vanish. 
Then, the type (e) Yukawa couplings vanish because of the 
anti-symmetric nature of the contraction of the $\SU(3)$ 
indices, and the type (a) Yukawa interactions simply do 
not exist because there is no particle like $\bar{U}^c$
at low-energy. Thus, the effective theory is only with 
the type (d) Yukawa interaction. This model is known as 
the 3--2 model, one of the most famous calculable models 
of dynamical supersymmetry breaking \cite{3-2}.

\section*{Acknowledgments}  

T.W. thanks Ron Donagi and Stefano Guerra for useful comments and 
fruitful discussion.
Communications with Martijn Wijnholt helped us improve 
section~\ref{ssec:gamma2G} in version 2.
M.Y. would like to thank Yukawa Institute for Theoretical Physics for
hospitality where he stayed during the final stages of this work.
This work is supported by PPARC (RT), 
by World Premier International Research Center Initiative 
(WPI Initiative), MEXT, Japan (YT, TW), and 
by JSPS fellowships for Young Scientists (MY).


\appendix

\section{Direct Image as a Pushforward from its Support}
\label{sec:push}

If a sheaf ${\cal E}$ on an algebraic variety $X$ is 
supported on a closed subvariety $i: Y \hookrightarrow X$, 
there exists a sheaf of Abelian group ${\cal F}$ on $Y$ 
such that ${\cal E} = i_* {\cal F}$ as a sheaf of Abelian 
group. It is not true in general, however, that there exists 
a sheaf of ${\cal O}_Y$-module ${\cal F}$ such that 
${\cal E} = i_* {\cal F}$ as a sheaf of ${\cal O}_X$-module. 

Any locally holomorphic functions on $X$ acts on a sheaf 
of the form $i_* {\cal F}$ by restricting them on $Y$ first, 
and then by multiplying them to ${\cal F}$. Thus, in order to 
see whether ${\cal E}$ on $X$ is given by a pushforward of 
a sheaf of ${\cal O}_Y$-module ${\cal F}$, one needs to 
make sure that any local sections of ideal sheaf of $Y$, 
${\cal I}_Y$ in  
\begin{equation}
 0 \rightarrow {\cal I}_Y \rightarrow {\cal O}_X \rightarrow 
   {\cal O}_Y \rightarrow 0,
\end{equation} 
acts trivially on ${\cal E}$. 

To get a feeling of when Fourier--Mukai transforms (\ref{eq:FM-rho(V)})
of bundles $\rho(V)$ and direct images $R^1\pi_{Z*}\rho(V)$ are 
given by pushforward of sheaves of modules, we explicitly calculate 
direct images for a simple case using \v{C}ech cohomology. 

\subsection{Warming Up}

We would like to calculate direct images associated with 
elliptic fibration $\pi_Z: Z \rightarrow B_2$. Before discussing 
$R^i \pi_{Z*} \rho(V)$, however, we begin with an elementary 
exercises of calculating sheaf cohomology on an elliptic curve $E$. 

Suppose that an elliptic curve $E$ is given by an equation 
\begin{equation}
 Z Y^2 = X^3 + f_0 X Z^2 + g_0 Z^3
\end{equation}
in $\P^2$, where $\left[X: Y: Z\right]$ are homogeneous coordinates 
of $\P^2$. $E$ is covered by two Affine open sets, $U_Z$ and $U_Y$.
$U_Z$ is $E$ in $\P^2 \backslash \{Z = 0\}$, and hence 
is $E \backslash \{e_0 = \infty\}$.
In $U_Z$, the defining equation of $E$ above can be written 
in terms of Affine coordinates $(x,y) = (X/Z, Y/Z)$.
$U_Y$ is obtained by removing three points in $E$ specified by 
intersection of $E$ and a hyperplane $Y=0$. This choice of 
hyperplane $Y=0$ is rather arbitrary; it could have been any
other hyperplane, as long as it is not $Z = 0$. When the $Y = 0$ 
hyperplane is used, then the three points $p_i$ ($i=1,2,3$) 
that are not contained in $E$ correspond to 
$(x,y) = (x_i, 0)$, with $x_i$'s three roots of 
$x^3 + f_0 x + g_0 = 0$.
Affine open sets $U_Z$ and $U_Y$ cover the entire $E$, 
and this open covering can be used to calculate \v{C}ech 
cohomology \cite{Hartshorne}.

Let us begin with calculation of $H^i(E; {\cal O}_E)$ ($i = 0,1$) 
in terms of \v{C}ech cohomology. In the \v{C}ech complex
\begin{equation}
 0 \rightarrow C^0 \rightarrow C^1 \rightarrow 0,
\end{equation} 
\begin{eqnarray}
 C^0 & = & \{ \varphi_Z \in \C (E)| \varphi \in \C [E \backslash e_0] \} 
    \oplus \{ \varphi_Y \in \C (E) | \varphi \in 
        \C [E \backslash \{p_i\}_{i=1,2,3}]\},  \\
 C^1 & = & \{ \varphi \in \C (E) | 
   \varphi \in \C [E \backslash \{e_0,p_1,p_2,p_3\}]\}. 
\end{eqnarray}
Here, $\C(E)$ means a set of rational functions on $E$, 
and $\C[U]$ a set of regular functions on $U \subset E$.
Now, $H^0(E;{\cal O}_E)$ is a subset of $C^0$ given by 
$\{(\varphi_Z, \varphi_Y)| \varphi_Z = \varphi_Y \in U_Z \cap U_Y \}$.
Thus, $\varphi_Y$ cannot have a pole at $p_{1,2,3}$ 
because $\varphi_Z$ does not, and $\varphi_Z$ cannot have a pole 
at $e_0$ because $\varphi_Y$ does not. 
Thus, $\varphi_Z = \varphi_Y$ should be chosen as a function 
that is regular everywhere in $E$, which means that they are 
a constant-valued function. This is how one can obtain 
a well-known result $H^0(E;{\cal O}_E) \simeq \C$ in 
\v{C}ech cohomology.

$H^1(E; {\cal O}_E)$ is generated by either one of functions
\begin{equation}
 \frac{y}{x-x_i} \qquad (i=1,2,3).
\label{eq:generator}
\end{equation} 
These functions are regular on $U_Z \cap U_Y$, and have poles 
of order one at $e_0$ and $p_i$, but nowhere else. 
They are elements of $C^1$. 
If the functions (\ref{eq:generator}) on $U_Z \cap U_Y$ are 
to be expressed as $\varphi_1 - \varphi_2$ with $\varphi_1 \in C^0_Z$ and 
$\varphi_2 \in C^0_Y$, the order-one pole at $e_0$ should 
come from $\varphi_1$ and the order-one pole at $p_i$ 
from $\varphi_2$. Both $\varphi_1$ and $\varphi_2$ have 
to be regular otherwise. Since any elliptic functions 
have at least a pole of order two or two poles of order one, 
no elliptic functions have properties required for $\varphi_1$ 
or $\varphi_2$. Thus, the functions (\ref{eq:generator}) 
belongs to the cokernel of $C^0 \rightarrow C^1$, and hence 
can be a generator of $H^1(E; {\cal O}_E)$.
On the other hand, difference between any two functions 
in (\ref{eq:generator}) does not have a pole at $e_0$, 
and hence it can be expressed as $\varphi_1 - \varphi_2$.
Thus, there is only one independent generator out of three 
functions in (\ref{eq:generator}).
This is how one can understand $h^1(E; {\cal O}_E) = 1$.

We will work on another exercise: $H^i(E; {\cal O}(p - e_0))$ 
($i = 0,1$). It is well known that both cohomology groups vanish for 
$p \neq e_0$, but it is an instructive exercise to reproduce this 
result, before taking on an even more difficult problem, 
calculation of $R^1 \pi_*$ in elliptic fibration given by $\pi$.
For ${\cal O}_E(p - e_0)$, 
\begin{eqnarray}
  C^0 & = & C^0_Z \oplus C^0_Y, \label{eq:Cech-C0-E2}\\
  C^0_Z & = & \{ \varphi_Z \in \C (E) | 
    \varphi_Z \in \C [E \backslash \{e_0, p \}], \; 
    v_p(\varphi_Z) \geq -1 \}, \\ 
  C^0_Y & = & \{ \varphi_Y \in \C (E) | 
    \varphi_Y \in \C [E \backslash \{ p, p_{1,2,3}\}], \;
    v_p(\varphi_Y) \geq -1, \; 
    v_{e_0}(\varphi_Y) \geq 1\}, \\
  C^1 & = & \{ \varphi \in \C (E) | 
   \varphi \in \C [E \backslash \{ e_0, p, p_{1,2,3}\}], \; 
   v_p (\varphi) \geq -1\}. \label{eq:Cech-C1-E2}
\end{eqnarray}
Here, $v_p(\varphi) = m$ means that a rational function 
$\varphi$ has a zero of order $m$ at a point $p$ if $m>0$, and 
$\varphi$ has a pole of order $(-m)$ at a point $p$ if $m < 0$.
Thus, global holomorphic section, $(\varphi_Z, \varphi_Y)$ such 
that $\varphi_Z - \varphi_Y = 0$ on $U_Z \cap U_Y$, has to be 
constant-valued function with $\varphi_Y (e_0) = 0$.
Thus, there is only one trivial holomorphic section, 
$\varphi_Z = \varphi_Y = 0$, and $h^0(E; {\cal O}_E(p - e_0)) = 0$. 

Let us now turn to $H^1(E; {\cal O}_E(p - e_0))$. 
Although the functions (\ref{eq:generator}) generate the cohomology 
group $H^1(E; {\cal O}_E)$, they are now decomposed into 
\begin{equation}
 \frac{y}{x - x_i} = \frac{y + y(p)}{x - x(p)} - 
   \frac{y (x(p) - x_i) + y(p) (x - x_i)}{(x - x(p))(x - x_i)} 
 =: \varphi_1 - \varphi_2.
\end{equation}
Here, $(x(p), y(p))$ are the coordinates of the point $p$.
$\varphi_1$ has poles of order one at $p$ and $e_0$, nowhere else.
Thus, $\varphi_1 \in C^0_Z$. $\varphi_2$ has poles of order one at 
$p$ and $p_i$, but $e_0$ is a zero of order one for $\varphi_2$. 
Thus, $\varphi_2 \in C^0_Y$. Thus, the functions (\ref{eq:generator})
do not generate the cokernel of $C^0 \rightarrow C^1$, 
and hence the cohomology group $H^1(E; {\cal O}_E(p-e_0))$ is trivial.
Note also that the decomposition $\varphi_1 \in C^0_Z$ and 
$\varphi_2 \in C^0_Y$ is actually unique. 

\subsection{$R^1 \pi_* {\cal O}(C - \sigma)$}

Fourier--Mukai transforms 
\begin{equation}
 R^1 p_{1*} \left[ p_2^*(\rho(V)) \otimes {\cal P}_B^{-1} \otimes 
   {\cal O}(- q^* K_{B_2})\right]
\end{equation}
for an elliptic fibration $p_1: Z \times_B Z \rightarrow Z$ and 
direct images
\begin{equation}
 R^1\pi_{Z*} \rho(V) 
\end{equation}
for an elliptic fibration $\pi_Z: Z \rightarrow B_2$ often vanishes 
apart from closed subsets $C_{\rho(V)} \hookrightarrow Z$ and 
$\bar{c}_{\rho(V)} \hookrightarrow B_2$. We address a question here, 
whether those sheaves with their support on closed subvarieties 
are expressed as pushforwards of sheaves of 
${\cal O}_{C_{\rho(V)}}$-modules and 
${\cal O}_{\bar{c}_{\rho(V)}}$-modules, respectively. 
 
We restrict our attention to cases where the support subvariety 
codimension-1 has its well-defined normal coordinate. Since the
question---whether the direct images are expressed as pushforwards
or not---is about a local property, the following argument is 
valid wherever a normal coordinate to the support is well defined 
locally. If the normal coordinate is well defined, then 
the question is answered by checking whether multiplication 
of the normal coordinate (seen as a function) upon the generators 
of direct images is trivial or not.

For a bundle $V$ given by spectral cover construction, 
$V$ is locally expressed as $\oplus_i (p_i - e_0)$, with 
$p_i$ varying over the coordinates of base manifold. 
$\rho(V)$ also share the same property. Multiplication of 
${\cal P}_B^{-1}$ in Fourier--Mukai transform does not change 
this structure, either. Thus, it is sufficient to deal with 
individual summand, all of which are of the form 
${\cal O}(C - \sigma)$ with $C$ describing a locus of $p_i$, 
varying over the base manifold. Because of the nature of 
our question we try to address, it is sufficient to 
maintain only the normal direction in the base manifold; 
a normal direction of $C_{\rho(V)}$ in $Z$, or 
that of $\bar{c}_{\rho(V)}$ in $B_2$. We will use $t$ 
as the coordinate of this transverse direction. 

$R^1 \pi_* {\cal O}(C - \sigma)$ is trivial around a 
point away from $C_{\rho(V)}$ or $\bar{c}_{\rho(V)}$.  
We only need to allow the coefficients in $C^0$ and $C^1$ in 
(\ref{eq:Cech-C0-E2}--\ref{eq:Cech-C1-E2}) to be holomorphic 
functions of $t$. Arguments there leading to 
$h^1(E; {\cal O}(p - e_0)) = 0$ does not need to be changed. 

On the support locus of $R^1 \pi_* {\cal O}(C - \sigma)$, 
$C \cdot \sigma$, things are different. Let us take 
an open set $U$ on the base manifold that contains some 
section of $C \cdot \sigma$. Then, 
\begin{equation}
 [R^1\pi_* {\cal O}(C - \sigma)] (U) = 
 H^1(\pi^{-1}(U); {\cal O}(C - \sigma)|_{\pi^{-1}(U)}).
\end{equation}
$\pi^{-1}(U)$ is covered by two open sets, $U_Z$ and $U_Y$; 
notion of $Z=0$ and $Y=0$ are well-defined in elliptic fibration. 
$C^0_Z$, for example, is given by 
\begin{equation}
 C^0_Z = {\rm Span}_{\C} \left\{ 1, x ,y , x^2 , \cdots \right\}\otimes 
   \C [t] + \varphi_1 \otimes (t \C [t] ).
\end{equation}
$\varphi_1$ is not contained in $C^0_Z$, because, as we see shortly, 
$v_{t=0}(\varphi_1) = -1$.  This is why only $t \times \C [t]$ 
are allowed as a coefficient of $\varphi_1$.
Similarly, $\varphi_2$ does not belong to $C^0_Y$ without being 
multiplied by the transverse coordinate $t$.
This means that 
\begin{equation}
 \frac{y}{x - x_i} = \varphi_1 - \varphi_2
\end{equation}
in $C^1$ is not expressed as an image from $C^0$;
the decomposition into $\varphi_1$ and $\varphi_2$ was unique, 
but neither $\varphi_1$ nor $\varphi_2$ belong to $C^0_Z$ 
or $C^0_Y$. Thus, these functions ($i=1,2,3$) generate 
the cohomology group $H^1(\pi^{-1}(U);{\cal O}(C - \sigma))$.
On the other hand, once it is multiplied by the transverse coordinate $t$, 
\begin{equation}
 t \times \frac{y}{x - x_i} = t \times \varphi_1 - t \times \varphi_2
\end{equation}
is in the image from $C^0$. Therefore, 
i) there is a non-vanishing direct image $R^1\pi_* {\cal O}(C - \sigma)$
that is localized on $C \cdot \sigma$, and 
ii) multiplication of the transverse coordinate $t$ annihilates it. 
Thus, the ideal sheaf of $C \cdot \sigma$ acts trivially on 
$R^1\pi_* {\cal O}(C - \sigma)$, and it is given by a pushforward 
of a sheaf on $C \cdot \sigma$ as a sheaf of ${\cal O}_{B_2}$-module. 
Furthermore, iii) the sheaf on $C \cdot \sigma$ is rank-1, 
(the same argument after (\ref{eq:generator}) is applied also here) 
and iv) since generator $y/(x - x_i)$ transforms like a section of 
${\cal L}^{\otimes 3}_H \otimes {\cal L}^{\otimes -2}_H \simeq {\cal L}_H$, 
its coefficient function transforms as 
${\cal L}^{-1}_H \simeq {\cal O}(K_{B_2})$.

We have yet to verify 
\begin{equation}
 v_{t=0}(\varphi_1) = v_{t=0}(\varphi_2) = -1.
\label{eq:pole}
\end{equation}
As a point $p$ approaches to $e_0$, $(x(p), y(p))$ goes to 
$(\infty,\infty)$. To see geometry of elliptic curve around the 
infinity point, it is better to use $(\xi, \zeta):= (X/Y, Z/Y)$ 
as the Affine coordinate system. The defining equation of 
elliptic curve becomes 
\begin{equation}
 \zeta = \xi^3 + f_0 \xi \zeta^2 + g_0 \zeta^3,
\end{equation}
and the infinity point $e_0$ corresponds to $(\xi,\zeta)=(0,0)$.
$\xi$ can be chosen as a local coordinate on $E$ around 
the infinity point $e_0$, and $\zeta \simeq \xi^3$ approximately.
Thus, as a point $p$ approaches $e_0$, 
\begin{eqnarray}
 x(p) & = & \left(\frac{\xi}{\zeta}\right)(p) \simeq \frac{1}{\xi(p)^2},
  \\
 y(p) & = & \frac{1}{\zeta(p)} \simeq \frac{1}{\xi(p)^3}.
\end{eqnarray}
By extracting the leading behavior of $\xi(p) \rightarrow 0$, 
we find that 
\begin{eqnarray}
 \varphi_1 & \simeq & - \frac{y(p)}{x(p)} \simeq - \frac{1}{\xi(p)}, \\
 \varphi_2 & \simeq & \frac{y(p)(x - x_i)}{x(p)(x-x_i)} \simeq 
    - \frac{1}{\xi(p)}.
\end{eqnarray}
Since we chose $t$ as the transverse coordinate to $C_{\rho(V)}$ 
or $\bar{c}_{\rho(V)}$, $\xi(p) \propto t$ around $C \cdot \sigma$.
Thus, we verified (\ref{eq:pole}).

\section{Appendices to Section \ref{sec:Examples}}
\label{sec:app2Examples}

\subsection{A Relation between $\chi(V)$ and $\chi(\wedge^2 V)$ 
}
In this section, prove \eqref{eq:conj-rel}, which gives a relation between $\chi(V)$ and $\chi(\wedge^2 V)$. This relation can be used as a consistency check of the computation, or as a shortcut to obtain $\chi(\wedge^2 V)$ from $\chi(V)$.

For a $\U(N)$ bundle $V$, one can show that  
\begin{eqnarray}
 {\rm c}_1 (\wedge^2 V)&=&(N-1) c_1(V),\\
 {\rm ch}_3 (V) & = & \frac{1}{2}c_3(V) - \frac{1}{2} c_2(V) c_1(V) 
 + \frac{1}{6}c_1(V)^3,  \\
 {\rm ch}_3 (\wedge^2 V) & = & (N-4) {\rm ch}_3 (V) - c_2(V) c_1(V)
  + \frac{1}{2} c_1(V)^3.
\end{eqnarray}
See e.g. an appendix B of \cite{Penn5}.

For a Calabi--Yau 3-fold $Z$ ($c_1(TZ)=0$), 
\begin{eqnarray}
 \chi(Z ; V) 
             & = & \int_Z {\rm ch}_3(V) + 
                   \int_Z \frac{c_2(TZ)}{12} c_1(V), \\
 \chi(Z; \wedge^2 V) 
        & = & \int_Z {\rm ch}_3(\wedge^2 V) +
         \int_Z \frac{c_2(TZ)}{12} c_1(\wedge^2 V) \nonumber \\
  & = & (N-4) \int_Z {\rm ch}_3(V) + (N-1)\int_Z \frac{c_2(TZ)}{12}
   c_1(V) \nonumber \\
  & & \qquad \qquad \qquad \qquad \qquad 
    - \int_Z c_2(V) c_1(V) + \frac{1}{2} \int_Z c_1(V)^3.
\end{eqnarray}
Since for elliptic fibration $\pi_Z: Z \rightarrow B$ we have
\begin{eqnarray}
 c_2(TZ) & = & \sigma \cdot 12 c_1(TB) + \cdots, \label{eq:ell-frmla-1} \\
 c_2(V) & = & \sigma \cdot \eta + \cdots, \label{eq:ell-frmla-2}\\
 c_1(V)^3 & = & 0 \qquad {\rm for~} c_1(V) = \pi^*_Z \pi_{C*} \gamma, 
   \label{eq:ell-frmla-3}
\end{eqnarray}
we finally obtain
\begin{eqnarray}
\chi(\wedge^2 V) = (N-4) \chi(V) + \left(3K_{B_2} + \eta\right) \cdot c_1(V),
\label{eq:conj-rel}
\end{eqnarray}
where $\chi(V) := - \chi(Z; V)$ and and 
$\chi(\wedge^2 V) := - \chi(Z; \wedge^2 V)$.

\subsection{Calculation of ${\rm deg} \; r|_D$ and ${\rm deg} \; j^* r$}
\label{ssec:r-D}

We present an explicit calculation of ${\rm deg} \; r|_D$ on 
$D$ from individual type (a) intersection points. In other words, 
we calculate the multiplicity of intersection of two curves 
$r$ and $D$ in the spectral surface $C_V$. 
The type (a) intersection points of $r$ and $D$ are also 
where the ramification divisor $r$ intersects the matter curve 
$\bar{c}_V = \sigma|_{C_V}$. We will also see below through 
explicit calculation that the multiplicity of the intersection 
of $r$ and $\bar{c}_V$ is 1, although that is already clear from 
an argument presented in the text. 

In order to find out the multiplicity of intersection of two 
curves on a surface, only local geometry of the surface 
matters. We will first describe local geometry of the spectral 
surface $C_V$ around a type (a) intersection point, and find 
the defining equations of the curves $r$, $D$ and $\bar{c}_V$. 
It is quite easy, then, to find out the multiplicity of intersection.

\subsubsection{For a ${\rm rank} \; V = 3$ Case}

Because the type (a) intersection points are always on the 
zero section $\sigma$, it is convenient to use the 
coordinates $(\xi,\zeta)$ in describing the direction 
of the elliptic fiber. Furthermore, since we focus 
on a local geometry of $C_V$, we can use $\zeta \sim \xi^3$.

Type (a) intersection points are found wherever both $a_N$
and $a_{N-1}$ vanish on the base 2-fold $B_2$. Here, 
$N := {\rm rank} \; V$. 
We choose a local patch of one of such points in $B_2$, and 
set coordinates in the patch so that $a_N = u$, and $a_{N-1} = -v$.
In a local patch of $Z$ with a set of coordinates $(u,v,\xi)$, 
the spectral surface of a rank-3 bundle $V$ is given by 
\begin{equation}
 u - v \xi + \xi^3 = 0,
\label{eq:local-eq-rk3}
\end{equation}
where we have set $a_0 = 1$ at the point; we do not lose 
generality by doing so, because the coordinates $u$ and $v$ 
can be rescaled if necessary. We study the ${\rm rank} \; V = 3$ 
case first. This defining equation was used when drawing the 
spectral surface $C_V$ in Figure~\ref{fig:rk3-CV}. 

\begin{figure}[t]
 \begin{center}
\begin{tabular}{cc}
\includegraphics[width=.3\linewidth]{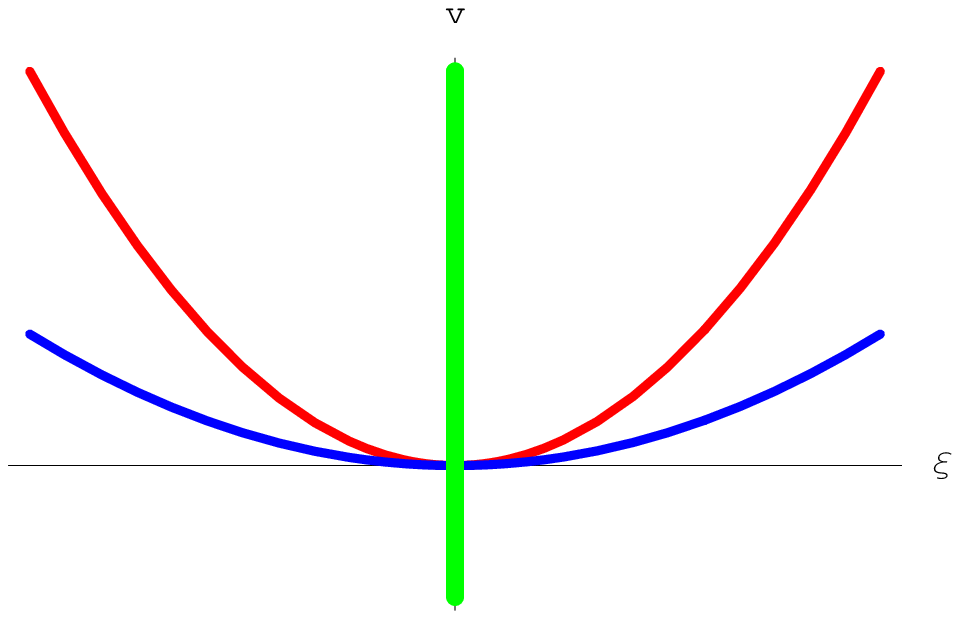} & 
\includegraphics[width=.3\linewidth]{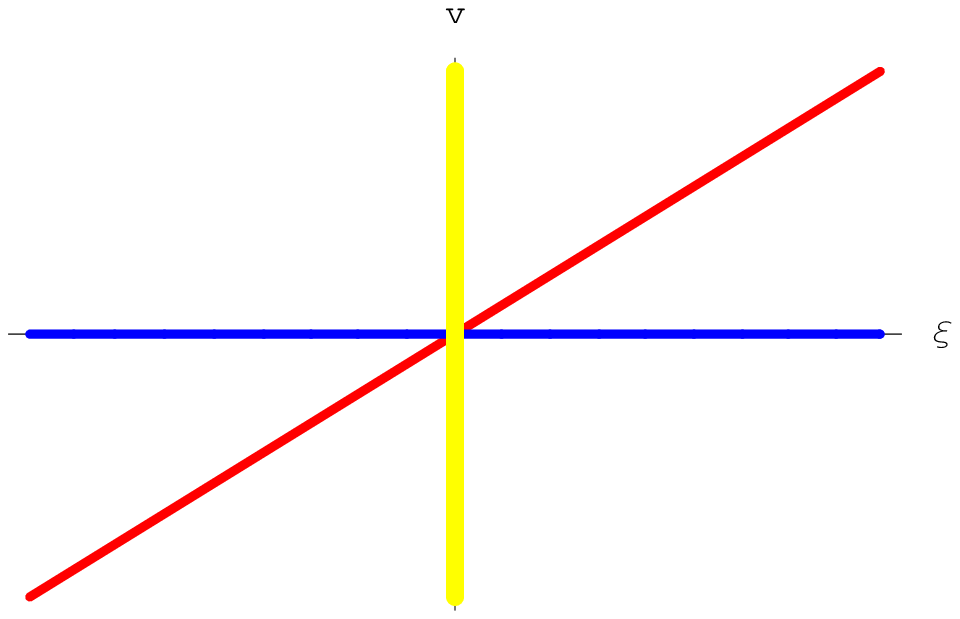} \\
${\rm rank} \; V = 3$ & ${\rm rank} \; V > 3$
\end{tabular}
 \caption{\label{fig:localonCV}This figure shows how the curves 
$r$, $D$ and $\bar{c}_V$ intersect on $C_V$ at a type (a) point. }
 \end{center}
\end{figure}

An appropriate choice of local coordinates on $C_V$ is $(\xi,v)$, 
whereas $(u,v)$ can be used for the base 2-fold $B_2$. 
The projection $\pi_C: C_V \rightarrow B_2$ 
is given by 
\begin{equation}
 \begin{array}{llll}
  \pi_C: & p  & \mapsto & b = \pi_C(p), \\
         & (\xi,v) & \mapsto & (u,v) = (v\xi - \xi^3,v).
 \end{array}
\end{equation}
Thus, 
\begin{equation}
 \pi_{C}^* (du \wedge dv) = (v - 3\xi^2) d\xi \wedge dv,
\end{equation}
and hence 
\begin{equation}
 r = {\rm div} \; (v-3\xi^2).
\end{equation}
Note also that 
\begin{equation}
 \pi_C^*(u) = \xi(v-\xi^2), \quad {\rm and} \quad 
 {\rm div} \; \pi_C^*(u) = {\rm div} \; \xi + {\rm div} \; (v-\xi^2)
 = \bar{c}_V + D.
\end{equation}
Thus, $r$ and $D$ intersects with multiplicity 2, while $r$ and 
$\bar{c}_V$ with multiplicity 1. See Figure~\ref{fig:localonCV}.

\subsubsection{For ${\rm rank} \; V \geq 4$ Cases}

For bundles with ${\rm rank} \; V \geq 4$, local defining 
equation of the spectral surface becomes 
\begin{equation}
 u - v \xi + \xi^2 = 0,
\end{equation}
where now local coordinates of $B_2$ around a type (a) point 
are chosen so that $a_N \sim u$ and $a_{N-1} \sim -v$, and 
we set $(a_{N-2} x^{(N-2)/2} + \cdots + a_0) \xi^{N-2} = 1$ 
at a type (a) intersection point without a loss of generality. 
The difference from (\ref{eq:local-eq-rk3}) is due to the 
fact that the $a_{N-2}$ term is absent in the equation determining 
the spectral surface of rank-3 bundles. The left panel of 
Figure~\ref{fig:rk4-CV} was drawn using the defining equation above.

We can use $(\xi,v)$ as the local coordinates on $C_V$, and 
the projection $\pi_C: C_V \rightarrow B_2$ is given by 
\begin{equation}
 \begin{array}{llll}
  \pi_C: & p & \mapsto & b = \pi_C(b), \\
         & (\xi,v) & \mapsto & (u,v) = (v \xi - \xi^2,v)
 \end{array}
\end{equation}
and the ramification divisor can be read out from 
\begin{equation}
 \pi_C^* (du \wedge dv) = (v - 2\xi) d \xi \wedge dv;
\end{equation}
now we have $r = {\rm div} \; (v-2\xi)$. 
Note also that $D = {\rm div} \; v$ and $\bar{c}_V = {\rm div} \; \xi$. 
Thus, the curves $r$ and $D$ intersect with multiplicity 1 at 
a type (a) intersection point for bundles with ${\rm rank} \; V > 3$. 
The two curves $\bar{c}_V$ and $r$ also intersect transversely.

\subsection{Geometry of $C_{\wedge^2 V}$ around the Pinch Points}
\label{ssec:pinch}

In this appendix, geometry of the spectral surface $C_{\wedge^2 V}$ 
for rank-4 bundles $V$ is discussed. 

For any point $b \in B_2$, the spectral surface $C_V$ for 
the bundle in the fundamental representation determines 
four points $\{ p_i, p_j, p_k, p_l\}$ in the elliptic fiber 
of $b$, $E_b$. Points on $B_2$ satisfying both 
$a_3 = 0$ and $R^{(4)}:= a_2^2 - 4 a_4 a_0 = 0$ are called 
type (c) points. We will focus on a local neighborhood $U \subset B_2$ 
of a type (c) point, and determine the behavior of $C_{\wedge^2 V}$
in $\pi_Z^{-1}(U) \subset Z$.

We can choose $\tilde{a}_3 \equiv a_3/a_4$ and 
$\tilde{R}^{(4)} \equiv R^{(4)}/a_4^2 = (a_2^2 - 4 a_4 a_0)/a_4^2$ 
as a set of local coordinates.
Then, the coordinates of the four points of $C_V$ in the 
fiber direction are determined as functions of the coordinates 
of the base manifold for small $(\tilde{a}_3, \tilde{R}^{(4)})$:
\begin{eqnarray}
 p_i: & &(x,y) \sim \left( 
  x_* + \frac{1}{2}\sqrt{\tilde{R}^{(4)} - 4 y_* \tilde{a}_3}, \; 
  + \left(y_* + \frac{3x_*^2 + f_0}{4y_*}
                   \sqrt{\tilde{R}^{(4)} -4y_* \tilde{a}_3}\right)\right), \\
 p_j: & &(x,y) \sim \left( 
  x_* + \frac{1}{2}\sqrt{\tilde{R}^{(4)} + 4 y_* \tilde{a}_3}, \; 
  - \left(y_* + \frac{3x_*^2 + f_0}{4y_*}
                   \sqrt{\tilde{R}^{(4)} +4y_* \tilde{a}_3}\right)\right), \\ 
 p_k: & &(x,y) \sim \left( 
  x_* - \frac{1}{2}\sqrt{\tilde{R}^{(4)} - 4 y_* \tilde{a}_3}, \; 
  + \left(y_* - \frac{3x_*^2 + f_0}{4y_*}
                   \sqrt{\tilde{R}^{(4)} -4y_* \tilde{a}_3}\right)\right), \\
 p_l: & &(x,y) \sim \left( 
  x_* - \frac{1}{2}\sqrt{\tilde{R}^{(4)} + 4 y_* \tilde{a}_3}, \;
  - \left(y_* - \frac{3x_*^2 + f_0}{4y_*}
                   \sqrt{\tilde{R}^{(4)} +4y_* \tilde{a}_3}\right)\right), 
\end{eqnarray}
where $p_i = p_k = (x_*,+ y_*)$ and $p_j = p_l = (x_*, -y_*)$ 
are the four points right on the type (c) point, and 
$x_* = - a_2/(2a_4)$ and $(y_*)^2 = x_*^3 + f_0 x_* + g_0$.
Only the leading order deviation from $(x_*, \pm y_*)$ for small 
$(\tilde{a}_3, \tilde{R}^{(4)})$ are maintained in the expressions above, and
higher order dependence on $(\tilde{a}_3, \tilde{R}^{(4)})$ is dropped.

Now that the coordinates in the fiber direction are given for the 
four points of $C_V$ in each fiber, one can determine the coordinates 
for the six points of $C_{\wedge^2 V}$ in each fiber. 
If two points $p_1$ and $p_2$ on an elliptic curve are given 
coordinate values $p_1 = (x_1, y_1)$ and $p_2 = (x_2, y_2)$, 
then the coordinates of their group-law sum 
$p_1 \boxplus p_2 = (x_{1 \boxplus 2}, y_{1 \boxplus 2})$ are given by
\begin{eqnarray}
 x_{1 \boxplus 2} & = & - x_1 - x_2 
   + \left(\frac{y_1 - y_2}{x_1 - x_2}\right)^2, \\
 y_{1 \boxplus 2} & = & - 
  \frac{(y_1 - y_2) x_{1 \boxplus 2} + (x_1 y_2 - x_2 y_1)}
       {(x_1 - x_2)}.
\end{eqnarray}
Using this addition theorem, one can calculate $(x,y)$ for 
all six $p_i \boxplus p_j$ $(1 \leq i < j \leq 4)$.

We know that two points $p_i \boxplus p_j$ and $p_k \boxplus p_l$ 
are on the zero section on the matter curve $\bar{c}_{\wedge^2 V}$
specified by $\tilde{a}_3 \propto a_3 = 0$. 
We also know that $p_k = p_i$ and 
$p_l = p_j$ on the type (c) points, and hence $p_i \boxplus p_l$ 
and $p_j \boxplus p_k$ are also on the zero section. 
We are interested in how those four points on $C_{\wedge^2 V}$ 
behave around the zero section in $\pi^{-1}_Z(U)$. 
For the purpose of describing geometry around the zero section,  
$\xi \sim x/y$ is more useful coordinate in the fiber direction 
than the pair $(x,y)$. After a little calculation, one finds that 
\begin{eqnarray}
 \xi(p_i \boxplus p_j) & \sim & 
   \frac{- \sqrt{\tilde{R}^{(4)} -4y_* \tilde{a}_3} 
         + \sqrt{\tilde{R}^{(4)} +4y_* \tilde{a}_3}}
        {4 y_*}, \\
 \xi(p_k \boxplus p_l) & \sim & 
   \frac{+ \sqrt{\tilde{R}^{(4)} -4y_* \tilde{a}_3} 
         - \sqrt{\tilde{R}^{(4)} +4y_* \tilde{a}_3}}
        {4 y_*}, \\
 \xi(p_i \boxplus p_l) & \sim & 
   \frac{- \sqrt{\tilde{R}^{(4)} -4y_* \tilde{a}_3} 
         - \sqrt{\tilde{R}^{(4)} +4y_* \tilde{a}_3}}
        {4 y_*}, \\
 \xi(p_j \boxplus p_k) & \sim & 
   \frac{+ \sqrt{\tilde{R}^{(4)} -4y_* \tilde{a}_3} 
         + \sqrt{\tilde{R}^{(4)} +4y_* \tilde{a}_3}}
        {4 y_*}, 
\end{eqnarray}
The geometry (i.e., $C_{\wedge^2 V}$) that those four points sweep 
is better parametrized by 
\begin{equation}
  w_{\pm} := 
 \sqrt{\tilde{R}^{(4)} \mp 4y_* \tilde{a}_3}
\end{equation}
With these two parameters, the spectral surface $C_{\wedge^2 V}$
is locally described by 
\begin{equation}
 \left(\xi, \tilde{a}_3 , \tilde{R}^{(4)} \right) \sim 
 \left( \frac{w_+ + w_-}{4 y_* }, 
  - \frac{w_+^2 - w_-^2}{8 y_*}, \frac{w_+^2 + w_-^2}{2}\right).
\end{equation}
The two panels in Figure~\ref{fig:rk4-CV} showing $C_{\wedge^2 V}$ 
were obtained in this way. The defining equation of $C_{\wedge^2 V}$ 
in the ambient space $Z$ with coordinates 
$(\xi, \tilde{a}_3, \tilde{R}^{(4)})$
is obtained by erasing $w_+$ and $w_-$. It is 
\begin{equation}
\tilde{a}_3^2 = \xi^2 \tilde{R}^{(4)} - 4 y_*^2 \xi^4.
\end{equation}
From this equation, one can see that $(\xi, \tilde{a}_3) = (0,0)$ 
is a double point in the $(\xi, \tilde{a}_3)$ plane for 
${}^\forall \tilde{R}^{(4)} \neq 0$, and hence the 
$(\xi, \tilde{a}_3) = (0,0)$ locus is a double curve. 
Furthermore, at the type (c) point, $\tilde{R}^{(4)} = 0$, 
$C_{\wedge^2 V}$ becomes more singular; it is called a pinch point.

On a generic point on the matter curve $\bar{c}_{\wedge^2 V}$, 
two branches of $C_{\wedge^2 V}$, $p_i \boxplus p_j$ and 
$p_k \boxplus p_l$ approach the zero section as 
$\xi \sim \pm \tilde{a}_3/\sqrt{\tilde{R}^{(4)}} \propto \tilde{a}_3$, 
where $\tilde{a}_3$ is the coordinate transverse to the matter curve. 
At a type (c) point, on the other hand, two points 
$p_i \boxplus p_j$ and $p_k \boxplus p_l$ approach as 
$\xi \propto \mp (w_+ - w_-)$, and two others 
$p_i \boxplus p_l$ and $p_k \boxplus p_j$ as 
$\xi \propto \mp (w_+ + w_-)$.
The transverse coordinate $\tilde{a}_3 \propto (w_+ + w_-)(w_+-w_-)$
contains both factors. It is necessary to know these behavior, 
when one tries to determine how the sheaf of 
${\cal O}_{B_2}$-module $R^1\pi_{Z*} \wedge^2 V$
responds to the action of the ideal sheaf of the matter curve
$\bar{c}_{\wedge^2 V}$.

Finally, let us study the geometry of $\widetilde{C}_{\wedge^2 V}$ 
introduced in section~\ref{sec:Idea}. $\widetilde{C}_{\wedge^2 V}$ 
is obtained by resolving two branches of $C_{\wedge^2 V}$ along 
the double-curve singularity into two disjoint components. 
Because we did not specify in section~\ref{sec:Idea} how to define
$\widetilde{C}_{\wedge^2 V}$ around the pinch points of 
$C_{\wedge^2 V}$, we will obtain $\widetilde{C}_{\wedge^2 V}$ 
along the double-curve singularity, and extrapolate it to the 
pinch points to see what happens there.

$\widetilde{C}_{\wedge^2 V}$ is obtained as a strict transform 
of $C_{\wedge^2 V}$, when the ambient space $Z$ is blown up 
with a center along the double curve singularity of $C_{\wedge^2 V}$.
%
%
Let $\tilde{Z}$ be the blow up of $Z$.
Since the double-curve locus is $(\xi, \tilde{a}_3) = (0,0)$, we consider 
a blow up of $Z$ centered at $(\xi,\tilde{a}_3) = (0,0)$.
The coordinate system $(\xi,u,v) \equiv (\xi, \tilde{a}_3, \tilde{R}^{(4)})$ 
of a local neighborhood 
$\pi^{-1}(U) \subset Z$ is replaced by those of two patches 
of $\nu_Z^{-1}(\pi^{-1}_Z(U)) \subset \tilde{Z}$, 
$(\xi, \tilde{u},v)$ in one patch and $(\tilde{\xi},u,v)$ in 
the other. The two patches are glued together under an 
identification 
\begin{equation}
 u = \xi \tilde{u}, \qquad \xi = u \tilde{\xi}.
\end{equation}
These relations also determine the map $\nu_Z: \tilde{Z} \rightarrow Z$. 

In the first patch of $\tilde{Z}$, the defining equation of 
$\widetilde{C}_{\wedge^2 V}$ becomes 
\begin{equation}
 \tilde{u}^2 = v - 4y_*^2 \xi^2.
\end{equation}
Although this blow-up was intended to resolve the double-curve 
singularity of $C_{\wedge^2 V}$ at $(\xi,u)=(0,0)$ and $v\neq 0$, 
this process also resolves the codimension-2 singularity at 
$(\xi,u,v) = (0,0,0)$; 
$\widetilde{C}_{\wedge^2 V}$ given by the equation above is 
smooth even at $(\xi,\tilde{u},v) = (0,0,0)$. We use the equation 
above as the definition of $\widetilde{C}_{\wedge^2 V}$ even 
at the pinch point $(\xi,u,v)=(0,0,0)$.

The zero section $\sigma \hookrightarrow Z$ is now replaced 
by $\sigma^* := \nu_Z^*(\sigma) \hookrightarrow \tilde{Z}$. 
It consists of two irreducible components. One of them does not 
appear in the first patch, because it is specified by 
$\tilde{\xi} = 1/\tilde{u} = 0$. The other component 
is the exceptional locus $E$ of this blow-up. 
It is given by $\xi = 0$. 
If the covering matter curve $\tilde{\bar{c}}_{\wedge^2 V}$ 
is defined by $\tilde{\bar{c}}_{\wedge^2 V} := 
\sigma^* \cdot \widetilde{C}_{\wedge^2 V}$, then it is specified by 
\begin{equation}
 \tilde{u}^2 = v, \qquad \xi = 0.
\end{equation}
$\tilde{u}$ can be chosen as a local coordinate of the covering matter 
curve $\tilde{\bar{c}}_{\wedge^2 V}$, while $v$ is the local coordinate 
of the matter curve $\bar{c}_{\wedge^2 V}$.
The map $\nu_{\bar{c}_{\wedge^2 V}}: \tilde{\bar{c}}_{\wedge^2 V}
\rightarrow \bar{c}_{\wedge^2 V}$ is clearly a degree-2 cover, 
$v = \tilde{u}^2$, and each type (c) point on 
$\bar{c}_{\wedge^2 V}$ $(v=0)$ is a branch point of this degree-2 cover.

\end{document}